\documentclass{amsart}
\usepackage{amsmath,amsfonts,amssymb,amsbsy,amscd,amsthm}
\usepackage{array}
\usepackage{bigstrut}
\usepackage{enumerate}
\usepackage{color}
\usepackage{latexsym}
\usepackage{url,hyperref}
\usepackage{fancybox}
\usepackage{lmodern}
\input xy
\xyoption{all} \xyoption{poly} \xyoption{knot}\xyoption{curve}
\def\sub{\begin{outline}\item}

\def\endsub{\end{outline}}
\DeclareMathAlphabet{\mathrm}    {OT1}{cmr}{m}{n}
\DeclareMathAlphabet{\mathrmbf}  {OT1}{cmr}{bx}{n}
\DeclareMathAlphabet{\mathrmit}  {OT1}{cmr}{m}{it}
\DeclareMathAlphabet{\mathrmbfit}{OT1}{cmr}{bx}{it}
\DeclareMathAlphabet{\mathsf}    {OT1}{cmss}{m}{n}
\DeclareMathAlphabet{\mathsfbf}  {OT1}{cmss}{bx}{n}
\DeclareMathAlphabet{\mathsfit}  {OT1}{cmss}{m}{sl}
\DeclareMathAlphabet{\mathttbf}  {OT1}{cmtt}{bx}{n}
\def\to{\rightarrow}
\def\To{\xrightarrow}
\def\from{\leftarrow}
\def\ZZ{{\mathbb Z}}
\def\RR{{\mathbb R}}

\def\ss{\subseteq}
\def\taking{\colon}

\def\To{\xrightarrow}
\def\From{\xleftarrow}

\def\cross{\times}
\def\Ob{{\bf Ob}}

\def\Set{{\bf Set}}

\def\id{\tn{id}}

\def\Cat{{\bf Cat}}

\def\im{{\bf im}}

\def\mcC{{\mathcal C}}
\def\mcD{{\mathcal D}}

\def\mcS{{\mathcal S}}

\def\surj{\twoheadrightarrow}
\def\inj{\hookrightarrow}
\def\rr{\raggedright}
\newcommand{\LMO}[1]{\bullet^{#1}}

\newcommand{\LA}[2]{\ar[#1]^-{\tn {#2}}}
\newcommand{\LAL}[2]{\ar[#1]_-{\tn {#2}}}
\newcommand{\obox}[3]{\stackrel{#1}{\fbox{\parbox{#2}{#3}}}}
\newcommand{\labox}[2]{\obox{#1}{1.6in}{#2}}
\newcommand{\mebox}[2]{\obox{#1}{1in}{#2}}
\newcommand{\smbox}[2]{\stackrel{#1}{\fbox{#2}}}
\newcommand{\fakebox}[1]{\tn{$\ulcorner$#1$\urcorner$}}
\newcommand{\sq}[4]{\xymatrix{#1\ar[r]\ar[d]&#2\ar[d]\\#3\ar[r]&#4}}

\newcommand{\comment}[1]{}
\def\hsp{\hspace{.2in}}
\def\lcone{^\triangleleft}
\def\rcone{^\triangleright}
\DeclareMathOperator{\colim}{colim}

\def\edge{\ar@{-}}
\def\ullimit{\ar@{}[rd]|(.3)*+{\lrcorner}}
\def\urlimit{\ar@{}[ld]|(.3)*+{\llcorner}}
\def\lllimit{\ar@{}[ru]|(.3)*+{\urcorner}}
\def\lrlimit{\ar@{}[lu]|(.3)*+{\ulcorner}}
\newtheorem{theorem}{Theorem}[subsection]

\theoremstyle{remark}
\newtheorem{remark}[theorem]{Remark}
\newtheorem{example}[theorem]{Example}
\newtheorem{warning}[theorem]{Warning}

\newtheorem{rules}[theorem]{Rules of good practice}
\theoremstyle{definition}
\newtheorem{definition}[theorem]{Definition}

\def\tn{\textnormal}

\begin{document}
\title{Ologs: a categorical framework for knowledge representation}
\author{David I. Spivak}
\address{Mathematics, MIT, Cambridge, MA 02139}
\email{dspivak@math.mit.edu}

\author{Robert E. Kent}
\address{Ontologos}
\email{rekent@ontologos.org}

\thanks{This project was supported by Office of Naval Research grant: N000141010841 and a generous contribution by Clark Barwick, Jacob Lurie, and the Massachusetts Institute of Technology Department of Mathematics}

\begin{abstract}
In this paper we introduce the olog, or ontology log, 
a category-theoretic model for knowledge representation (KR). 
Grounded in formal mathematics, 
ologs can be rigorously formulated and cross-compared in ways that other KR models (such as semantic networks) cannot. 
An olog is similar to a relational database schema; 
in fact an olog can serve as a data repository if desired. 
Unlike database schemas, which are generally difficult to create or modify, 
ologs are designed to be user-friendly enough 
that authoring or reconfiguring an olog is a matter of course rather than a difficult chore. 
It is hoped that learning to author ologs is much simpler than learning a database definition language, 
despite their similarity. 
We describe ologs carefully and illustrate with many examples. 
As an application we show that any primitive recursive function can be described by an olog. 
We also show that ologs can be aligned or connected together into a larger network using functors. 
The various methods of information flow and institutions can then be used to integrate local and global world-views. 
We finish by providing several different avenues for future research.
\end{abstract}

\maketitle
\tableofcontents
\raggedbottom


\section{Introduction}\label{sec:intro}

Scientists have a pressing need to organize their experiments, their data, their results, and their conclusions into a framework such that this work is reusable, transferable, and comparable with the work of other scientists. In this paper, we will discuss the ``ontology log" or {\em olog} as a possibility for such a framework. Ontology is the study of what something {\em is}, i.e the nature of a given subject, and ologs are designed to record the results of such a study. The structure of ologs is based on a branch of mathematics called category theory. An olog is roughly a category that models a given real-world situation. 

The main advantages of authoring an olog rather than writing a prose description of a subject are that \begin{itemize}\item an olog gives a precise formulation of a conceptual world-view,\item an olog can be formulaically converted into a database schema,\item an olog can be extended as new information is obtained,\item an olog written by one author can be easily and precisely referenced by others,\item an olog can be input into a computer and ``meaningfully stored", and\item different ologs can be compared by functors, which in turn  generate automatic terminology translation systems.\end{itemize}  The main disadvantage to using ologs over prose, aside from taking more space on the page, is that writing a good olog demands a clarity of thought that ordinary writing or conversation can more easily elide. However, the contemplation required to write a good olog about a subject may have unexpected benefits as well.

A category is a mathematical structure that appears much like a directed graph: it consists of objects (often drawn as nodes or dots, but here drawn as boxes) and arrows between them. The feature of categories that distinguishes them from graphs is the ability to declare an equivalence relation on the set of paths. A functor is a mapping from one category to another that preserves the structure (i.e. the nodes, the arrows, and the equivalences). If one views a category as a kind of language (as we shall in this paper) then a functor would act as a kind of translating dictionary between languages. There are many good references on category theory, including \cite{LS}, \cite{Sic}, \cite{Pie}, \cite{BW1}, \cite{Awo}, and \cite{Mac}; the first and second are suited for general audiences, the third and fourth are suited for computer scientists, and the fifth and sixth are suited for mathematicians (in each class the first reference is easier than the second).

A basic olog, defined in Section \ref{sec:basic ologs}, is a category in which the objects and arrows have been labeled by English-language phrases that indicate their intended meaning. The objects represent types of things, the arrows represent functional relationships (also known as aspects, attributes, or observables), and the commutative diagrams represent facts. Here is a simple olog about an amino acid called arginine (\cite{W1}):  \begin{align}\label{dia:arginine}\fbox
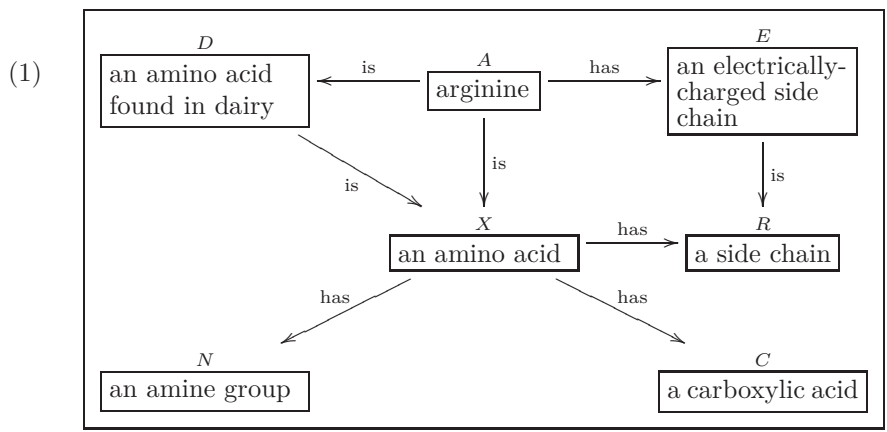
{\xymatrix{\obox{D}{1in}{\rr an amino acid found in dairy}\LAL{dr}{is}&\obox{A}{.5in}{arginine}\LA{r}{has}\LAL{l}{is}\LA{d}{is}&\obox{E}{.9in}{\rr an electrically-charged side chain}\LA{d}{is}\\&\obox{X}{.9in}{an amino acid}\LAL{dl}{has}\LA{dr}{has}\LA{r}{has}&\smbox{R}{a side chain}\\\mebox{N}{an amine group}&&\mebox{C}{a carboxylic acid}}}\end{align}  

The idea of representing information in a graph is not new. For example the Resource Descriptive Framework (RDF) is a system for doing just that \cite{CM}. The key difference between a category and a graph is the consideration of paths, and that two paths from $A$ to $B$ may be declared identical in a category (see \cite{Spi-Cats}). For example, we can further declare that in Diagram (\ref{dia:arginine}), the diagram \begin{align}\label{dia:comm sq}\sq{A}{E}{X}{R}\end{align} {\em commutes}, i.e. that the two paths $\xymatrix@1{A\ar@/^.3pc/[r]\ar@/_.3pc/[r]&R}$ are equivalent, which can be translated as follows. Let $A$ be a molecule of arginine. On the one hand $A$, being an amino acid, has a side chain; on the other hand $A$ has an electrically-charged side-chain, which is of course a side chain. We seem to have associated {\em two} side-chains to $A$, but in fact they both refer to the same physical thing, the same side-chain. Thus, the two paths $A\to R$ are deemed equivalent. The fact that this equivalence may seem trivial is not an indictment of the category idea but instead reinforces its importance --- we must be able to indicate obvious facts within a given situation because what is obvious is the most essential.

While many situations can be modeled using basic ologs (categories), we often need to encode more structure.  For this we will need so-called sketches. An olog will be defined as a finite limit, finite colimit sketch (see \cite{BW2}), meaning we have the ability to encode objects (``types"), arrows (``aspects"), commutative diagrams (``facts"), as well as finite limits (``layouts") and finite colimits (``groupings").

Throughout this paper, whenever we refer to ``the author" of an olog we am referring to the fictitious person who created it. We will refer to ourselves, David Spivak and Robert Kent, as ``we" so as not to confuse things. 

\begin{warning}\label{warn:world-view}

The author of an olog has a world-view, some fragment of which is captured in the olog. When person A examines the olog of person B, person A may or may not ``agree with it."  For example, person B may have the following olog $$\fbox
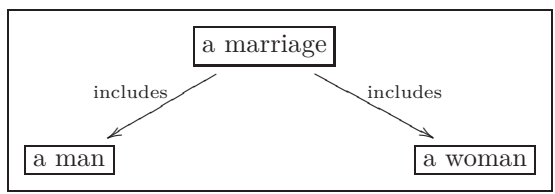
{\xymatrix{&\fbox{a marriage}\LA{dr}{ includes}\LAL{dl}{includes }\\\fbox{a man}&&\fbox{a woman}}}$$ which associates to each marriage a man and a woman. Person A may take the position that some marriages involve two men or two women, and thus see B's olog as ``wrong."  Such disputes are not ``problems" with either A's olog or B's olog, they are discrepancies between world-views. Hence, throughout this paper, a reader R may see a displayed olog and notice a discrepancy between R's world-view and our own, but R should not worry that this is a problem. This is not to say that ologs need not follow rules, but instead that the rules are enforced to ensure that an olog is structurally sound, rather than that it ``correctly reflects reality," whatever that may mean.

\end{warning}

\subsection{Plan of this paper}

In this paper, we will define ologs and give several examples. We will state some rules of ``good practice" which help one to author ologs that are meaningful to others and easily extendable. We will begin in Section \ref{sec:basic ologs} by laying out the basics: types as objects, aspects as arrows, and facts as commutative diagrams. In Section \ref{sec:instances}, we will explain how to attach ``instance" data to an olog and hence realize ologs as database schemas. In Section \ref{sec:connecting ologs}, we will discuss meaningful constraints betweeen ologs that allow us to develop a higher-dimensional web of information called an information system, and we will discuss how the various parts of such a system interact via information channels. In Sections \ref{sec:expressive I} and \ref{sec:expressive II}, we will extend the olog definition language to include ``layouts" and ``groupings", which make for more expressive ologs; we will also describe two applications, one which explicates the computation of the factorial function, and the other which defines a notion from pure mathematics (that of pseudo-metric spaces). Finally, in Section \ref{sec:further}, we will discuss some possible directions for future research.

For the remainder of the present section, we will explain how ologs relate to existing ideas in the field of knowledge representation.

\subsection{The semantic advantage of ologs: modularity}

The difference between ologs and prose is modularity: small conceptual pieces can form large ideas, and these pieces work best when they are reusable. The same phenomenon is true throughout computer science and mathematics. In programming languages, modularity brings not only vast efficiency to the writing of programs but enables an ``abstraction barrier" that keeps the ideas clean. In mathematics, the most powerful results are often simple lemmas that are reusable in a wide variety of circumstances. 

Web pages that consist of prose writing are often referred to as {\em information silos.}  The idea is that a silo is a ``big tube of stuff" which is not organized in any real way. Links between web pages provide some structure, but such a link does not carry with it a precise method to correlate the information within the two pages. Similarly in science, one author may reference another paper, but such a reference carries very little structure --- it just points to a silo.   

Ologs can be connected with links which are much richer than the link between two silos could possibly be. Individual concepts and connections within one olog can be ``functorially aligned" with concepts and connections in another. A functor creates a precise connection between the work of one author and the work of another so that the precise nature of the comparison is not left to the reader's imagination but explicitly specified. The ability to incorporate mathematical precision into the sharing of ideas is a central feature of ologs.

\subsection{Relation to other models}

There are many languages for knowledge representation (KR). For example, there are database languages such as SQL, ontology languages such as RDF and OWL, the language of Semantic Nets, and others (see \cite{Bor}). One may ask what makes the olog concept different or better than the others. 

The first response is that ologs are closely related to the above ideas. Indeed, all of these  KR models can be ``categorified" (i.e. phrased in the language of category theory) and related by functors, so that many of the ideas align and can be transferred between the different systems. In fact, as we will make clear in Section \ref{sec:instances}, ologs are almost identical to the categorical model of databases presented in \cite{Spi-FDM}. 

However, ologs have advantages over many existing KR models. The first advantage arises from the notion of commutative diagrams (which allow us to equate different paths through the domain, see Section \ref{sec:facts}) and of limits and colimits (which allow us to lay out and group things, see Sections \ref{sec:expressive I} and \ref{sec:expressive II}). The additional expressivity of ologs give them a certain semantic clarity and interoperability that cannot be achieved with graphs and networks in the usual sense. 
The second advantage arises from the notion of olog morphisms, 
which allow the definition of meaningful constraints between ologs. With this in hand, we can integrate a set of similar ologs into a single information system, and go on to define information fusion. This will be discussed further Section \ref{sec:connecting ologs}.

In the remainder of this section we will provide a few more details on the relationship between ologs and each of the above KR models: databases, RDF/OWL, and semantic nets. The reader who does not know or care much about other systems of knowledge representation can skip to Section \ref{sec:acknowledgements}.

\subsubsection{Ologs and Databases}

A database is a system of tables, each table of which consists of a header of columns and a set of rows. A table represents a type of thing $T$, each column represents an attribute of $T$, and each row represents an example of $T$. An attribute is itself a ``type of thing", so each column of a table points to another table. 

The relationship between ologs and databases is that every box $B$ in an olog represents a type of thing and every arrow $B\to X$ emanating from $B$ represents an attribute of $B$ (whose results are of type $X$). Thus the boxes and arrows in an olog correspond to tables and their columns in a database. The rows of each table in a database will correspond to ``instances" of each type in an olog. Again, this will be made more clear in Section \ref{sec:instances} or one can see \cite{Spi-FDM} or \cite{K:DBS}. 

The point is that every olog can serve as a database schema, and the schemas represented by ologs range from simple (just objects and arrows) to complex (including commutative diagrams, products, sums, etc.). However, whereas database schemas are often prescriptive (``you must put your data into this format!"), ologs are usually descriptive (``this is how I see things"). One can think of an olog as an interface between people and databases: an olog is human readable, but it is also easily converted to a database schema upon which powerful applications can be put to work. Of course, if one is to use an olog as a database schema, it will become prescriptive. However, since the intention of each object and arrow is well-documented (as its label), schema evolution would be straightforward. Moreover, the categorical structure of ologs allows for {\em functorial data migration} by which one can transfer the instance data from an older schema to the current one (see \cite{Spi-FDM}).

\subsubsection{Ologs and RDF / OWL}

In \cite{Spi-FDM}, the first author explained how a categorical database can be converted into an RDF triple store using the Grothendieck construction. The main difference between a categorical database schema (or an olog) and an RDF schema is that one cannot specify commutativity in an RDF schema. Thus one cannot express things like ``the woman parent of a person $x$ is the mother of $x$."  Without this expressivity, it is hard to enforce much rigor, and thus RDF data tends to be too loose for many applications. 

OWL schemas, on the other hand, can express many more constraints on classes and properties. We have not yet explored the connection, nor compared the expressive power, of ologs and OWL. However, they are significantly different systems, most obviously in that OWL relies on logic where ologs rely on category theory. 

\subsubsection{Semantic Nets}

On the surface, ologs look the most like semantic networks, or concept webs, but there are important differences between the two notions. First, arrows in a semantic network need not indicate functions; they can be relations. So there could be an arrow \fakebox{a father}$\To{\tn{has}}$\fakebox{a child} in a semantic network, but not in an olog (see Section \ref{sec:relations} for how the same idea is expressible in an olog). There is a nice category of sets and relations, often denoted {\bf Rel}, but this category is harder to reason about than is the ordinary category of sets and functions (often denoted $\Set$). Thus, as mentioned above, semantic networks are categorifiable (using {\bf Rel}), but this underlying formalism does not appear to play a part in the study or use of semantic networks. However, some attempt to integrate category theory and neural nets has been made, see \cite{HC}.

Moreover, commutative diagrams and other expressive abilities held by ologs are not generally part of the semantic network concept (see \cite{Sow}). For these reasons, semantic networks tend to be brittle: minor changes can have devastating effects. For example, if two semantic networks are somehow synced up and then one is changed, the linkage must be revised or may be altogether broken. Such a disaster is often avoided if one uses categories: because different paths can be equivalent, one can simply add new ideas (types and aspects) without changing the semantic meaning of what was already there.
As section \ref{sec:CG} demonstates with an extended example,
conceptual graphs, which are a popular formalism for semantics nets,
can be linearized to ologs, thereby gaining in precision and expressibility.

\subsection{Acknowledgements}\label{sec:acknowledgements}

\subsubsection{David Spivak's acknowledgments}

I would like to thank Mathieu Anel and Henrik Forssell for many pleasant and quite useful conversations. I would also like to thank Micha Breakstone for his help on understanding the relationship between ologs and linguistics. Finally I would like to thank Dave Balaban for helpful suggestions on this document itself.

\subsubsection{Robert Kent's acknowledgments}

I would like to thank the participants in the Standard Upper Ontology working group
for many interesting, spirited, rewarding and enlightening discussions
about knowledge representation in general and ontologies in particular;
I especially want to thank Leo Obrst, Marco Schorlemmer and John Sowa from that group.
I want to thank Jon Barwise 
for leading the development of the theory of information flow.
I want to thank Joseph Goguen 
for leading the development of the theory of institutions,
and for pointing out the common approach to knowledge representation
used by both the Information Flow Framework and the theory of institutions.


\section{Types, aspects, and facts}\label{sec:basic ologs}

In this section we will explain basic ologs, which involve types, aspects, and facts. A basic olog is a category in which each object and arrow has been labeled by text; throughout this paper we will assume that text to be written in English. 

The purpose of this section is to show how one can convert a real-world situation into an olog. It is probably impossible to explain this process precisely in words. Instead, we will explain mainly by example. We will give ``rules of good practice" that lead to good ologs. While these rules are not strictly necessary, they help to ensure that the olog is properly formulated. As the Dalai Lama says, ``Learn the rules so you know how to break them properly."

\subsection{Types}

A type is an abstract concept, a distinction the author has made. We represent each type as a box containing a {\em singular indefinite noun phrase.}   Each of the following four boxes is a type: \begin{align}\label{dia:types}\xymatrix{\fbox{a man}&\fbox{an automobile}\\\obox{}{1.5in}{a pair $(a,w)$, where $w$ is a woman and $a$ is an automobile}&\obox{}{1.5in}{a pair $(a,w)$ where $w$ is a woman and $a$ is a blue automobile owned by $w$}}\end{align}

Each of the four boxes in (\ref{dia:types}) represents a type of thing, a whole class of things, and the label on that box is what one should call {\em each example} of that class. Thus \fakebox{a man} does not represent a single man, but the set of men, each example of which is called ``a man"\footnote{In other words, types in ologs are intentional, rather than extensional --- the label on a type describes its intention. The extension of a type will be captured by {\em instance data}; see Section \ref{sec:instances}\;.}. Similarly, the bottom right-hand box in (\ref{dia:types}) represents an abstract type of thing, which probably has more than a million examples, but the label on the box indicates a common name for each such example. 

Typographical problems emerge when writing a text-box in a line of text, e.g. the text-box \fbox{a man} seems out of place here, and the more in-line text-boxes one has in a given paragraph, the worse it gets. To remedy this, we will denote types which occur in a line of text with corner-symbols, e.g. we will write \fakebox{a man} instead of \fbox{a man}.


\subsubsection{Types with compound structures}

Many types have compound structures; i.e. they are composed of smaller units. Examples include \begin{align}\label{dia:compound}\xymatrix{\obox{}{.7in}{\rr a man and a woman}&\obox{}{1.4in}{\rr a food $f$ and a child $c$ such that $c$ ate all of $f$}&\labox{}{a triple $(p,a,j)$ where $p$ is a paper, $a$ is an author of $p$, and $j$ is a journal in which $p$ was published}}\end{align}  It is good practice to declare the variables in a ``compound type", as we did in the last two cases of (\ref{dia:compound}). In other words, it is preferable to replace the first box above with something like $$\obox{}{.8in}{a man $m$ and a woman $w$}\hsp\tn{or}\hsp\obox{}{1.1in}{\rr a pair $(m,w)$ where $m$ is a man and $w$ is a woman}$$ so that the variables $(m,w)$ are clear.

\begin{rules}\label{rules:types}

A type is presented as a text box. The text in that box should 
\begin{enumerate}[(i)]\item begin with the word ``a" or ``an";\item refer to a distinction made and recognizable by the author;\item refer to a distinction for which instances can be documented;\item not end in a punctuation mark;\item declare all variables in a compound structure. \end{enumerate}

\end{rules}

The first, second, and third rules ensure that the class of things represented by each box appears to the author as a well-defined set; see Section \ref{sec:instances} for more details. The fourth and fifth rules encourage good ``readability" of arrows, as will be discussed next in Section \ref{sec:aspects}. 

We will not always follow the rules of good practice throughout this document. We think of these rules being followed ``in the background" but that we have ``nicknamed" various boxes. So \fakebox{Steve} may stand as a nickname for \fakebox{a thing classified as Steve} and \fakebox{arginine} as a nickname for \fakebox{a molecule of arginine}.

\subsection{Aspects}\label{sec:aspects}

An aspect of a thing $x$ is a way of viewing it, a particular way in which $x$ can be regarded or measured. For example, a woman can be regarded as a person; hence ``being a person" is an aspect of a woman. A man has a height (say, taken in inches), so ``having a height (in inches)" is an aspect of a man. In an olog, an aspect of $A$ is represented by an arrow $A\to B$, where $B$ is the set of possible ``answers" or results of the measurement. For example when observing the height of a man, the set of possible results is the set of integers, or perhaps the set of integers between 20 and 120. \begin{align}\label{dia:aspect 1}\xymatrix{\fbox{a woman}\LA{r}{is}&\fbox{a person}}\end{align}\begin{align}\label{dia:aspect 2}\xymatrix{\fbox{a man}\LA{rrr}{has as height (in inches)}&&&\fbox{an integer between 20 and 120}}\end{align}

We will formalize the notion of aspect by saying that aspects are functional relationships.\footnote{In type theory, what we here call aspects are called {\em functions}. Since our types are not fixed sets (see Section \ref{sec:instances}), we preferred a term that was less formal.} Suppose we wish to say that a thing classified as $X$ has an aspect $f$ whose result set is $Y$.  This means there is a functional relationship called $f$ between $X$ and $Y$, which can be denoted $f\taking X\to Y$.  We call $X$ the {\em domain of definition} for the aspect $f$, and we call $Y$ the {\em set of result values} for $f$. For example, a man has a height in inches whose result is an integer, and we could denote this by $h\taking M\to{\bf Int}$. Here, $M$ is the domain of definition for height and ${\bf Int}$ is the set of result values. 

A set may always be drawn as a blob with dots in it. If $X$ and $Y$ are two sets, then a {\em a function from $X$ to $Y$}, denoted $f\taking X\to Y$ can be presented by drawing arrows from dots in blob $X$ to dots in blob $Y$. There are two rules: \begin{enumerate}[(i)]\item each arrow must emanate {\em from} a dot in $X$ and point {\em to} a dot in $Y$;\item each dot in $X$ must have precisely {\em one} arrow emanating from it.\end{enumerate}  Given an element $x\in X$, the arrow emanating from it points to some element $y\in Y$, which we call {\em the image of $x$ under $f$} and denote $f(x)=y$. 

Again, in an olog, an aspect of a thing $X$ is drawn as a labeled arrow pointing from $X$ to a ``set of result values."   Let us concentrate briefly on the arrow in (\ref{dia:aspect 1}). The domain of definition is the set of women (a set with perhaps 3 billion elements); the set of result values is the set of persons (a set with perhaps 6 billion elements).  We can imagine drawing an arrow from each dot in the ``woman" set to a unique dot in the ``person" set.  No woman points to two different people, nor to zero people --- each woman is exactly one person --- so the rules for a functional relationship are satisfied. Let us now concentrate briefly on the arrow in (\ref{dia:aspect 2}). The domain of definition is the set of men, the set of result values is the set of integers $\{20,21,22,\ldots,119,120\}$. We can imagine drawing an arrow from each dot in the ``man" set to a single dot in the ``integer" set. No man points to two different heights, nor can a man have no height: each man has exactly one height. Note however that two different men can point to the same height.

\subsubsection{Invalid aspects}

We tried above to clarify what it is that makes an aspect ``valid", namely that it must be a ``functional relationship."  In this subsection we will present two arrows which on their face may appear to be aspects, but which on closer inspection are not functional (and hence are not valid as aspects). 
 
Consider the following two arrows: \begin{align}\tag{7*}\xymatrix{\fbox{a person}\LA{r}{has}&\fbox{a child}}\end{align}\vspace{-.13in}\begin{align}\tag{8*}\xymatrix{\fbox{a mechanical pencil}\LA{r}{uses}&\fbox{a piece of lead}}\end{align}\setcounter{equation}{8}  A person may have no children or may have more than one child, so the first arrow is invalid: it is not functional because it does not satisfy rule (2) above. Similarly, if we drew an arrow from each mechanical pencil to each piece of lead it uses, it would not satisfy rule (2) above. Thus neither of these is a valid aspect.

Of course, in keeping with Warning \ref{warn:world-view}, the above arrows may not be wrong but simply reflect that the author has a strange world-view or a strange vocabulary. Maybe the author believes that every mechanical pencil uses exactly one piece of lead. If this is so, then $\fakebox{a mechanical pencil}\To{\tn{uses}}\fakebox{a piece of lead}$ is indeed a valid aspect!   Similarly, suppose the author meant to say that each person {\em was once} a child, or that a person has an inner child. Since every person has one and only one inner child (according to the author), the map $\fakebox{a person}\To{\tn{has as inner child}}\fakebox{a child}$ is a valid aspect. We cannot fault the author for such a view, but note that we have changed the name of the label to make its intention more explicit.

\subsubsection{Reading aspects and paths as English phrases}

Each arrow (aspect) $X\To{f} Y$ can be read by first reading the label on its source box (domain of definition) $X$, then the label on the arrow $f$, and finally the label on its target box (set of result values) $Y$. For example, the arrow \begin{align}\label{dia:first author}\fbox{\xymatrix{\smbox{}{a book}\LA{rrr}{has as first author}&&&\smbox{}{a person}}}\end{align} is read ``a book has as first author a person", a valid English sentence.

\comment{

\begin{remark}

Note that the map in (\ref{dia:first author}) is a valid aspect, but that a similarly benign-looking map $\fakebox{a book}\To{\tn{has as author}}\fakebox{a person}$ would not be valid, because it is not functional. The authors of an olog must be vigilant about this type of mistake because it is easy to miss and it can corrupt the olog.

\end{remark}

}

Sometimes the label on an arrow can be shortened or dropped altogether if it is obvious from context. We will discuss this more in Section \ref{sec:facts} but here is a common example from the way we write ologs. \begin{align}\label{dia:pair of integers}\fbox
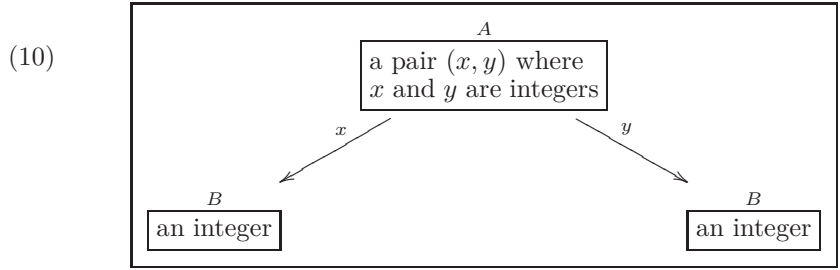
{\xymatrix{&\obox{A}{1.2in}{\rr a pair $(x,y)$ where $x$ and $y$ are integers}\ar[dl]_x\ar[dr]^y\\\smbox{B}{an integer}&&\smbox{B}{an integer}}}\end{align}  Neither arrow is readable by the protocol given above (e.g. ``a pair $(x,y)$ where $x$ and $y$ are integers $x$ an integer" is not an English sentence), and yet it is obvious what each map means. For example, given the pair $(8,11)$ which belongs in box $A$, application of arrow $x$ would yield $8$ in box $B$. The label $x$ can be thought of as a nickname for the full name ``yields, via the value of $x$," and similarly for $y$. We do not generally use the full name for fear that the olog would become cluttered with text.

\comment{ 

\begin{remark}

Unlabeled arrows can be generically labeled ``is functionally assigned."  For example one could read \fakebox{a hammer}$\to$\fakebox{a manufacturer} as ``a hammer is functionally assigned a manufacturer". However, if there are many unlabeled arrows in a given olog then the author must somehow differentiate between them. For this reason it is good practice to formally assign such an arrow the label ``is functionally assigned by $m$" where $m$ is some specific integer. This will be made precise in Definition \ref{def:basic olog}.

\end{remark}

} 

One can also read paths through an olog by inserting the word ``which" after each intermediate box. For example the following olog has two paths of length 3 (counting arrows in a chain): \small\begin{align}\label{olog:paths}\fbox
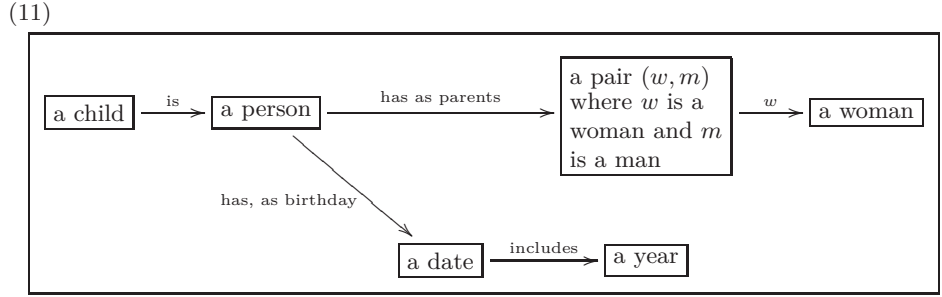
{\xymatrix{\fbox{a child}\LA{r}{is}&\fbox{a person}\LA{rr}{has as parents}\LAL{dr}{has, as birthday}&&\obox{}{.8in}{\rr a pair $(w,m)$ where $w$ is a woman and $m$ is a man}\LA{r}{$w$}&\fbox{a woman}\\&&\fbox{a date}\LA{r}{includes}&\fbox{a year}}}\end{align}  \normalsize The top path is read ``a child is a person, which has as parents a pair $(w,m)$ where $w$ is a woman and $m$ is a man, which yields, via the value of $w$, a woman."  The reader should read and understand the content of the bottom path.

\subsubsection{Converting non-functional relationships to aspects}\label{sec:relations}

There are many relationships that are not functional, and these cannot be considered aspects. Often the word ``has" indicates a relationship --- sometimes it is functional as in $\fakebox{a person}\To{\tn{ has }}\fakebox{a stomach}$, and sometimes it is not, as in $\fakebox{a father}\To{\tn{has}}\fakebox{a child}$. (Obviously, a father may have more than one child.)  A quick fix would be to replace the latter by $\fakebox{a father}\To{\tn{has}}\fakebox{a set of children}$.  This is ok, but the relationship between \fakebox{a child} and \fakebox{a set of children} then becomes an issue to deal with later. There is another way to indicate such ``non-functional" relationships.

In mathematics, a relation between sets $A_1, A_2$, and so on through $A_n$ is defined to be a subset of the Cartesian product $$R\ss A_1\cross A_2\cross\cdots\cross A_n.$$  The set $R$ represents those sequences $(a_1,a_2,\ldots,a_n)$ that are so-related. In an olog, we represent this as follows $$\fbox
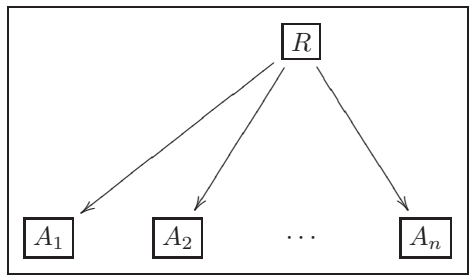
{\xymatrix{&&\fbox{$R$}\ar[ddll]\ar[ddl]\ar[ddr]\\\\\fbox{$A_1$}&\fbox{$A_2$}&\cdots&\fbox{$A_n$}}}$$  For example, $$\fbox
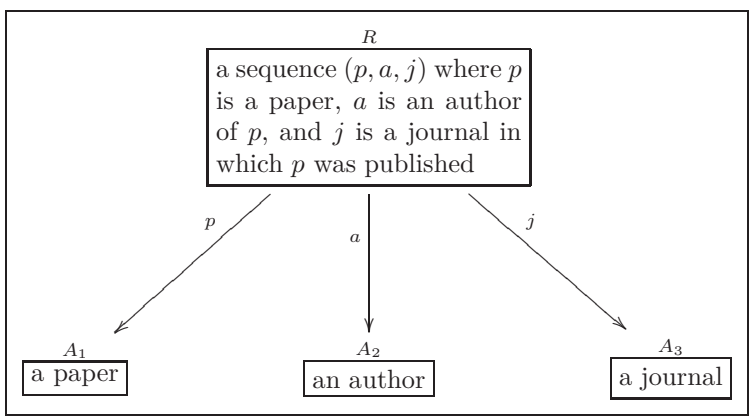
{\xymatrix{&\labox{R}{a sequence $(p,a,j)$ where $p$ is a paper, $a$ is an author of $p$, and $j$ is a journal in which $p$ was published}\ar[ddl]_p\ar[dd]_a\ar[ddr]^j\\\\\smbox{A_1}{a paper}&\smbox{A_2}{an author}&\smbox{A_3}{a journal}}}$$  Whereas $A_1\cross A_2\cross A_3$ includes all possible triples $(p,a,j)$ where $a$ is a person, $p$ is a paper, and $j$ is a journal, it is obvious that not all such triples are found in $R$. Thus $R$ represents a proper subset of $A_1\cross A_2\cross A_3$.

\comment{

A functional relationship is a special kind of relation and as such could be written in the format above. For example the arrow $\fakebox{a child}\To{\tn{has}}\fakebox{a father}$ is a functional relationship and can be replaced by \begin{align}\label{olog:functional relation}\fbox{\xymatrix{&\obox{R}{1.7in}{a pair $(c,f)$ where $c$ is a child and $f$ is the father of $c$}\ar[dl]_c\ar[dr]^f\\\smbox{A_1}{a child}&&\smbox{B}{a father}}}\end{align}  Here $R$ is called the {\em graph} of the functional relationship $\fakebox{a child}\To{\tn{has}}\fakebox{a father}$. But the same relation encodes that every father has a set of children. 

}

\begin{rules}\label{rules:aspects}

An aspect is presented as a labeled arrow, pointing from a source box to a target box. The arrow text should

\begin{enumerate}[(i)]
\item begin with a verb;
\item yield an English sentence, when the source-box text followed by the arrow text followed by the target-box text is read;
\item refer to a functional dependence: each instance of the source type should give rise to a specific instance of the target type;
\end{enumerate}

\end{rules}

\subsection{Facts}\label{sec:facts}

In this section we will discuss facts and their relationship to ``path equivalences."  It is such path equivalences, which exist in categories but do not exist in graphs, that make category theory so powerful. See \cite{Spi-Cats} for details.

Given an olog, the author may want to declare that two paths are equivalent. For example consider the two paths from $A$ to $C$ in the olog \begin{align}\label{olog:commute}\fbox
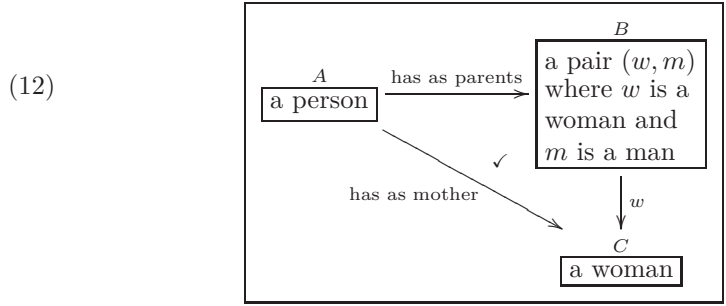
{\xymatrix{\smbox{A}{a person}\LA{rr}{has as parents}\LAL{drr}{\parbox{.8in}{has as mother}}&&\obox{B}{.8in}{\rr a pair $(w,m)$ where $w$ is a woman and $m$ is a man}\ar@{}[dll]|(.4){\checkmark}\LA{d}{$w$}\\&&\smbox{C}{a woman}}}\end{align}  We know as English speakers that a woman parent is called a mother, so these two paths $A\to C$ should be equivalent. A more mathematical way to say this is that the triangle in Olog (\ref{olog:commute}) {\em commutes}. 

A {\em commutative diagram} is a graph with some declared path equivalences. In the example above we concisely say ``a woman parent is equivalent to a mother."  We declare this by defining the diagonal map in (\ref{olog:commute}) to be {\em the composition} of the horizontal map and the vertical map. 

We generally prefer to indicate a commutative diagram by drawing a check-mark, $\checkmark$, in the region bounded by the two paths, as in Olog (\ref{olog:commute}). Sometimes, however, one cannot do this unambiguously on the 2-dimensional page. In such a case we will indicate the commutative diagrams (fact) by writing an equation. For example to say that the diagram $$\xymatrix{A\ar[r]^f\ar[d]_h&B\ar[d]^g\\C\ar[r]_i&D}$$ commutes, we could either draw a checkmark inside the square or write the equation $f;g=h;i$ above it. Either way, it means that ``$f$ then $g$" is equivalent to ``$h$ then $i$". 

\comment{ 

The following is an example used in the paper \cite{Sp}.

\begin{example}\label{ex:employee}

Suppose one is running a department store in which every employee works in a specified department, every employee has a manager, and every department has a secretary (who is an employee). The secretary of any department works in that department, and the manager of any employee works in the same department as that employee. Finally, every employee has a first and last name and every department has a name as well.

This is captured in the olog:
\begin{align}\label{dia:basic cat} \fbox{\parbox{3.9in}{\underline{has as manager;works in=works in}\\\underline{has as secretary;works in=$\id_{\fakebox{a department}}$}\\\\\xymatrix@=40pt{\smbox{E}{an employee}\ar@<.5ex>[rr]^{\tn{works in}}\ar@{{}{-}*{\curlyvee}}`u[]`r[][]_(0){\parbox{.8in}{\footnotesize\tn{has as manager\\~}}}\ar@/_1pc/[dd]_{\tn{has as first name}}\ar@/^1pc/[dd]^{\tn{has as last name}}&&\smbox{D}{a department}\ar@<.5ex>[ll]^{\tn{has as secretary}}\ar@/^1pc/[ddll]^{\tn{has as name}}\\\\\smbox{S}{a string of letters}}}}\end{align}  At the top are two underlined equations, or facts. The first fact states that every employee works in the same department that his or her manager works in. The second fact states that the secretary of any given department works in that department. Note that some diagrams do not commute. For example, the first name of the secretary of a department is not equal to the name of the department, even though both are paths $D\to S$. 

\end{example}

} 

\subsubsection{More complex facts}

Recording real-world facts in an olog can require some creativity. Whereas a fact like ``the brother of ones father is ones uncle" is recorded as a simple commutative diagram, others are not so simple. We will try to show the range of expressivity of commutative diagrams in the following two examples.

\setcounter{theorem}{1}\begin{example}\label{ex:truck car}

How would one record a fact like ``a truck weighs more than a car"? We suggest something like this:\small$$\fbox
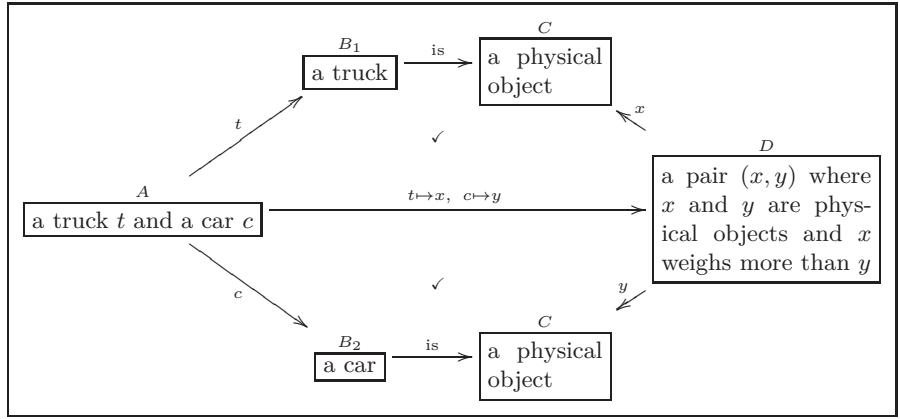
{\xymatrix@=10pt{&\smbox{B_1}{a truck}\LA{rr}{is}&\ar@{}[d]|{\checkmark}&\obox{C}{.6in}{a physical object}\\\smbox{A}{a truck $t$ and a car $c$}\ar[ur]^t\ar[dr]_c\ar[rrrr]^{t \mapsto x,\;\;c \mapsto y}&&&&\obox{D}{1.1in}{a pair $(x,y)$ where $x$ and $y$ are physical objects and $x$ weighs more than $y$}\ar[ul]_-x\ar[dl]_-y\\&\smbox{B_2}{a car}\LA{rr}{is}&\ar@{}[u]|{\checkmark}&\obox{C}{.6in}{a physical object}}}$$\normalsize  where both top and bottom commute. This olog exemplifies the fact that simple sentences sometimes contain large amounts of information. While the long map may seem to suffice to convey the idea ``a truck weighs more than a car," the path equivalences (declared by check-marks) serve to ground the idea in more basic types. These other types tend to be useful for other purposes, both within the olog and when connecting it to others.

\end{example}

\comment{

\begin{example}

There is a difference between one person liking another person, and two people being friends --- the second is assumed symmetric. To capture this one could use the olog $$\fbox{\parbox{2.9in}{\begin{center}\underline{$\tn{flip};\tn{flip}=\id_1$}\hsp\underline{$\tn{flip};x=y$}\hsp\underline{$\tn{flip};y=x$}\end{center}\xymatrix@=12pt{&\mebox{1}{a pair of people $(x,y)$ where $x$ and $y$ are friends}\ar[dddd]_{\tn{flip}}\ar[ddl]_x\ar[ddr]^y\\\\\smbox{2}{a person}\ar@{}[r]|-{\checkmark}&&\smbox{2}{a person}\ar@{}[l]|-{\checkmark}\\\\&\mebox{1}{a pair of people $(x,y)$ where $x$ and $y$ are friends}\ar[uul]_y\ar[uur]^x}}}$$    The fact that every object is written twice in this olog might be confusing; it was done so as to ease readability of the olog. Actually ``flip" is a map from 1 to itself, and $x$ and $y$ have the same domain of definition as well as the same set of result values. In the precise definition of basic ologs, given in Definition \ref{def:basic olog}, this way of drawing ologs will be made precise.

\end{example}

}

\setcounter{subsubsection}{2}\subsubsection{Specific facts at the olog level}

Another fact one might wish to record is that ``John Doe's weight is 150 lbs."  This is established by declaring that the following diagram commutes:\begin{align}\label{olog:weight}\fbox
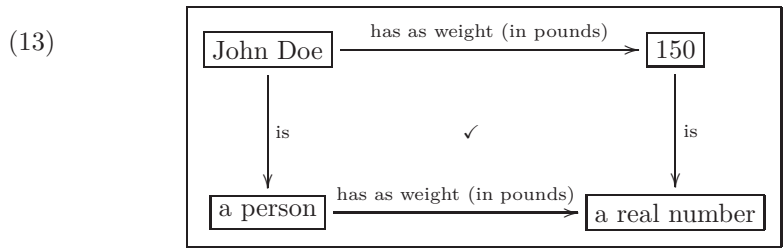
{\xymatrix@=18pt{\fbox{John Doe}\ar@{}[ddrrrr]|{\checkmark}\LA{rrrr}{has as weight (in pounds)}\LA{dd}{is}&&&&\fbox{150}\LA{dd}{is}\\\\\fbox{a person}\LA{rrrr}{has as weight (in pounds)}&&&&\fbox{a real number}}} \end{align}  If one only had the top line, it would be less obvious how to connect its information with that of other ologs. (See Section \ref{sec:connecting ologs} for more on connecting different ologs).

\comment{

\begin{remark}

In Section \ref{sec:mass nouns} that a proper noun like \fakebox{John Doe} should be replaced by something like \fakebox{an instance of John Doe}. But then the top map in Olog (\ref{olog:weight}) becomes suspect: have all instances of John Doe had the same weight?  This is a case where using rules of good practice helps expose inaccuracies that could cause problems later. To better reflect the intended meaning, the \fakebox{John Doe} box should be replaced by something like \fakebox{John Doe, on January 1, 2011}.

\end{remark}

}

Note that the top line in Diagram (\ref{olog:weight}) might also be considered as existing at the ``data level" rather than at the ``olog level."  In other words, one could see John Doe as an ``instance" of \fakebox{a person}, rather than as a type in and of itself, and similarly see 150 as an instance of \fakebox{a real number}. This idea of an olog having a ``data level" is the subject of the Section \ref{sec:instances}.

\setcounter{theorem}{3}\begin{rules}\label{rules:facts}

A fact is the declaration that two paths (having the same source and target) in an olog are equivalent. Such a fact is either presented as a checkmark between the two paths (if such a check-mark is unambiguous) or by an equation. Every such equivalence should be declared; i.e. no fact should be considered too obvious to declare.

\end{rules}


\section{Instances}\label{sec:instances}

The reader at this point hopefully sees an olog as a kind of ``concept map," and it is one, albeit a concept map with a formal structure (implicitly coming from category theory) and specific rules of good practice. In this section we will show that one can also load an olog with data. Each type can be assigned a set of instances, each aspect will map the instances of one type to instances of the other, and each fact will equate two such mappings. We give examples of these ideas in Section \ref{sec:instances of taf}. 

In Section \ref{sec:relationship olog db}, we will show that in fact every olog can also serve as the layout for a database. In other words, given an olog one can immediately generate a {\em database schema}, i.e. a system of tables, in any reasonable data definition language such as that of SQL. The tables in this database will be in one-to-one correspondence with the types in the olog. The columns of a given table will be the aspects of the corresponding type, i.e. the arrows whose source is that type. Commutative diagrams in the olog will give constraints on the data.

In fact, this idea is the basic thesis in \cite{Spi-FDM}, even though the word olog does not appear in that paper. There it was explained that a category $\mcC$ naturally can be viewed as a database schema and that a functor $I\taking\mcC\to\Set$, where $\Set$ is the category of sets, is a database state. Since an olog is a drawing of a category, it is also a drawing of a database schema. The current section is about the ``states" of an olog, i.e. the kinds of data that can be captured by it. 

\subsection{Instances of types, aspects, and facts}\label{sec:instances of taf}

Recall from Section \ref{sec:basic ologs} that basic ologs consist of types, displayed as boxes; aspects, displayed as arrows; and facts, displayed as equations or check-marks. In this section we discuss the instances of these three basic constructions.  The rules of good practice (\ref{rules:types}, \ref{rules:aspects}, and \ref{rules:facts}) were specifically designed to simplify the process of finding instances.

\subsubsection{Instances of types}\label{sec:instances of types}

According to Rules \ref{rules:types}, each box in an olog contains text which should refer to {\bf a distinction made and recognizable by the author for which instances can be documented.}  For example if my olog contains a box \begin{align}\label{dia:petting}\obox{}{1.3in}{a pair $(p,c)$ where $p$ is a person, $c$ is a cat, and $p$ has petted $c$}\end{align} then I must have some concept of when this situation occurs. Every time I witness a new person-cat petting, I document it. Whether this is done in my mind, in a ledger notebook, or on a computer does not matter; however using a computer would probably be the most self-explanatory. Imagine a computer program in which one can create ologs. Clicking a text box in an olog results in it ``opening up" to show a list of documented instances of that type. If one is reading the CBS news olog and clicks on the box \fakebox{an episode of 60 Minutes}, he or she should see a list of all episodes of the TV show ``60 Minutes." If we wish to document a new person-cat petting incident we click on the box in (\ref{dia:petting}) and add this new instance.

\subsubsection{Instances of aspects}

According to Rules \ref{rules:aspects}, each arrow in an olog should be labeled with text that refers to a functional relationship between the source box and the target box. A functional relationship $f\taking A\to B$ between finite sets $A$ and $B$ can always be written as a 2-column table: the first column is filled with the instances of type $A$ and the second column is filled with their $f$-values, which are instances of type $B$. 

For example, consider the aspect \begin{align}\label{dia:moon1}\fbox{a moon}\To{\tn{orbits}}\fbox{a planet}\end{align}  We can document some instances of this relationship using the following table: \begin{align}\label{dia:moon2}\begin{tabular}{| c || c |}\hline\multicolumn{2}{| c |}{\bf orbits}\\\hline{\bf a moon}&{\bf a planet}\\\hline\hline The Moon&Earth\\\hline Phobos&Mars\\\hline Deimos&Mars\\\hline Ganymede & Jupiter\\\hline Titan & Saturn\\\hline\end{tabular}\end{align}  Clearly, this table of instances can be updated as more moons are discovered by the author (be it by telescope, conversation, or research).

The correspondence between aspect (\ref{dia:moon1}) and Table (\ref{dia:moon2}) makes it clear that ologs can serve to hold data which exemplifies the author's world-view. In Section \ref{sec:relationship olog db}, we will show that ologs (which have many aspects and facts) can serve as bona fide database schemas.

\subsubsection{Instances of facts}

Recall the following olog: \begin{align*}\tag{\ref{olog:commute}}\fbox
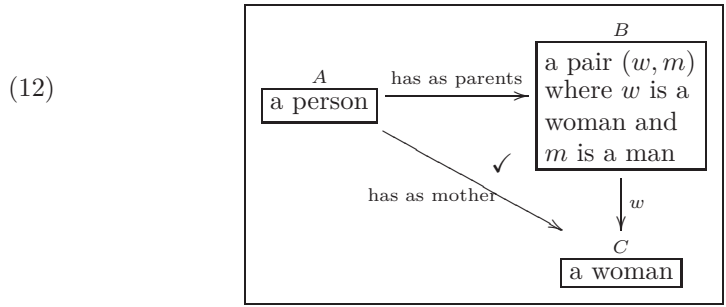
{\xymatrix{\smbox{A}{a person}\LA{rr}{has as parents}\LAL{drr}{has as mother}&&\obox{B}{.8in}{\rr a pair $(w,m)$ where $w$ is a woman and $m$ is a man}\ar@{}[dl]|{\parbox{.5in}{\checkmark\\~}}\LA{d}{$w$}\\&&\smbox{C}{a woman}}}\end{align*}  and consider the following instances of the three aspects in it: $$\begin{tabular}{| c || c |}\hline\multicolumn{2}{| c |}{\bf has as parents}\\\hline{\bf a person}&{\bf a pair $(w,m)$ ...}\\\hline\hline Cain&(Eve, Adam)\\\hline Abel&(Eve, Adam)\\\hline Chelsey&(Hillary, Bill)\\\hline\end{tabular}\hsp\begin{tabular}{| c || c |}\hline\multicolumn{2}{| c |}{\bf $w$}\\\hline{\bf a pair $(w,m)$ ...}&{\bf a woman}\\\hline\hline (Eve, Adam)&Eve\\\hline (Hillary, Bill)&Hillary\\\hline (Margaret, Samuel)&Margaret\\\hline (Emily, Kris)&Emily\\\hline\end{tabular}$$\vspace{-.32in}\begin{align}\label{dia:instances of facts}~\end{align}\vspace{-.1in}$$\begin{tabular}{| c || c |}\hline\multicolumn{2}{| c |}{\bf has as mother}\\\hline{\bf a person}&{\bf a woman}\\\hline\hline Cain&Eve\\\hline Abel &Eve\\\hline Chelsey &Hillary\\\hline\end{tabular}$$

When we declare that the diagram in (\ref{olog:commute}) commutes (using the check-mark), we are saying that for every instance of \fakebox{a person} (of which we have three: Cain, Abel, and Chelsey), the two paths to \fakebox{a woman} give the same answers. Indeed, for Cain the two paths are: \begin{enumerate}[(i)]\item Cain $\mapsto$ (Eve, Adam) $\mapsto$ Eve; \item Cain $\mapsto$ Eve;\end{enumerate} and these answers agree. If one changed any instance of the word ``Eve" to the word ``Steve" in one of the tables in (\ref{dia:instances of facts}), some pair of paths would fail to agree. Thus the ``fact" that the diagram in (\ref{olog:commute}) commutes ensures that there is some internal consistency between the meaning of parents and the meaning of mother, and this consistency must be born out at the instance level.

All of this will be formalized in Section \ref{sec:instance data}.

\subsection{The relationship between ologs and databases}\label{sec:relationship olog db}

Recall from Section \ref{sec:instances of types} that we can imagine creating an olog on a computer. The user creates boxes, arrows, and compositions, hence creating a category $\mcC$. Each text-box $x$ in the olog can be ``clicked" by the computer mouse, an action which allows the user to ``view the contents" of $x$. The result will be a set of things, which we might call $I(x)\in\Set$, whose elements are things of type $x$. So clicking on the box \fakebox{a man} one sees $I(\fakebox{a man})$, the set of everything the author has documented as being a man. For each aspect $f\taking x\to y$ of $x$, the user can see a function from the set $I(x)$ to $I(y)$, perhaps as a 2-column table as in (\ref{dia:instances of facts}). 

The type $x$ may have many aspects, which we can put together into a single multi-column table. Its columns are the aspects of $x$, and its rows are the elements of $I(x)$. Consider the following olog, taken from \cite{Spi-FDM} where it was presented as a database schema. \begin{align}\label{olog:employee} \fbox
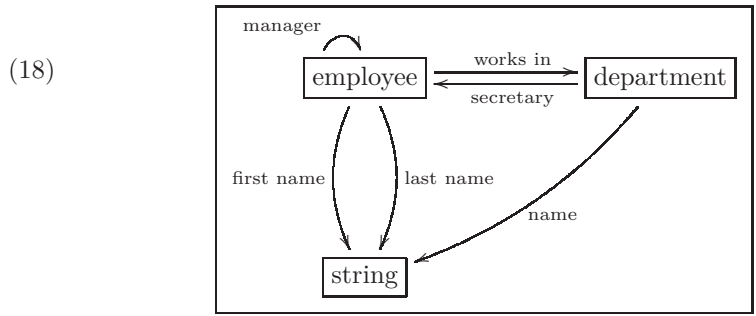
{\xymatrix{\fbox{employee}\ar@<.5ex>[rr]^{\tn{works in}}\ar@(l,u)[]^{\tn{manager}}\ar@/_1pc/[dd]_{\tn{first name}}\ar@/^1pc/[dd]^{\tn{last name}}&&\fbox{department}\ar@<.5ex>[ll]^{\tn{secretary}}\ar@/^1pc/[ddll]^{\tn{name}}\\\\\fbox{string}}}\end{align}  The type \fakebox{Employee} has four aspects, namely {\tt manager} (valued in \fakebox{Employee}), {\tt works in} (valued in \fakebox{department}), and {\tt first name} and {\tt last name} (valued in \fakebox{string}). As a database, each type together with its aspects form a multi-column table, as in the following example.

\begin{example}\label{ex:instances of employee}

We can convert Olog (\ref{olog:employee}) into a database schema. Each box represents a table, each arrow out of a box represents a column of that table. Here is an example state of that database. 

 \begin{align}\label{dia:flb}\xymatrix{\parbox{3.5in}{\begin{tabular}{| l || l | l | l | l |}\hline\multicolumn{5}{| c |}{\bf employee}\\\hline {\bf Id}&{\bf first name}&{\bf last name}&{\bf manager}&{\bf works in}\\\hline 101&David&Hilbert&103&q10\\\hline 102&Bertrand&Russell&102&x02\\\hline 103&Alan&Turing&103&q10\\\hline\end{tabular}\\~\vspace{.1in}\\\begin{tabular}{| l || l | l |}\hline\multicolumn{3}{| c |}{\bf department}\\\hline {\bf Id}&{\bf name}&{\bf secretary}\\\hline q10&Sales&101\\\hline x02&Production&102\\\hline\end{tabular}}&\parbox{.5in}{\begin{tabular}{| l ||}\hline\multicolumn{1}{| c |}{\bf string}\\\hline{\bf Id}\\\hline a\\\hline b\\\hline\vdots\\\hline z\\\hline aa\\\hline ab\\\hline\vdots\\\hline\end{tabular}}}\end{align}  Note that every arrow $f\taking x\to y$ of Olog (\ref{olog:employee}) is represented in Database (\ref{dia:flb}) as a column of table $x$, and that every cell in that column can be found in the Id column of table $y$. For example, every cell in the ``works in" column of table {\bf employee} can be found in the Id column of table {\bf department}.
 
\end{example}

The point is that ologs can be drawn to represent a world-view (as in Section \ref{sec:basic ologs}), but they can also store data.  Rules 1,2, and 3 in  \ref{rules:types} align the construction of an olog with the ability to document instances for each of its types. 

\setcounter{subsubsection}{1}\subsubsection{Instance data as a set-valued functor}\label{sec:instance data}

Let $\mcC$ be an olog. Section \ref{sec:instances} so far has described instances of types, aspects, and facts and how all of these come together into a set of interconnected tables. The assignment of a set of instances to each type and a function to each aspect in $\mcC$, such that the declared facts hold, is called an assignment of {\em instance data} for $\mcC$. More precisely, instance data on $\mcC$ is a functor $\mcC\to\Set$, as in Definition \ref{def:set-valued functor}. 

\setcounter{theorem}{2}

\begin{definition}\label{def:set-valued functor}

Let $\mcC$ be a category (olog)
with underlying graph $|\mcC|$, 
and let $\Set$ denote the category of sets. 
An {\em instance of $\mcC$} 
(or {\em an assignment of instance data for $\mcC$}) 
is a functor $I\taking\mcC\to\Set$. 
That is, 
it consists of 
\begin{itemize}
\item a set $I(x)$ for each object (type) $x$ in $\mcC$,
\item a function $I(f)\taking I(x)\to I(y)$ for each arrow (aspect) $f\taking x\to y$ in $\mcC$, and 
\item for each fact (path-equivalence or equation) 
\footnote{If we let  
$f = f_{1} {\,;\,} f_{2} {\,;\,} \cdots {\,;\,} f_{n}$ and 
$f' = f'_{1} {\,;\,} f'_{2} {\,;\,} \cdots {\,;\,} f'_{m}$,
then we often write $(f = f') \colon i \rightarrow j$ to denote the fact that these paths are equivalent.}
$$f_{1} {\,;\,} f_{2} {\,;\,} \cdots {\,;\,} f_{n} = f'_{1} {\,;\,} f'_{2} {\,;\,} \cdots {\,;\,} f'_{m}$$ 
declared in $\mcC$, 
an equality of functions 
$$I(f_{1}) {\,;\,} I(f_{2}) {\,;\,} \cdots {\,;\,} I(f_{n}) = I(f'_{1}) {\,;\,} I(f'_{2}) {\,;\,} \cdots {\,;\,} I(f'_{m}).$$
\end{itemize}

\end{definition}

For more on this viewpoint of categories and functors, the reader can consult \cite{Spi-Cats}.


\section{Communication between ologs}\label{sec:connecting ologs}

The world is inherently heterogeneous. 
Different individuals 
\footnote{By an individual we mean 
either an individual person acting on their own, a community acting as a single entity, a software agent, etc.
Later in this section
we will use the notion of a community acting as a distributed collection of linked, yet independent, individuals.}
in the world naturally have different world-views 
--- each individual has its own perspective on the world. 
The conceptual knowledge (information resources) of an individual represents its world-view, 
and is encoded in an ontology log, or olog, containing the concepts, relations, and observations 
that are important to that individual.
An olog is a formal specification of 
an individual's world-view in a language representing the concepts and relationships used by that individual. 
In addition to the formulation of an expressive language,
a specification needs to contain axioms (facts) that constrain the possible interpretations of that language. 

Since the ologs of different individuals are encoded in different languages, 
the important need to merge disparate ologs into a more general representation 
is difficult, time-consuming and expensive. 
The solution is to develop appropriate communication between individuals to allow interoperability of their ologs.
Communication can occur between individuals when there is some commonality between their world-views. 
It is this commonality that allows one individual to benefit from the knowledge and experience of another. 
In this section we will discuss how to formulate these channels of communication, 
thereby describing a generalized and practical technique for merging ologs.

The mathematical concept that makes it all work is that of a functor. 
A functor is a mapping from one category to another that preserves all the declared structure.
Whereas in Definition \ref{def:set-valued functor} 
we defined a functor from an olog to $\mathrmbf{Set}$, 
here we will be discussing functors from one olog to another. 

Suppose we have two ologs, $\mcC$ and $\mcD$, that represent the world-views of two individuals. 
A functor $F\taking\mcC\to\mcD$ is basically a way of matching each type (box) of $\mcC$ to a type of $\mcD$, 
and each aspect (arrow) in $\mcC$ to an aspect (or path of aspects) in $\mcD$. 
Once ologs are aligned in this way, communication can occur: 
the two individuals know what each other is talking about. 
In fact, 
mathematically we can show that 
instance data held in $\mcC$ can be transformed (in coherent ways) to instance data held in $\mcD$, 
and vice versa (see \cite{Spi-FDM}).
In simple terms, 
once individuals understand each other in a certain domain (be it social, mathematical, etc.), 
they can communicate their views about it.

While the basic idea is not hard, the details can be a bit technical. This section is written in a more formal and logical style, and is decidedly more difficult than the others. For this section only, we assume the reader is familiar with the notion of fibered categories, colimits in the category $\Cat$ of categories, etc. We return to our more informal style in Section \ref{sec:expressive I}, where we discuss how an individual can author a more expressive olog.

\subsection{Categories and their presentations}

We never defined categories in this paper, but we defined ologs and said that the two notions amounted to the same thing. Thus, we implied that a category consists of the following: a set of objects, a set of arrows (each pointing from one object to another), and a congruence relation on paths.\footnote{A congruence relation on paths is an equivalence relation on paths
that respects endpoints and is closed under composition from left and right 
(see the axioms in \ref{tab:entailment:axioms}).}
This differs from the standard definition of categories (see \cite{Mac}), which replaces
our congruence relation with a composition rule and associativity law (obtained by taking the categorical quotient). One could say that an olog is a presentation of a category by generators (objects and arrows) and relations (path congruences). Any category can be resolved and presented in such a way, which we will call a specification. Likewise any functor can be resolved and presented as a morphism between specifications.
\footnote{We take an agnostic approach to foundations here. 
With the presentation form, we show how categories and functors are definable in terms of sets and functions,
indicating how category theoretic concepts could be defined in terms of set theory. 
However, we fully understand that $\mathbf{Set}$, the category of sets and functions, is but one example of a topos, 
indicating how set theoretic concepts could be defined in terms of category theory.}

In fact, this presentation form for categories (and the analogous one for functors) 
is preferable for our work on communication between ologs, 
because it separates the strictly graphical part of an olog (its types and aspects, regarded as the olog language) 
from the propositional part (its facts, regarded as the olog formalism). 
This presentation form is standard in the institutions \cite{GB:INS} and information flow \cite{BS:IF} communities, 
since it separates the mechanism of flow from the content of flow; in this case the formal content. 
Our work here applies the general theories of institutions and information flow to the specific logical system that underlies categories and functors,\footnote{For the expert, this refers to the sketch logical system {\ttfamily Sk}, 
in its various manifestations.}
demonstrating how this logical system can be used for knowledge representation.
Using the presentation forms for categories and functors, we show how communication  between individuals is effected by the flow of information along channels. 

\subsection{The architecture underlying information systems}

We think of a community of people, businesses, etc. in terms of the ologs of each individual participant together with the information channels that connect them. These channels are functors between ologs, which allow communication to occur. The heterogeneity of multiple differing world-views 
connected through such links can lead to a flexibility and robustness of interaction. 
For example, heterogeneity allows for multiple schemas to be employed in the design of database systems in particular, 
and multiple languages to be employed in the design of knowledge representation systems in general.

For any olog, consider the underlying graph of types and aspects. 
We regard this graph as being the language of the olog, 
\footnote{Section \ref{sec:CG} indicates how natural languages can be encoded into ologs.}
with the facts of the olog being a subset of all the possible assertions that one can make within this language.
Any two ologs with the same underlying graph of types and aspects have the same language,
and since the facts of each olog are expressed in the same language,
they can be ``understood'' by each other without translation.
As such, 
we think of the collection of all ologs with the same language (underlying graph) as forming a homogeneous {\em context},
with the ologs ordered in a specialization-generalization hierarchy. 

Whereas an olog represents (the world-view of) a single individual,
an information system (of ologs) represents a community of separate, independent and distributed individuals.
Here we consider an information system to be a diagram of ologs of some shape $\mathrmbf{I}$;
that is,
a collection of ologs and constraints indexed by a base category $\mathrmbf{I}$.
The parts of the system represent 
either the ologs of the various individuals in the system 
or common grounds needed for communication between the individuals.
Each part of the system specifies its world-view as facts expressed in terms of its language.
The system is heterogeneous, since each part has a separate language for the expression of its world-view. 
The morphisms between the parts are the alignment (constraint) links defining the common grounds. 

As will be made clear in a moment, 
there is an underlying distributed system consisting of 
the language (underlying graph) for each component part of the information system and 
a translation (graph morphism) for each alignment link. 
We can think of this distributed system
as an underlying system of languages linked by translating dictionaries.
This distributed system determines an information channel 
with core language (graph) and component translation links (graph morphisms)
along which the specifications of each component part can flow to the core.
We can think of this core as a universal language for the whole system
and the channel as a translation mechanism from parts to whole.
At the core, 
the direct flow of the component specifications are joined together (unioned) and allowed to interact through entailment. 
The result of this interaction can then be distributed back to the component parts,
thereby allowing the separate parts of an information system to interoperate.

In this section, we will make all this clear and rigorous. 
As mentioned above, we will work with category presentations (here called {\em specifications}) rather than categories. 
We will discuss the homogeneous contexts called fibers in detail and give the axioms of satisfaction. 
We will then discuss how morphisms between graphs (the translating dictionaries between the  ologs) allow for direct and inverse information flow between these homogeneous fiber contexts. 
Finally, we discuss specifications (also known as {\em theories}) and the lattice of theories construction for ontologies.

In Section \ref{sec:alignment} 
we will discuss how the information in ologs can be aligned by the use of common grounds. 
This alignment will result in the creation of {\em information systems}, 
which are systems of ologs connected together along functors. 
We will discuss how to take the information contained in each olog of a heterogeneous system and integrate it all into a single whole, called the fusion olog. Finally we will discuss how the consequence of bringing all this information together, and allowing it to interact, can be transferred back to each part of the system (individual olog) as a set of local facts entailed by remote ologs, allowing for a kind of interoperability between ologs.
In Section \ref{sec:CG} we will discuss conceptual graphs and their relationship to ologs.

\subsubsection{Fibers}\label{sec:fibers}

A graph $G$ contains types as nodes and aspects as edges.
The graphs underlying an olog is considered its {\em language}.
Any category $\mathcal{C}$ has an underlying graph $|\mathcal{C}|$.
In particular, 
$|\mathrmbf{Set}|$ is the graph underlying the category of sets and functions.
Olog (12) has an underlying graph containing the three types \fakebox{person}, \fakebox{person-pair} and \fakebox{woman} 
and the three aspects `has a parent', `woman' and `has as mother'.
Olog (17) has an underlying graph containing the three types \fakebox{employee}, \fakebox{department}, and \fakebox{string} 
and the six aspects `manager', `works in', `secretary', `name', `first name' and `last name'.
Let $\mathrmbfit{eqn}(G)$ denote the set of all facts (equations) 
that are possible to express using the types and aspects of $G$.
A $G$-specification
is a set $E \subseteq \mathrmbfit{eqn}(G)$ consisting of some of the facts expressible in $G$.
The singleton set with the one fact 
that ``the female parent of a person is his/her mother'' 
is a specification for the graph of Olog (12).
The set with the two facts 
that ``the manager has the same department as any employee'' 
and ``the secretary of a department is an employee in that department''
is a specification for the graph of Olog (17).
Let $\mathrmbfit{spec}(G)$ 
denote the collection of all $G$-specifications 
ordered by inclusion $E_{1} \subseteq E_{2}$.

\subsubsection{Satisfaction}

It will be useful here to define an instance of a graph $G$,
instead of an instance of a category $\mathcal{C}$.
An instance of a graph populates the graph by assigning instance data to it.
An instance of a graph $G$ is a graph morphism $D \colon G \to |\mathrmbf{Set}|$ 
mapping each type $x$ in $G$ to a set $D(x)$ of instances and
mapping each aspect $e \colon x \to y$ in $G$ to an instance function $D(e) \colon D(x) \to D(y)$.
Using database terminology,
we also call $D$ a key diagram,
since it gives the set of row identifiers (primary keys) of tables and the cell contents defined by key maps.

A key diagram $D \colon G \rightarrow |\mathrmbf{Set}|$ 
satisfies (is a model of) a $G$-fact $\epsilon \in \mathrmbfit{eqn}(G)$
%
%
(see Definition \ref{def:set-valued functor}),
symbolized $D \models_{G} \epsilon$, 
when we have an equality of functions $D^\ast(\epsilon_{0}) = D^\ast(\epsilon_{1})$.
We also say that
$\epsilon$ (holds in) is true when interpreted in $D$. 
An identity
$(f =_{G} f) \colon i \rightarrow j$
holds in all key diagrams (hence, is a tautology),
and vice-versa
for any set $A \in |\mathrmbf{Set}|$
a constant key diagram
$\Delta(A) \colon G \to |\mathrmbf{Set}|$
satisfies any fact $\epsilon \in \mathrmbfit{eqn}(G)$. 
A key diagram $D \colon G \rightarrow |\mathrmbf{Set}|$ satisfies (is a model of) a $G$-specification $E$, 
symbolized $D \models_{G} E$, 
when it satisfies every fact in the specification.
%
%
For any graph $G$,
a $G$-specification $E$ entails a $G$-fact $\epsilon$,
denoted by $E \vdash_{G} \epsilon$,
when any model of the specification satisfies the fact.
The consequence $E^{\scriptscriptstyle\bullet}$ of a $G$-specification $E$ is the set of all entailed equations.
The consequence operator $(-)^{\scriptscriptstyle\bullet}$ is a closure operator,
and the consequence of a specification is a congruence.
For any $G$-specification $E$,
entailment satisfies the following axioms.
\begin{gather}
\mbox{\scriptsize\begin{tabular}{r@{\hspace{10pt}}p{250pt}}
(basic)
&
If
$E$ contains the equation $\epsilon$,
then
$E$ entails $\epsilon$.
\\
(reflexive)
&
$E$ entails the equations $(f =_{G} f) \colon i \rightarrow j$
for any path $f \colon i \rightarrow j$. 
\\
(symmetric)
&
If
$E$ entails the equation $(f_{1} =_{G} f_{2}) \colon i \rightarrow j$,
then
$E$ entails the equation $(f_{2} =_{G} f_{1}) \colon i \rightarrow j$.
\\
(transitive)
&
If
$E$ entails the two equations $(f_{1} =_{G} f_{2}) \colon i \rightarrow j$ and $(f_{2} =_{G} f_{3}) \colon i \rightarrow j$,
then
$E$ entails the equation $(f_{1} =_{G} f_{3}) \colon i \rightarrow j$.
\\
(compositional)
&
If
$E$ entails the two equations $(f_{1} =_{G} f_{2}) \colon i \rightarrow j$ and $(g_{1} =_{G} g_{2}) \colon j \rightarrow k$,
then 
$E$ entails the equation $(f_{1} {\,;\,} g_{1} =_{G} f_{2} {\,;\,} g_{2}) \colon i \rightarrow k$.
\\
(bi-closed)
&
If
$E$ entails the equation $(g_{1} =_{G} g_{2}) \colon j \rightarrow k$, 
then 
$E$ entails the equations 
$(f {\,;\,} g_{1} =_{G} f {\,;\,} g_{2}) \colon i \rightarrow k$ 
and $(g_{1} {\,;\,} h =_{G} g_{2} {\,;\,} h) \colon j \rightarrow l$ 
for any left composable path $f \colon i \rightarrow j$ and any right composable path $h \colon k \rightarrow l$.
\end{tabular}}
\label{tab:entailment:axioms}
\end{gather}
These are converted to inference rules in Table~\ref{entailment:inference:rules}.
To construct $E^{\scriptscriptstyle\bullet}$, 
we first take the reflexive, symmetric, and transitive closure $E^\ast$ of $E$
(so that $E^\ast$ is a $G$-specification and also the smallest equivalence relation containing $E$),
and then we get $E^{\scriptscriptstyle\bullet}$
by closing up under composition on left and right.
We extend specification inclusion with the entailment order,
where $E_{1} \leq_{G} E_{2}$
when $E_{1}$ entails each equation in $E_{2}$;
that is,
when $E_{1}^{\scriptscriptstyle\bullet} \supseteq E_{2}$
or equivalently when  
$E_{1}^{\scriptscriptstyle\bullet} \supseteq E_{2}^{\scriptscriptstyle\bullet}$.
The statement ``$E_{1} \leq_{G} E_{2}$'' asserts that
$E_{1}$ is at least as specialized as $E_{2}$.
The entailment order ${\langle{\mathrmbfit{spec}(G),\leq_{G}}\rangle}$,
which is a specialization-generalization order, 
represents a local version of
the ``lattice of theories'' construction of Sowa \cite{S:KR} (see Section \ref{sec:LOT}).
The opposite entailment order
$\mathrmbfit{fbr}(G) 
= {\langle{\mathrmbfit{spec}(G),\geq_{G}}\rangle}$
is called the fiber order.\footnote{For consistency in discussion,
we follow the terminology of formal concept analysis \cite{GW:FCA}, information flow \cite{BS:IF} and the theory of institutions \cite{GB:INS}.
This includes the polarity induced by concept lattices and the directionality of infomorphisms.}
%
In the lattice\footnote{This is a complete preorder,
loosely called a ``lattice''.}
$\mathrmbfit{spec}(G)$, 
the meet is union $\wedge = \cup$ and the join is intersection $\vee = \cap$;
whereas
in the lattice $\mathrmbfit{fbr}(G)$, the join is union $\vee = \cup$ and the meet is intersection $\wedge = \cap$.
Any specification $E$ is entailment equivalent to its consequence $E \cong E^{\scriptstyle\bullet}$.
A specification $E$ is closed when it is equal to its consequence $E = E^{\scriptstyle\bullet}$.
There is a one-one correspondence between closed $G$-specifications and categories over graph $G$.
%
%
The conceptual intent
of a key diagram $D$, 
implicit in satisfaction,
is the closed specification $\mathrmbfit{int}(D)$ consisting of all facts satisfied by the key diagram. 
Hence,
$D \models_{G} E$
iff
$E \subseteq \mathrmbfit{int}(D)$
iff
$\mathrmbfit{int}(D) \leq_{G} E$.\footnote{This is the first step in the algebraization of Tarski's ``semantic definition of truth'' \cite{K:AT}.}

\subsubsection{Elementary flow}

A graph morphism $H \colon G_{1} \rightarrow G_{2}$ maps the types and aspects of $G_{1}$ to the types and aspects of $G_{2}$.
Graph morphisms are the translations between ologs.
A functor $\mathcal{F} \colon \mathcal{C}_{1} \rightarrow \mathcal{C}_{2}$ has 
an underlying graph morphism $|\mathcal{F}| \colon |\mathcal{C}_{1}| \rightarrow |\mathcal{C}_{2}|$.
For any graph morphism $H \colon G_{1} \rightarrow G_{2}$,
there is a fact function
$\mathrmbfit{eqn}(H) \colon \mathrmbfit{eqn}(G_{1}) \rightarrow \mathrmbfit{eqn}(G_{2})$ 
that maps a $G_{1}$-equation $(f_{1} =_{G_{1}} f'_{1}) \colon i_{1} \rightarrow j_{1}$
to the $G_{2}$-equation $(H^\ast(f_{1}) =_{G_{2}} H^\ast(f'_{1})) \colon H(i_{1}) \rightarrow H(j_{1})$,
and a key diagram functor
$\mathrmbfit{dgm}(H) \colon \mathrmbfit{dgm}(G_{2}) \rightarrow \mathrmbfit{dgm}(G_{1})$
that maps 
a key diagram $D_{2} \colon G_{2} \rightarrow |\mathrmbf{Set}|$
to the key diagram $H \circ D_{2} \colon G_{2} \rightarrow |\mathrmbf{Set}|$.\footnote{The composition of graph morphisms is written in diagrammatic order.}
The fact function is the fundamental unit of information (formal) flow for ologs, and
the key diagram functor is the fundamental unit of semantic flow for ologs.\footnote{This is so, at the abstraction of institutions \cite{K:SC}.}
Formal flow is adjoint to semantic flow
--- satisfaction is invariant under flow:
$\mathrmbfit{dgm}(H)(D_{2}) \models_{G_{1}} \epsilon_{1}$
\underline{iff}
$D_{2} \models_{G_{2}} \mathrmbfit{eqn}(H)(\epsilon_{1})$
for any graph morphism $H \colon G_{1} \rightarrow G_{2}$,
source fact $\epsilon_{1}$ and
target diagram $D_{2}$.
%
%
Specifications can be moved along graph morphisms by extending the fact (equation) function.
For any graph morphism $H \colon G_{1} \rightarrow G_{2}$,
define the {\em direct flow} operator
$\mathrmbfit{dir}(H) = {\wp}\mathrmbfit{eqn}(H)
: \mathrmbfit{spec}(G_{1}) \rightarrow \mathrmbfit{spec}(G_{2})$\footnote{The symbol $\wp$ denotes the power-set operator.}
and the {\em inverse flow} operator
$\mathrmbfit{inv}(H) = 
\mathrmbfit{eqn}(H)^{-1}((\mbox{-})^{\scriptscriptstyle\bullet})
: \mathrmbfit{spec}(G_{2}) \rightarrow \mathrmbfit{spec}(G_{1})$.
Direct and inverse flow are adjoint monotonic functions 
${\langle{\mathrmbfit{dir}(H) \dashv \mathrmbfit{inv}(H)}\rangle} 
\colon \mathrmbfit{fbr}(G_{1}) \rightarrow \mathrmbfit{fbr}(G_{2})$
w.r.t. fiber order:
$\mathrmbfit{dir}(H)(E_{1}) \geq_{G_{2}} E_{2}
\text{ \underline{iff} } 
E_{1} \geq_{G_{1}} \mathrmbfit{inv}(H)(E_{2})$.
%
%
For 
any graph morphism $H \colon G_{1} \rightarrow G_{2}$,
any $G_{1}$-specification $E_{1}$, and
any $G_{2}$-specification $E_{2}$, 
entailment satisfies the following axioms.
\begin{center}
{\scriptsize
\begin{tabular}{r@{\hspace{10pt}}p{250pt}}
(direct flow)
&
If
$E_{1}$ entails the equation 
$(f =_{G_{1}} f') \colon i \rightarrow j$,
then
$\mathrmbfit{dir}(H)(E_{1})$ entails the equation 
$(H^\ast(f_{1}) =_{G_{2}} H^\ast(f'_{1})) \colon H(i_{1}) \rightarrow H(j_{1})$.
\\
(inverse flow)
&
If
$E_{2}$ entails the equation 
$(H^\ast(f) =_{G_{2}} H^\ast(f')) \colon H(i) \rightarrow H(j)$,
then
$\mathrmbfit{inv}(H)(E_{2})$ 
entails the equation 
$(f =_{G_{1}} f') \colon i \rightarrow j$.
\end{tabular}}
\end{center}
These are converted to inference rules in Table~\ref{entailment:inference:rules}.
A graph morphism $H \colon G_{1} \rightarrow G_{2}$ defines a consequence operator 
${(\mbox{-})}^{{\scriptscriptstyle\blacklozenge}_{H}}
= \mathrmbfit{dir}(H) \circ \mathrmbfit{inv}(H)$
on the fiber preorder $\mathrmbfit{fbr}(G_{1})$,
where $E_{1} \geq_{G_{1}} E_{1}^{\scriptstyle\bullet} \geq_{G_{1}} E_{1}^{{\scriptscriptstyle\blacklozenge}_{H}}$.
%
%


\subsubsection{Specifications}
%
A specification $\mathcal{S} = {\langle{G,E}\rangle}$ is an indexed notion
consisting of a graph $G$ and a $G$-specification $E \in \mathrmbfit{spec}(G)$.
It is sometimes convenient to use the symbol `$\mathcal{S}$' in place of `$E$';
for example,
to say that ``$\mathcal{S} \in \mathrmbfit{spec}(G)$''.
A category $\mathcal{C}$ can be resolved and presented as 
a specification $\mathrmbfit{spec}(\mathcal{C}) = {\langle{G,E}\rangle}$ 
consisting of the underlying graph $G = |\mathcal{C}|$ containing the types and aspects of $\mathcal{C}$ and 
the collection $E$ of all facts that hold in $\mathcal{C}$.
In the other direction,
any specification $\mathcal{S}$ induces a (quotient) category $\mathrmbfit{cat}(\mathcal{S})$.
Olog (12) and Olog (17) are described as specifications in Section \ref{sec:fibers}. 
A specification morphism $H \colon {\langle{G_{1},E_{1}}\rangle} \rightarrow {\langle{G_{2},E_{2}}\rangle}$
is a graph morphism $H \colon G_{1} \rightarrow G_{2}$ that preserves entailment:
$E_{1} \vdash_{G_{1}} \epsilon_{1}$ implies $E_{2} \vdash_{G_{2}} \mathrmbfit{eqn}(H)(\epsilon_{1})$
for any $\epsilon_{1} \in \mathrmbfit{eqn}(G_{1})$;
or equivalently
that satisfies the adjointness conditions,
$\mathrmbfit{dir}(H)(E_{1}) \geq_{G_{2}} E_{2}
\text{ \underline{iff} } 
E_{1} \geq_{G_{1}} \mathrmbfit{inv}(H)(E_{2})$.
Being a graph morphism, 
it maps types to types and aspects to aspects.
Moreover, 
it also maps facts in $E_{1}$ to facts in $E_{2}$;
that is, it preserves all the declared structure. 
%
%
A functor $\mathcal{F} \colon \mathcal{C}_{1} \rightarrow \mathcal{C}_{2}$
can be resolved and presented as a specification morphism 
$\mathcal{F} \colon \mathrmbfit{spec}(\mathcal{C}_{1}) \rightarrow \mathrmbfit{spec}(\mathcal{C}_{2})$. 
Hence, the presentation form for a functor does exactly what the functor does.
The fibered category of specifications $\mathrmbf{Spec}$
has specifications as objects and specification morphisms as morphisms.
Thus,
it is defined in terms of information flow.
There is an underlying graph functor
$\mathrmbfit{gph} \colon \mathrmbf{Spec} \rightarrow \mathrmbf{Gph}$ from specifications to graphs 
${\langle{G,E}\rangle} \mapsto G$.
The subcategory over any fixed graph $G$ is the fiber $\mathrmbfit{fbr}(G)$;
because of the opposite orientation,
we say that
``the category of specifications points downward in the concept lattice''. 
Throughout this section we identify ologs with specifications and olog morphisms with specification morphisms.

\subsubsection{The lattice of theories construction}\label{sec:LOT}

Sowa's ``lattice of theories'' construction (LOT) describes a modular framework for ontologies \cite{S:KR}.
The Olog formalism follows the approach to LOT described in \cite{IFF:LOT}.\footnote{The IFF term `theory' is replaced by the Olog term 'specification' or 'olog'.}
In the Olog formalism, 
LOT is 
locally represented by the entailment preorders $\mathrmbfit{spec}(G)$,
and globally represented by the category of specifications $\mathrmbf{Spec}$.
We follow the discussion in section 6.5 ``Theories, Models and the World'' of Sowa \cite{S:KR}. 
From each olog (specification) in the ``lattice of theories'', 
the entailment ordering defines paths 
to the more generalized ologs above and the more specialized ologs below.
Sowa defines four ways for moving along paths from one olog to another: contraction, expansion, revision and analogy.
\begin{description}
\item[Contraction] 
Any olog can be contracted or reduced to a smaller, simpler olog, 
moving upward in the preorder $\mathrmbfit{spec}(G)$, 
by deleting one or more facts.
%
\item[Expansion] 
Any olog can be expanded, 
moving downward in the preorder $\mathrmbfit{spec}(G)$, 
by adding one or more facts.
%
\item[Revision] 
A revision step is composite, 
moving crosswise in the preorder $\mathrmbfit{spec}(G)$;
it uses a contraction step to discard irrelevant details, 
followed by an expansion step to added new facts.
\item[Analogy] 
Unlike contraction and expansion, 
which move to nearby ologs in an entailment preorder $\mathrmbfit{spec}(G)$, 
analogy moves to an olog in a remote entailment preorder in the category $\mathrmbf{Spec}$
via the flow
along an underlying graph morphism  $H \colon G_{1} \rightarrow G_{2}$
by systematically renaming the types and aspects that appear in the facts:
any olog $E_{1}$ in $\mathrmbfit{spec}(G_{1})$
is moved (by systematic renaming) to the olog
$\mathrmbfit{dir}(H)(E_{1})$ in $\mathrmbfit{spec}(G_{2})$.
\end{description}
According to Sowa, 
the various methods used in nonmonotonic logic and the operators for belief revision 
correspond to movement through the lattice of theories.

\subsection{Alignment and integration of information systems}\label{sec:alignment}

\subsubsection{Common ground}

Given the world-views of two individuals,
as represented by ologs 
$\mathcal{S}_{1} = {\langle{G_{1},E_{1}}\rangle}$ and $\mathcal{S}_{2} = {\langle{G_{2},E_{2}}\rangle}$,
there is little hope that one of them completely contains the other 
(even after allowing for renaming of types and aspects), 
and there is correspondingly little chance of finding a meaningful 
olog morphism between the two.
Instead, 
in order to communicate
the two individuals could attempt to find a common ground, 
a third olog $\mathcal{S} = {\langle{G,E}\rangle}$ 
and meaningful morphisms\footnote{Roughly speaking,
an olog morphism $F \colon \mathcal{C} \to \mathcal{D}$ is {\em meaningful} when for each type $X$ in $\mcC$, 
every intended instance of $X$ in $\mcC$ would be considered an instance of $F(X)$ by the author of $\mcD$ 
(in which case we say the intention for types is respected), 
and in a similar way the intention for aspects is respected.  
Precisely speaking,
if $I\taking\mcC\to\Set$ and $J\taking\mcD\to\Set$ are instance data for $\mcC$ and $\mcD$, 
then $F$ is meaningful relative to $I$ and $J$ 
if one can exhibit a natural transformation $\mu \taking I \Rightarrow F \circ J$ as in \cite{Spi-FDM}.}
$H_{1} \colon  \mathcal{S} \to \mathcal{S}_{1}$ and $H_{2} \colon \mathcal{S} \to \mathcal{S}_{2}$.\footnote{A common ground olog is also called a reference ontology in knowledge representation.}
This connection is a 1-dimensional knowledge network 
$\mathcal{S}_{1} \xleftarrow{H_{1}} \mathcal{S} \xrightarrow{H_{2}} \mathcal{S}_{2}$
of shape
$\bullet \leftarrow \bullet \rightarrow \bullet$
called a span (in $\mathrmbf{Spec}$), 
where each node is an olog and each edge is a morphism between ologs. 
%
%
The requirements of this span are that
$\mathrmbfit{dir}(H_{1})(E) \geq_{G_{1}} E_{1}$
and
$\mathrmbfit{dir}(H_{2})(E) \geq_{G_{2}} E_{2}$,
two requirements involving \underline{local} flow.
Equivalently, that
$E \geq_{G} \mathrmbfit{inv}(H_{1})(E_{1}) \vee_{G} \mathrmbfit{inv}(H_{2})(E_{2})$.
The latter precise expression can be rendered in natural language as
``the world-view of the common ground is contained in the combined world-views of the two individuals''.
The various local direct/inverse flows allow world-views to be compared. 
Such a common ground can be expanded and improved over time. 
The basic idea is that one individual can attempt to explain a new idea (type, aspect or fact) to another
in terms of the common ground. 
Then the other individual can either interpret this idea as they already have, 
learn from it (i.e. freely add it to their olog), or reject it. 
We view an olog morphism $H_{1} \colon \mathcal{S}_{1} \to \mathcal{S}_{2}$ 
as an atomic constraint (alignment) link between $\mathcal{S}_{1}$ and $\mathcal{S}_{2}$.\footnote{This is so, at the abstraction of institutions \cite{K:SC}.}
We view a common ground span 
$\mathcal{S}_{1} \xleftarrow{H_{1}} \mathcal{S} \xrightarrow{H_{2}} \mathcal{S}_{2}$ 
as a molecular constraint between $\mathcal{S}_{1}$ and $\mathcal{S}_{2}$,
which is weakest when $\mathcal{S} = \emptyset$ 
and strongest when $\mathcal{S}_{1} = \mathcal{S} = \mathcal{S}_{2}$. 

\subsubsection{Systems of ologs}

In the general case, 
more than two individuals will share a common ground. 
For example, 
companies that do business together may have a common-ground olog as part of a legal contract;
or, 
the various participants at a conference will have some common understanding of the topic of that conference. 
In fact, 
for any finite set of ologs $\mathbb{X} = \{ \mcS_{1}, \mcS_{2}, \ldots, \mcS_{n} \}$, 
there should be a common ground world-view (even if empty), 
say $\mathcal{S}_{\mathbb{X}}$. 
If $\mathbb{Y} \subseteq \mathbb{X}$ is a subset, 
then there should be a map $\mathcal{S}_{\mathbb{X}} \to \mathcal{S}_{\mathbb{Y}}$ 
because any common understanding held by the individuals in $\mathbb{X}$ 
is held by the individuals in $\mathbb{Y}$. 
For example,
the triangular-shaped diagram 
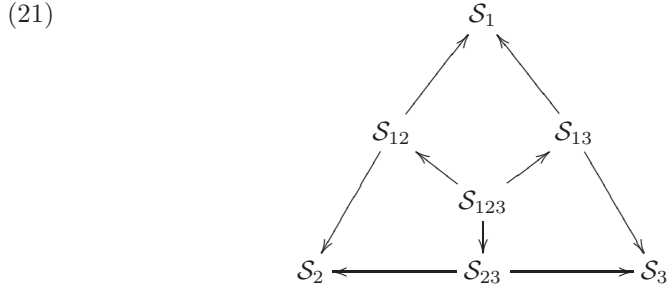
\begin{align}\label{dia:triangle}\xymatrix@=12pt{
						&	                             &	\mcS_{1}														&															&	\\
						&	                             &																			&															&	\\
						&	\mcS_{12}	\ar[uur]	\ar[ddl] &																			&	\mcS_{13}	\ar[ddr]	\ar[uul]	\\
						&	                             &	\mcS_{123}	\ar[d]	\ar[ul]	\ar[ur]																	\\
	\mcS_{2}	&	&	\mcS_{23}	\ar[ll]	\ar[rr]	 &																			&	\mcS_{3}
}\end{align}
represents three individuals $\{1,2,3\}$, 
their ologs $\{\mcS_{1},\mcS_{2},\mcS_{3}\}$, 
their pair-wise commonality ologs $\{\mcS_{12},\mcS_{13},\mcS_{23}\}$, and 
their three-way commonality olog $\mcS_{123}$. 
This diagram, which stands for the interaction between individuals $\{1,2,3\}$, does not stand alone, 
but is part of an intricate web of other ologs and alignment constraints. 
In particular,
individuals 1 and 3 may be part of some different interacting group, say of individuals $\{1,3,6,7\}$, 
and hence the right edge of the diagram would be part of some tetrahedron-shaped diagram with vertices $\{1,3,6,7\}$. 
If we take the point-of-view that 
``a collection of ologs representing the world-views of various individuals'' 
is a system, 
then we can think of the ologs as being the types of that system,
the morphisms connecting the ologs as being the aspects of that system, with
the shape of a system being its underlying graph.
In essence,
we can apply ologs to themselves.
In the system represented by diagram (\ref{dia:triangle}),
there are seven types $\{\mcS_{1},\mcS_{2},\mcS_{3},\mcS_{12},\mcS_{13},\mcS_{23},\mcS_{123}\}$
and nine aspects $\{\cdots,\mcS_{123}\to\mcS_{13},\dots\}$,
and the shape looks like this
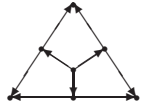
\begin{center}
\begin{tabular}{c}
\setlength{\unitlength}{0.4pt}
\begin{picture}(120,120)(0,-10)
\put(0,0){\circle*{6}}
\put(120,0){\circle*{6}}
\put(60,88){\circle*{6}}
\put(30,45){\vector(2,3){30}}
\put(30,45){\vector(-2,-3){30}}
\put(90,45){\vector(-2,3){30}}
\put(90,45){\vector(2,-3){30}}
\put(60,0){\vector(-1,0){60}}
\put(60,0){\vector(1,0){60}}
\put(60,26){\begin{picture}(0,0)(0,0)
\put(30,20){\circle*{6}}
\put(-30,20){\circle*{6}}
\put(0,0){\circle*{6}}
\put(0,-26){\circle*{6}}
\put(0,0){\vector(3,2){30}}
\put(0,0){\vector(-3,2){30}}
\put(0,0){\vector(0,-1){26}}
\end{picture}}
\end{picture}
\end{tabular}
\end{center}
In addition,
we can introduce certain facts to represent the meaning of that system
and then enforce those facts. 

A {\em distributed system} is a diagram (functor)
$\mathcal{G} \colon \mathrmbf{I} \rightarrow \mathrmbf{Gph}$
of shape $\mathrmbf{I}$
within the ambient category $\mathrmbf{Gph}$.
As such,
it consists of an indexed family 
$\{ G_{n} \mid n \in \mathrmbf{I} \}$
of graphs together with an indexed family 
$\{ G_{e} \colon G_{n} \rightarrow G_{m} \mid (e \colon n \rightarrow m) \in \mathrmbf{I} \}$
of graph morphisms.
Let $\mathrmbf{Dist}(\mathrmbf{I})$ denote 
the collection of distributed systems of shape $\mathrmbf{I}$.
An {\em information system} is a diagram
$\mathcal{S} \colon \mathrmbf{I} \rightarrow \mathrmbf{Spec}$
of shape $\mathrmbf{I}$
within the ambient category $\mathrmbf{Spec}$.
As such,
it consists of an indexed family 
$\{ \mathcal{S}_{n} = {\langle{G_{n},E_{n}}\rangle} \mid n \in \mathrmbf{I} \}$
of ologs together with an indexed family 
$\{ \mathcal{S}_{e} \colon \mathcal{S}_{n} \rightarrow \mathcal{S}_{m} \mid (e \colon n \rightarrow m) \in \mathrmbf{I} \}$
of olog morphisms.
Some of these ologs might represent the world-views of various individuals,
whereas others could be common grounds;
also included might be portals between individual ologs and common grounds,
as in the CG example of Section \ref{sec:CG}.
%
%
Let $\mathrmbf{Info}(\mathrmbf{I})$ denote 
the collection of information systems of shape $\mathrmbf{I}$.
An information system $\mathcal{S}$ with component ologs $\mathcal{S}_{n} = {\langle{G_{n},E_{n}}\rangle}$
has an underlying distributed system $\mathcal{G}$ of the same shape with component graphs $G_{n}$ for $n \in \mathrmbf{I}$.
%
%
For any distributed system $\mathcal{G}$,
let $\mathrmbfit{info}_{\mathrmbf{I}}(\mathcal{G})$ denote 
the collection of information systems over $\mathcal{G}$ of shape $\mathrmbf{I}$.
There is a pointwise entailment order $\mathcal{S} \leq^{\mathrmbf{I}}_{\mathcal{G}} \mathcal{S}'$
on $\mathrmbfit{info}_{\mathrmbf{I}}(\mathcal{G})$
when component ologs satisfy the same entailment ordering $E_{n} \leq_{G_{n}} E'_{n}$ 
for $n \in \mathrmbf{I}$,
and by taking the coproduct
there is a pointwise entailment order on
$\mathrmbf{Info}(\mathrmbf{I}) = \coprod_{\mathcal{G} \in \mathrmbf{Dist}(\mathrmbf{I})}\mathrmbfit{info}_{\mathrmbf{I}}(\mathcal{G})$.
A constant distributed system 
$\Delta(G) \in \mathrmbf{Dist}(\mathrmbf{I})$
is a distributed system
$\Delta(G) \colon \mathrmbf{I} \rightarrow \mathrmbf{Gph}$
with the same language $G$ for any index $n \in \mathrmbf{I}$.
Any constant distributed system defines join and meet monotonic functions
$\bigvee^{\mathrmbf{I}}_{G},\bigwedge^{\mathrmbf{I}}_{G} 
: \mathrmbfit{info}_{\mathrmbf{I}}(\Delta(G)) \rightarrow \mathrmbfit{fbr}(G)$
mapping an information system 
$\mathcal{S} \in \mathrmbfit{info}_{\mathrmbf{I}}(\Delta(G))$ 
to the join and meet ologs
$\bigvee\mathcal{S} = \bigcup_{n \in \mathrmbf{I}} E_{n}$ and
$\bigwedge\mathcal{S} = \bigcap_{n \in \mathrmbf{I}} E_{n}$
in $\mathrmbfit{fbr}(G)$.
The join monotonic function is adjoint to the constant monotonic function
$\Delta^{\mathrmbf{I}}_{G} 
: \mathrmbfit{fbr}(G) \rightarrow \mathrmbfit{info}_{\mathrmbf{I}}(\Delta(G))$
that distributes an olog $\mathcal{S}' \in \mathrmbfit{fbr}(G)$
to the various locations $n \in \mathrmbf{I}$
forming a constant information system $\Delta(\mathcal{S}') \in \mathrmbfit{info}_{\mathrmbf{I}}(\Delta(G))$,
since
$\bigvee\mathcal{S} \geq_{G} \mathcal{S}'$
iff
$\mathcal{S} \geq^{\mathrmbf{I}}_{\Delta(G)} \Delta(\mathcal{S}')$
for any system $\mathcal{S} \in \mathrmbfit{info}_{\mathrmbf{I}}(\Delta(G))$ 
and any olog $\mathcal{S}' \in \mathrmbfit{fbr}(G)$.

\subsubsection{System morphisms}

Just as ologs are linked by morphisms,
information systems are also linked by morphisms.
For these there is the new complication of shape.
In this paper we define fixed-shape system moorphisms,
but a more general definition would allow the shape to vary. 
%
A distributed system morphism 
$\theta \colon \mathcal{G} \Rightarrow \mathcal{G}'$
in $\mathrmbf{Dist}(\mathrmbf{I})$
consists of a collection 
$\{ \theta_{n} \colon G_{n} \rightarrow G'_{n} \mid n \in \mathrmbf{I} \}$
of component graph morphisms,
which are systematically coordinated in the sense that they satisfy the naturality conditions
$G_{e} \circ \theta_{m} = \theta_{n} \circ G'_{e}$
for any indexing link $e \colon n \rightarrow m$ in $\mathrmbf{I}$.
A direct flow operator 
$\mathrmbfit{dir}_{\mathrmbf{I}}(\theta) :
\mathrmbfit{info}_{\mathrmbf{I}}(\mathcal{G}) \rightarrow \mathrmbfit{info}_{\mathrmbf{I}}(\mathcal{G}')$
along $\theta$ can be define,
which maps an information system
$\mathcal{S} \in \mathrmbfit{info}_{\mathrmbf{I}}(\mathcal{G})$
to an information system
$\mathrmbfit{dir}_{\mathrmbf{I}}(\theta)(\mathcal{S}) \in \mathrmbfit{info}_{\mathrmbf{I}}(\mathcal{G}')$
defined by
$\mathrmbfit{dir}_{\mathrmbf{I}}(\theta)(\mathcal{S})_{n}
= \mathrmbfit{dir}(\theta_{n})(E_{n})$
for $n \in \mathrmbf{I}$.\footnote{Well-defined, since
$\mathrmbfit{dir}(G'_{e})(\mathrmbfit{dir}(\theta_{n})(E_{n}))
= \mathrmbfit{dir}(\theta_{m})(\mathrmbfit{dir}(G_{e})(E_{n}))
\geq_{m} \mathrmbfit{dir}(\theta_{m})(E_{m})$.}
An inverse flow operator
$\mathrmbfit{inv}_{\mathrmbf{I}}(\theta) :
\mathrmbfit{info}_{\mathrmbf{I}}(\mathcal{G}') \rightarrow \mathrmbfit{info}_{\mathrmbf{I}}(\mathcal{G})$
can similarly be defined.
Direct and inverse flow are adjoint monotonic functions
$
{\langle{\mathrmbfit{dir}_{\mathrmbf{I}}(\theta) \dashv \mathrmbfit{inv}_{\mathrmbf{I}}(\theta)}\rangle} :
\mathrmbfit{info}_{\mathrmbf{I}}(\mathcal{G}) \rightarrow \mathrmbfit{info}_{\mathrmbf{I}}(\mathcal{G}')$, since
$\mathrmbfit{dir}_{\mathrmbf{I}}(\theta)(\mathcal{S}) \geq^{\mathrmbf{I}}_{\mathcal{G}'} \mathcal{S}'$
iff
$\mathcal{S} \geq^{\mathrmbf{I}}_{\mathcal{G}} \mathrmbfit{inv}_{\mathrmbf{I}}(\theta)(\mathcal{S}')$.
An information system morphism 
$\theta \colon \mathcal{S} \Rightarrow \mathcal{S}'$
in $\mathrmbf{Info}(\mathrmbf{I})$
consists of a collection 
$\{ \theta_{n} \colon \mathcal{S}_{n} \rightarrow \mathcal{S}'_{n} \mid n \in \mathrmbf{I} \}$
of component olog morphisms,
which are systematically coordinated and preserve alignment
in the sense that they satisfy the naturality conditions
$\mathcal{S}_{e} \circ \theta_{m} = \theta_{n} \circ \mathcal{S}'_{e}$
for any indexing link $e \colon n \rightarrow m$ in $\mathrmbf{I}$;
equivalently,
$\theta$ is a morphism between the underlying distributed systems
$\theta \colon \mathcal{G} \Rightarrow \mathcal{G}'$
and the direct flow of $\mathcal{S}$ is at least as general as $\mathcal{S}'$:
$\mathrmbfit{dir}_{\mathrmbf{I}}(\theta)(\mathcal{S}) \geq^{\mathrmbf{I}}_{\mathcal{G}'} \mathcal{S}'$.
The ordering $\mathcal{S} \geq^{\mathrmbf{I}}_{\mathcal{G}} \mathcal{S}'$
is an information system morphism 
$\theta \colon \mathcal{S} \Rightarrow \mathcal{S}'$
with identity component translations
$\theta_{n} = \mathrmit{id}_{G_{n}}$ for each index $n \in \mathrmbf{I}$.
\subsubsection{Channels}

We continue with our systems point-of-view.
Since we have represented the whole system as a diagram $\mathcal{S}$ of parts (ologs) $\mathcal{S}_{n}$
with part-part relations (alignment constraints) $\mathcal{S}_{n} \rightarrow \mathcal{S}_{m}$,
we also want to represent the whole system as an olog $\mathcal{C}$
with part-whole relations $\mathcal{S}_{n} \rightarrow \mathcal{C}$.\footnote{The theory of part-whole relations is called mereology. 
It studies how parts are related to wholes, and how parts are related to other parts within a whole.}
An {\em information channel} 
${\langle{\gamma \colon \mathcal{M} \Rightarrow \Delta(C), C}\rangle}$
consists of an indexed family
$\{ \gamma_{n} \colon G_{n} \rightarrow C \mid n \in \mathrmbf{I} \}$
of graph morphisms called flow links 
with a common target graph $C$ called the core of the channel.
A channel ${\langle{\gamma,C}\rangle}$ covers a distributed system $\mathcal{G}$ of shape $\mathrmbf{I}$
when the part-whole relationships respect the alignment constraints 
(are consistent with the part-part relationships):
$\gamma_{n} = G_{e} \circ \gamma_{m}$ for each indexing morphism $e \colon n \rightarrow m$ in $\mathrmbf{I}$.
A covering channel is a 
distributed system morphism
$\gamma \colon \mathcal{G} \Rightarrow 
\Delta(C)$
in $\mathrmbf{Dist}(\mathrmbf{I})$
from distributed system $\mathcal{G}$
to constant distributed system $\Delta(C) \colon \mathrmbf{I} \rightarrow \mathrmbf{Gph}$.
Such a channel defines a direct flow operator
$\mathrmbfit{dir}_{\mathrmbf{I}}(\gamma) :
\mathrmbfit{info}_{\mathrmbf{I}}(\mathcal{G}) \rightarrow \mathrmbfit{info}_{\mathrmbf{I}}(\Delta(C))$
and an inverse flow operator
$\mathrmbfit{inv}_{\mathrmbf{I}}(\gamma) :
\mathrmbfit{info}_{\mathrmbf{I}}(\Delta(C)) \rightarrow \mathrmbfit{info}_{\mathrmbf{I}}(\mathcal{G})$.
For any two covering channels ${\langle{\gamma',C'}\rangle}$ and ${\langle{\gamma,C}\rangle}$
over the same distributed system $\mathcal{G}$,
a refinement $H \colon {\langle{\gamma',C'}\rangle} \rightarrow {\langle{\gamma,C}\rangle}$
is a graph morphism between cores $H \colon C' \rightarrow C$
that respects the part-whole relationships of the two channels:
$\gamma'_{n} \circ H = \gamma_{n}$ for $n \in \mathrmbf{I}$.
In such a situation, we say 
the channel ${\langle{\gamma',C'}\rangle}$ 
is a refinement of the channel ${\langle{\gamma,C}\rangle}$.
A channel ${\langle{\iota,\coprod\mathcal{G}}\rangle}$ 
is a minimal cover\footnote{Information flow terminology \cite{BS:IF}.} 
or optimal(ly refined covering) channel of a distributed system $\mathcal{G}$
when it covers $\mathcal{G}$ and for any other covering channel ${\langle{\gamma,C}\rangle}$
there is a unique refinement $[\gamma,C] \colon \coprod\mathcal{G} \rightarrow C$
from ${\langle{\iota,\coprod\mathcal{G}}\rangle}$ to ${\langle{\gamma,C}\rangle}$.

\subsubsection{System flow}

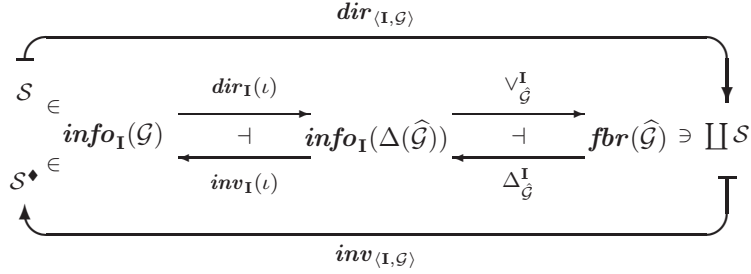
\begin{figure}
\begin{center}
\setlength{\unitlength}{0.83pt}
\begin{picture}(240,125)(0,0)
\put(0,60){\begin{picture}(0,0)(0,0)
\put(0,0){\makebox(0,0){\normalsize{$\mathrmbfit{info}_{\mathrmbf{I}}(\mathcal{G})$}}}
\put(120,0){\makebox(0,0){\normalsize{$\mathrmbfit{info}_{\mathrmbf{I}}(\Delta(\widehat{\mathcal{G}}))$}}}
\put(235,0){\makebox(0,0){\normalsize{$\mathrmbfit{fbr}(\widehat{\mathcal{G}})$}}}
\put(280,0){\makebox(0,0){\small{$\coprod\mathcal{S}$}}}
\put(261,0){\makebox(0,0){\footnotesize{$\ni$}}}
\put(-40,20){\makebox(0,0){\small{$\mathcal{S}$}}}
\put(-27,13){\makebox(0,0){\footnotesize{$\in$}}}
\put(-27,-13){\makebox(0,0){\footnotesize{$\in$}}}
\put(-40,-20){\makebox(0,0){\small{$\mathcal{S}^{\scriptscriptstyle\blacklozenge}$}}}
\put(120,55){\makebox(0,0){\small{$\mathrmbfit{dir}_{{\langle{\mathrmbf{I},\mathcal{G}}\rangle}}$}}}
\put(120,-55){\makebox(0,0){\small{$\mathrmbfit{inv}_{{\langle{\mathrmbf{I},\mathcal{G}}\rangle}}$}}}
\put(60,22){\makebox(0,0){\footnotesize{$\mathrmbfit{dir}_{\mathrmbf{I}}(\iota)$}}}
\put(60,-22){\makebox(0,0){\footnotesize{$\mathrmbfit{inv}_{\mathrmbf{I}}(\iota)$}}}
\put(185,22){\makebox(0,0){\footnotesize{$\vee^{\mathrmbf{I}}_{\hat{\mathcal{G}}}$}}}
\put(185,-22){\makebox(0,0){\footnotesize{$\Delta^{\mathrmbf{I}}_{\hat{\mathcal{G}}}$}}}
\put(30,10){\vector(1,0){60}}
\put(60,0){\makebox(0,0){\small{$\dashv$}}}
\put(90,-10){\vector(-1,0){60}}
\put(155,10){\vector(1,0){60}}
\put(185,0){\makebox(0,0){\small{$\dashv$}}}
\put(215,-10){\vector(-1,0){60}}
\thicklines
\put(-44,35){\line(1,0){8}}
\put(120,35){\oval(320,20)[t]}
\put(280,35){\vector(0,-1){20}}
\put(276,-19){\line(1,0){8}}
\put(280,-19){\line(0,-1){16}}
\put(120,-35){\oval(320,20)[b]}
\put(-40,-31){\vector(0,1){0}}
\end{picture}}
\end{picture}
\end{center}
\caption{System Flow}
\label{fig:system:flow}
\end{figure}

In order to represent an information system 
$\mathcal{S} = \{ \mathcal{S}_{n} \xrightarrow{\mathcal{S}_{e}} \mathcal{S}_{m} \}$
as a single olog $\coprod\mathcal{S}$, 
called the fusion of $\mathcal{S}$,
with part-whole relations $\mathcal{S}_{n} \rightarrow \coprod\mathcal{S}$,
we follow the colimit theorem of \cite{TBG:IC}
by recognizing the following three properties.
\begin{itemize}
\item 
Optimal channels exist for any distributed system $\mathcal{G}$.
%
\item 
$\mathrmbfit{fbr}(G)$ is a complete preorder for any graph $G$,
loosely called a ``lattice''.
\item 
For any graph morphism $H \colon G_{1} \rightarrow G_{2}$,
direct and inverse flow are adjoint monotonic functions
${\langle{\mathrmbfit{dir}(H),\mathrmbfit{inv}(H)}\rangle}
 \colon \mathrmbfit{fbr}(G_{1}) \rightarrow \mathrmbfit{fbr}(G_{2})$.
\end{itemize}
Let $\mathcal{G} \in \mathrmbf{Dist}(\mathrmbf{I})$ be a distributed system 
of shape $\mathrmbf{I}$
with optimal channel ${\langle{\iota,\coprod\mathcal{G}}\rangle}$.
The optimal core $\widehat{\mathcal{G}}=\coprod\mathcal{G}$ is called the sum of the distributed system $\mathcal{G}$,
and the optimal channel components (graph morphisms)
$\{ \iota_{n} \colon G_{n} \rightarrow \coprod\mathcal{G} \mid n \in \mathrmbf{I} \}$
are called flow links.
There is a {\em direct system flow} monotonic function (see Figure \ref{fig:system:flow})
$\mathrmbfit{dir}_{{\langle{\mathrmbf{I},\mathcal{G}}\rangle}}
= \mathrmbfit{dir}_{\mathrmbf{I}}(\iota) \cdot {\vee^{\mathrmbf{I}}_{\hat{\mathcal{G}}}} 
\colon \mathrmbfit{info}_{\mathrmbf{I}}(\mathcal{G}) \to \mathrmbfit{fbr}(\widehat{\mathcal{G}})$.
Direct system flow has two steps:
(i) direct (fixed shape) system flow of an information system along the optimal channel
($\mathrmbf{Dist}(\mathrmbf{I})$-morphism)
$\iota \colon \mathcal{G} \Rightarrow \Delta(\widehat{\mathcal{G}})$
and 
(ii) lattice join 
combining the contributions of the parts into a whole.
In the opposite direction,
there is an {\em inverse system flow} monotonic function (see Figure \ref{fig:system:flow})
$\mathrmbfit{inv}_{{\langle{\mathrmbf{I},\mathcal{G}}\rangle}}
= {\Delta^{\mathrmbf{I}}_{\hat{\mathcal{G}}}}  \cdot \mathrmbfit{inv}_{\mathrmbf{I}}(\iota)
\colon \mathrmbfit{fbr}(\widehat{\mathcal{G}}) \to \mathrmbfit{info}_{\mathrmbf{I}}(\mathcal{G})$.
Inverse system flow has two steps:
(i) mapping an olog with core language $\widehat{\mathcal{G}}$
to a constant information system over $\Delta(\widehat{\mathcal{G}})$ with shape $\mathrmbf{I}$ 
by distributing the olog to the locations $n \in \mathrmbf{I}$, and
(ii) inverse (fixed shape) system flow of this constant information system back along the optimal channel
$\iota \colon \mathcal{G} \Rightarrow \Delta(\widehat{\mathcal{G}})$.
Direct system flow is adjoint to inverse system flow
${\langle{\mathrmbfit{dir}_{{\langle{\mathrmbf{I},\mathcal{G}}\rangle}} \dashv \mathrmbfit{inv}_{{\langle{\mathrmbf{I},\mathcal{G}}\rangle}}}\rangle} 
\colon \mathrmbfit{info}_{\mathrmbf{I}}(\mathcal{G}) \rightarrow \mathrmbfit{fbr}(\widehat{\mathcal{G}})$,
since the composition components are adjoint.
For 
any distributed system $\mathcal{G} \in \mathrmbf{Dist}(\mathrmbf{I})$
with optimal core $\widehat{\mathcal{G}}=\coprod\mathcal{G}$,
any information system $\mathcal{S} \in \mathrmbfit{info}_{\mathrmbf{I}}(\mathcal{G})$, and
any olog $\widehat{\mathcal{S}} \in \mathrmbfit{fbr}(\widehat{\mathcal{G}})$,
entailment satisfies the following axioms.
\begin{center}
{\scriptsize
\begin{tabular}{r@{\hspace{10pt}}p{250pt}}
(direct flow)
&
If
$E_{n}$ entails the equation 
$(f =_{G_{n}} f') \colon i \rightarrow j$,
then
$\mathrmbfit{dir}_{{\langle{\mathrmbf{I},\mathcal{G}}\rangle}}(\mathcal{S})$ entails the equation 
$(\iota_{n}^\ast(f) =_{\hat{\mathcal{G}}} \iota_{n}^\ast(f')) \colon \iota_{n}(i) \rightarrow \iota_{n}(j)$
for any $n \in \mathrmbf{I}$.
\\
(inverse flow)
&
If
$\widehat{\mathcal{S}}$ entails the equation 
$(\iota_{n}^\ast(f) =_{\hat{\mathcal{G}}} \iota_{n}^\ast(f')) \colon \iota_{n}(i) \rightarrow \iota_{n}(j)$,
then
$\mathrmbfit{inv}_{{\langle{\mathrmbf{I},\mathcal{G}}\rangle}}(\widehat{\mathcal{S}})_{n}$ 
entails the equation 
$(f =_{G_{n}} f') \colon i \rightarrow j$
for any $n \in \mathrmbf{I}$.
\end{tabular}}
\end{center}
These are converted to inference rules in Table~\ref{entailment:inference:rules}.
Information flow can be used 
to compute the fusion olog for an information system and 
to define the consequence of an information system. 
Fusion is direct system flow, and
consequence is the composition of direct and inverse system flow.
Let
$\mathcal{S} \in \mathrmbfit{info}_{\mathrmbf{I}}(\mathcal{G})$
be any information system. 
The fusion 
$\coprod\mathcal{S}
= \mathrmbfit{dir}_{{\langle{\mathrmbf{I},\mathcal{G}}\rangle}}(\mathcal{S})
= {\langle{\coprod\mathcal{G},\bigvee_{n \in \mathrmbf{I}} \mathrmbfit{dir}(\iota_{n})(E_{n})}\rangle}
\in  \mathrmbfit{fbr}(\widehat{\mathcal{G}})$ 
is an olog that represents the whole system in a centralized fashion
\cite{K:SI},\cite{K:SC}.
The consequence 
$\mathcal{S}^{\scriptscriptstyle\blacklozenge}_{{\langle{\mathrmbf{I},\mathcal{G}}\rangle}}
= \mathrmbfit{inv}_{{\langle{\mathrmbf{I},\mathcal{G}}\rangle}}(\mathrmbfit{dir}_{{\langle{\mathrmbf{I},\mathcal{G}}\rangle}}(\mathcal{S}))
= \mathrmbfit{inv}_{{\langle{\mathrmbf{I},\mathcal{G}}\rangle}}(\coprod\mathcal{S})
= \{ \mathrmbfit{inv}(\iota_{n})(\coprod\mathcal{S}) \mid n \in \mathrmbf{I} \} 
\in \mathrmbfit{info}_{\mathrmbf{I}}(\mathcal{G})$
is an information system that represents the whole system in a distributed fashion
\cite{K:SC}.
It is inverse flow of the fusion olog along the optimal channel,
transfering the entailed facts of the whole system to the component parts.

The consequence operator $(\mbox{-})^{\scriptscriptstyle\blacklozenge}$,
which is defined on information systems,
is a closure operator 
on the complete preorder $\mathrmbfit{info}_{\mathrmbf{I}}(\mathcal{G})$, 
%
%
and by taking the coproduct
it is a closure operator on the complete preorder 
$\mathrmbf{Info}(\mathrmbf{I}) = \coprod_{\mathcal{G} \in \mathrmbf{Dist}(\mathrmbf{I})}\mathrmbfit{info}_{\mathrmbf{I}}(\mathcal{G})$ :
(increasing) $\mathcal{S} \geq \mathcal{S}^{\scriptscriptstyle\blacklozenge}$,
(monotonic)  $\mathcal{S} \geq \mathcal{S}'$ implies $\mathcal{S}^{\scriptscriptstyle\blacklozenge} \geq \mathcal{S}'^{\scriptscriptstyle\blacklozenge}$ and
(idempotent) $\mathcal{S}^{\scriptscriptstyle\blacklozenge\blacklozenge} = \mathcal{S}^{\scriptscriptstyle\blacklozenge}$.
\footnote{By allowing system shape to vary,
channels can be generalized to morphisms of distributed systems.
Then a notion of relative fusion (direct system flow) can be defined in terms of left Kan extension,
and a notion of relative system consequence can be defined as 
the composition of direct followed by inverse system flow.}
\underline{Pointwise} entailment order $\leq$ on $\mathrmbf{Info}(\mathrmbf{I})$
is only a preliminary order,
since it does not incorporate interactions between system component parts.
\underline{System} entailment order $\preceq$ on $\mathrmbf{Info}(\mathrmbf{I})$
is defined by 
$\mathcal{S}_{1} \preceq \mathcal{S}_{2}$
when
$\mathcal{S}_{1}^{\scriptscriptstyle\blacklozenge} \leq \mathcal{S}_{2}^{\scriptscriptstyle\blacklozenge}$;
equivalently,
$\mathcal{S}_{1}^{\scriptscriptstyle\blacklozenge} \leq \mathcal{S}_{2}$.
Pointwise order is stronger than system entailment order: 
$\mathcal{S}_{1} \leq \mathcal{S}_{2}$ implies $\mathcal{S}_{1} \preceq \mathcal{S}_{2}$.
This is a specialization-generalization order.
Any information system $\mathcal{S}$ is entailment equivalent to its consequence 
$\mathcal{S} \cong \mathcal{S}^{\scriptscriptstyle\blacklozenge}$.
An information system $\mathcal{S}$ is closed when it is equal to its consequence 
$\mathcal{S} = \mathcal{S}^{\scriptscriptstyle\blacklozenge}$.

The whole effect of taking the system consequence may be greater than the sum of its parts,
in the sense that
$\mathcal{S}_{n} 
\geq_{n} 
\mathcal{S}_{n}^{{\scriptscriptstyle\blacklozenge}_{\iota_{n}}}
\geq_{n} \bigvee_{m}\mathrmbfit{inv}(\iota_{n})(\mathrmbfit{dir}(\iota_{m})(\mathcal{S}_{m})) 
\geq_{n} \mathcal{S}^{\scriptscriptstyle\blacklozenge}_{n}$ 
for any $n \in \mathrmbf{I}$,
since separate parts may have a productive interaction at the channel core.
A final part of an information system is a part with no non-trivial constraint links \underline{from} it. 
(The graphical subsystem beneath) nonfinal parts are necessary for the alignment of information systems,
resulting in the equivalencing of types and aspects through quotienting.
However,
because of 
the covering condition $\iota_{n} = G_{e} \circ \iota_{m}$ and
the entailment order $\mathrmbfit{dir}(G_{e})(E_{n}) \geq_{m} E_{m}$
for constraint links $\mathcal{S}_{e} \colon \mathcal{S}_{n} \to \mathcal{S}_{m}$,
only the fact(ual) content of final parts of information systems are necessary to compute the system fusion and consequence.
\begin{table}
\begin{center}
{\scriptsize 
\renewcommand{\arraystretch}{1.5}
\begin{tabular}{l@{\hspace{30pt}}l@{\hspace{20pt}}l}
{\bfseries equivalence:}
&
(reflexive)
&
{\begin{tabular}{c} \\ \hline $(f =_{G} f) \colon i \rightarrow j$ \end{tabular}}
\\
&
(symmetric)
& 
{\begin{tabular}{c} $(f_{1} =_{G} f_{2}) \colon i \rightarrow j$ 
\bigstrut \\ \hline 
$(f_{2} =_{G} f_{1}) \colon i \rightarrow j$ \end{tabular}}
\\
&
(transitive)
& 
{\begin{tabular}{c} $(f_{1} =_{G} f_{2}) \colon i \rightarrow j$, $(f_{2} =_{G} f_{3}) \colon i \rightarrow j$ 
\bigstrut \\ \hline 
$(f_{1} =_{G} f_{3}) \colon i \rightarrow j$ \end{tabular}}
\\
{\bfseries algebra:}
&
(compositional)
& 
{\begin{tabular}{c} $(f_{1} =_{G} f_{2}) \colon i \rightarrow j$, $(g_{1} =_{G} g_{2}) \colon j \rightarrow k$ 
\bigstrut \\ \hline 
$(f_{1} {\;;\;} g_{1} =_{G} f_{2} {\;;\;} g_{2}) \colon i \rightarrow k$ \end{tabular}}
\\
&
(bi-closed)
& 
{\begin{tabular}{c} $(g_{1} =_{G} g_{2}) \colon j \rightarrow k$ 
\bigstrut \\ \hline 
$(f {\,;\,} g_{1} =_{G} f {\;;\;} g_{2}) \colon i \rightarrow k$, $(g_{1} {\;;\;} h =_{G} g_{2} {\,;\,} h) \colon j \rightarrow l$ \end{tabular}}
\\ && \\
{\bfseries morphic flow:}
&
(direct)
& 
{\begin{tabular}{c} $(f_{1} =_{G_{1}} f'_{1}) \colon i_{1} \rightarrow j_{1}$
\bigstrut \\ \hline 
$(H^\ast(f_{1}) =_{G_{2}} H^\ast(f'_{1})) \colon H(i_{1}) \rightarrow H(j_{1})$
\end{tabular}}
\\
&
(inverse)
& 
{\begin{tabular}{c} 
$(H^\ast(f_{1}) =_{G_{2}} H^\ast(f'_{1})) \colon H(i_{1}) \rightarrow H(j_{1})$
\bigstrut \\ \hline 
$(f_{1} =_{G_{1}} f'_{1}) \colon i_{1} \rightarrow j_{1}$
\end{tabular}}
\\ && \\
{\bfseries system flow:}
&
(direct)
& 
{\begin{tabular}{c} $(f =_{G_{n}} f') \colon i \rightarrow j$
\bigstrut \\ \hline 
$(\iota_{n}^\ast(f) =_{\hat{\mathcal{G}}} \iota_{n}^\ast(f')) \colon \iota_{n}(i) \rightarrow \iota_{n}(j)$
\end{tabular}}
\\
&
(inverse)
& 
{\begin{tabular}{c} $(\iota_{n}^\ast(f) =_{\hat{\mathcal{G}}} \iota_{n}^\ast(f')) \colon \iota_{n}(i) \rightarrow \iota_{n}(j)$
\bigstrut \\ \hline 
$(f =_{G_{n}} f') \colon i \rightarrow j$
\end{tabular}}
\\ && \\
\end{tabular}}
\end{center}
\caption{Inference Rules}
\label{entailment:inference:rules}
\end{table}

\subsubsection{General examples}

Here are some examples of system fusion/consequence.
\begin{itemize}
\item 
An information system $\mathcal{S}$ with a constant underlying distributed system,
$G_{i} = G$ for all $n \in \mathrmbf{I}$,
gathers together all the component parts of the information system and forms their consequence.
It has identity flow links
$\{ \iota_{n} = \mathrmit{id}_{G} \colon G \rightarrow G = \coprod\mathcal{G} \mid n \in \mathrmbf{I} \}$,
component join fusion 
$\coprod\mathcal{S} 
= \bigvee_{n \in \mathrmbf{I}}\mathcal{S}_{n}
= {\langle{G,\bigcup_{n \in \mathrmbf{I}}E_{n}}\rangle}$,
and constant system consequence
$\mathcal{S}^{\scriptscriptstyle\blacklozenge}_{n}
= \left(\bigvee_{n' \in \mathrmbf{I}}\mathcal{S}_{n'}\right)^{\scriptscriptstyle\bullet}$
for all $n \in \mathrmbf{I}$.
\item 
A discrete information system $\mathcal{S} = \{ \mathcal{S}_{n}={\langle{G_{n},E_{n}}\rangle} \mid n \in \mathrmbf{I} \}$
with no constraint links $G_{e} \colon \mathcal{S}_{n} \rightarrow \mathcal{S}_{m}$ for $n \neq m$,
has coproduct injection flow links
$\iota_{n} \colon G_{n} \rightarrow \mbox{\Large$+$}_{n \in \mathrmbf{I}}\,G_{n}$,
non-restricting fusion,
and inverse flow projecting back to individual component consequence
$\mathcal{S}^{\scriptscriptstyle\blacklozenge}_{n}
= \mathcal{S}_{n}^{\scriptscriptstyle\bullet}$
for all $n \in \mathrmbf{I}$.
No alignment (constraint) links means no interaction.
%
%
\item 
An information system 
$\mathcal{S} = \{ \mathcal{S}_{1} \xleftarrow{H_{1}} \mathcal{S} \xrightarrow{H_{2}} \mathcal{S}_{2} \}$ 
consisting of a single common ground $\mathcal{S}={\langle{G,E}\rangle}$
between two component ologs 
$\mathcal{S}_{1}={\langle{G_{1},E_{1}}\rangle}$ and
$\mathcal{S}_{2}={\langle{G_{2},E_{2}}\rangle}$,
with underlying distributed system (span)
$\mathcal{G} = \{ G_{1} \xleftarrow{H_{1}} G \xrightarrow{H_{2}} G_{2} \}$, 
has pushout injection flow links
$G_{1} \xrightarrow{\iota_{1}} \coprod\mathcal{G} \xleftarrow{\iota_{2}} G_{2}$,
direct image union fusion
$\coprod\mathcal{S}
= {\langle{\coprod\mathcal{G},\mathrmbfit{dir}(\iota_{1})(E_{1})\cup\mathrmbfit{dir}(\iota_{2})(E_{2})}\rangle}$,
and system consequence components 
$\mathcal{S}^{\scriptscriptstyle\blacklozenge}_{n} 
= {\langle{G_{n},
\mathrmbfit{inv}(\iota_{n})(\mathrmbfit{dir}(\iota_{1})(E_{1})\cup\mathrmbfit{dir}(\iota_{2})(E_{2}))}\rangle}$ 
for $n = 1,2$.
The flow links will quotient any types and aspects that are connected through the common ground
allowing for the approprate interaction in the fusion consequence
$(\mathrmbfit{dir}(\iota_{1})(E_{1})\cup\mathrmbfit{dir}(\iota_{2})(E_{2}))^{\scriptscriptstyle\bullet}$,
then the inverse flow will reconnect this with the component types and aspects.
\end{itemize}

\subsection{Conceptual graphs}\label{sec:CG}

The conceptual graph formalism (CG)
for knowledge representation \cite{S:KR},
was initially formulated to represent database systems (DBS),
but is now used in natural language processing (NLP) and first-order logic (FOL).
%
%
Verbs in NLP can often be represented relationally by star(-shaped conceptual) graphs.
For example,
the sentence ``John is going to Boston by bus'' might be represented by the conceptual graph
{\scriptsize
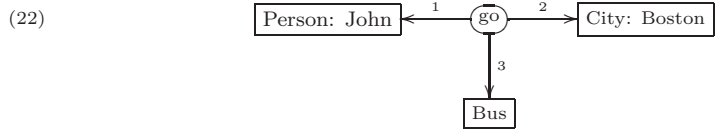
\begin{align}\label{dia:cg:eg}\xymatrix{
*+[F]{\text{\footnotesize{Person: John}}}&*+[F-:<13pt>]{\text{go}}\ar[l]_-{1}\ar[d]^-{3}\ar[r]^-{2}&*+[F]{\text{City: Boston}}\\
&*+[F]{\text{Bus}}&
}\end{align}}\hspace{-3pt}
In a sentence of natural language,
thematic roles are semantic descriptions of 
the way (the entities described by) a noun phrase functions with respect to (the action of) the verb. 
These entities are the participants in the occurrent expressed by the verb.
For the action of `going' in the above sentence
there are three participants and hence three thematic roles.
`John' plays the role of the agent of the action,
a `Bus' is the instrument used in the action and
`Boston' is the destination of the action.
Translations using thematic roles
can be used to align two ontologies with respect to a common ground.
A CG-style translation of conceptual graph (\ref{dia:cg:eg})
would replace the verb relation `going' with a concept `Go'
and replace the edges that form the signature of the `going' relation with binary relations for the three roles
`agent', `instrument' and `destination'.
{\scriptsize
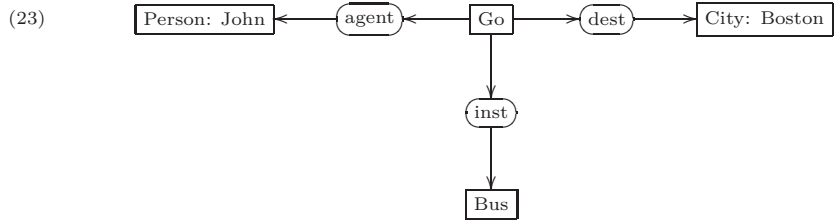
\begin{align}\label{dia:cg:trans}\xymatrix{
*+[F]{\text{Person: John}} & *+[F-:<13pt>]{\text{agent}} \ar[l]  
& *+[F]{\text{Go}} \ar[l]\ar[d]\ar[r] & *+[F-:<13pt>]{\text{dest}} \ar[r] & *+[F]{\text{City: Boston}}\\
&& *+[F-:<13pt>]{\text{inst}}\ar[d]&&\\
&& *+[F]{\text{Bus}}&&
}\end{align}}\hspace{-3pt}
However,
the case relations that semantically describe the thematic roles should be viewed as functional in nature;
that is,
for any instance of the action of a sentence's verb 
there is a unique entity described by a noun phrase of the sentence.
When this semantics is respected,
the translation to thematic roles becomes a process of ``linearization'',
which is best described abstractly as:
(1) the identification of relation types with entity types,
(2) the translation of a sorted multiarity relation to a span of functions, one function for each role, and
(3) the functional interpretation of thematic roles.

The Olog formalism,
which also represents DBS and NLP,
is a version of equational logic.
Both the Olog and CG formalisms were designed as graphical representations.
However,
the CG formalism is binary and relational,
whereas the Olog formalism is unary and functional.
The CG formalism is binary since it has two kinds of type, concepts and relations;
it is relational in the way it interprets edges.
The Olog formalism is unary since it has only one kind of type, the abstract concept;
it is functional in the way it interprets aspects (edges).
However,
much of the semantics of the CG formalism can be transformed to the Olog formalism by the process of linearization\footnote{The linearization process works for any binary/relational knowledge representation,
such as CGs, entity-relationship data modelling \cite{JRW:ERA},
relational database systems \cite{K:DBS} or the Information Flow Framework \cite{IFF}.
In the entity-relationship data modelling,
$n$-ary relationship links are replaced by $n$-ary spans of aspects and attributes are included as types.},
thereby gaining in efficiency and conciseness.
For example,
conceptual graph (\ref{dia:cg:eg})
can be linearized to the olog graph\footnote{\fakebox{1} is 
the universal type to which all types have a unique aspect.}
{\scriptsize
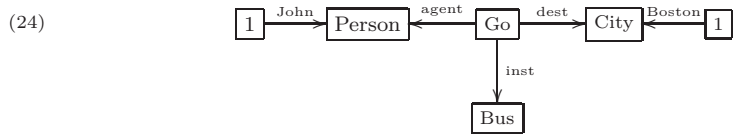
\begin{align}\label{dia:olog:trans}\xymatrix{
*+[F]{\text{\footnotesize{1}}}\ar[r]^-{\text{John}}&*+[F]{\text{\footnotesize{Person}}}&*+[F]{\text{Go}}\ar[l]_-{\text{agent}}\ar[d]^-{\text{inst}}\ar[r]^-{\text{dest}}&*+[F]{\text{City}}&*+[F]{\text{1}}\ar[l]_-{\text{Boston}}
\\
&&*+[F]{\text{Bus}}&&
}\end{align}}\hspace{-3pt}
Since olog aspects are interpreted functionally, the functional nature of thematic roles is respected. 
In this manner,
the olog formalism could be used to replace the CG representation of ontologies.
For example,
a community (acting as an individual) could build its ontology $\mathcal{C}$ from ground up
by aligning it with some top-level reference ontology $\mathcal{T}$ 
(such as in the appendix of \cite{S:KR}), 
thereby importing some formal semantics from $\mathcal{T}$.
The following fragment demonstrates how this works.

Assume that ontology $\mathcal{T}$
contains the concept of ``spatial process'' 
as represented by
the general concept type \fakebox{Spatial-Process} with aspects
$\fakebox{Spatial-Process}\xrightarrow{\text{agent}}\fakebox{Agent}$, 
$\fakebox{Spatial-Process}\xrightarrow{\text{inst}}\fakebox{Vehicle}$ and 
$\fakebox{Spatial-Process}\xrightarrow{\text{dest}}\fakebox{Location}$.
At some stage assume that the community ontology $\mathcal{C}$
has specified
the concept type orderings
$\fakebox{Person}\leq\fakebox{Agent}$,
$\fakebox{Bus}\leq\fakebox{Vehicle}$ and
$\fakebox{City}\leq\fakebox{Location}$
with corresponding injective aspects
$\fakebox{Person}\xrightarrow{\text{is}}\fakebox{Agent}$,
$\fakebox{Bus}\xrightarrow{\text{is}}\fakebox{Vehicle}$ and
$\fakebox{City}\xrightarrow{\text{is}}\fakebox{Location}$.
At the next stage it could
define
a concept type \fakebox{C}
with aspects
$\fakebox{C}\xrightarrow{\text{person}}\fakebox{Person}$, 
$\fakebox{C}\xrightarrow{\text{bus}}\fakebox{Bus}$ and 
$\fakebox{C}\xrightarrow{\text{city}}\fakebox{City}$,
and link it 
with the reference ontology concept \fakebox{Spatial-Process}
by specifying a connecting aspect
$\fakebox{C}\xrightarrow{\text{process}}\fakebox{Spatial-Process}$
and
asserting the facts
`$\mbox{person} {\;;\;} \mbox{is} = \mbox{process} {\;;\;} \mbox{agent}$', 
`$\mbox{bus} {\;;\;} \mbox{is} = \mbox{process} {\;;\;} \mbox{vehicle}$' and 
`$\mbox{city} {\;;\;} \mbox{is} = \mbox{process} {\;;\;} \mbox{location}$'.\footnote{The symbol `$;$' denotes concatenation or formal composition.}
In the more expressive ologs with joins (Section~\ref{sec:expressive I}),
the process concept of ``going to city by bus'' can then be defined
as the pullback of the ``spatial process'' concept:
here,
the concept type \fakebox{Go}
with aspects
$\fakebox{Go}\xrightarrow{\text{person}}\fakebox{Person}$, 
$\fakebox{Go}\xrightarrow{\text{bus}}\fakebox{Bus}$ and 
$\fakebox{Go}\xrightarrow{\text{city}}\fakebox{City}$
is pulled back along the above injective aspects,
resulting in the
injective aspect $\fakebox{Go}\xrightarrow{\text{is}}\fakebox{Spatial-Process}$
with corresponding concept type ordering
$\fakebox{Go}\leq\fakebox{Spatial-Process}$.
As a result,
the concept \fakebox{C} has the new mediating aspect
$\mbox{C}\xrightarrow{\text{going}}\mbox{Go}$,
which satisfies the fact
`$\mbox{going} {\;;\;} \mbox{is} = \mbox{process}$'.
In this manner the community ontology $\mathcal{C}$ has been enlarged.
{\scriptsize
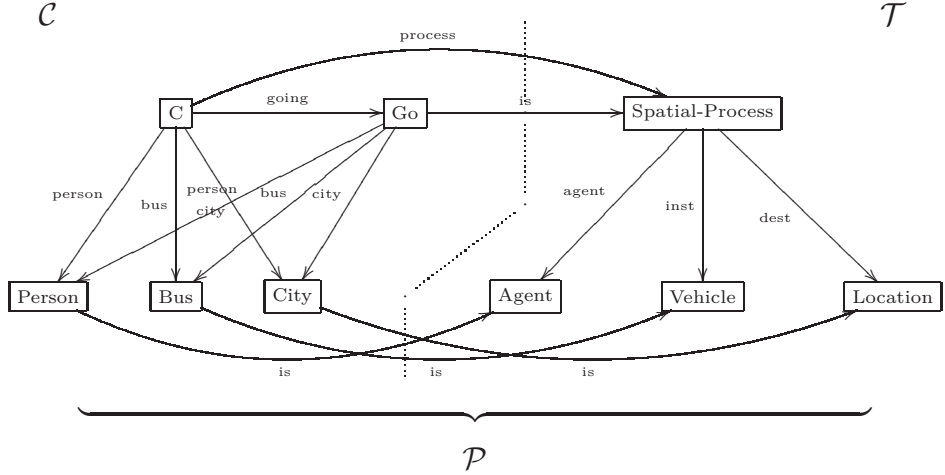
\begin{gather*}
\xymatrix{
{\mbox{\Large$\mathcal{C}$}}&&&&&&{\mbox{\Large$\mathcal{T}$}}
\\
&*+[F]{\text{C}}\ar[ddl]_-{\text{person}}\ar[dd]_-{\text{bus}}\ar[ddr]_-{\text{city}}\ar[rr]^-{\text{going}}\ar@/^2pc/[rrrr]^-{\text{process}}&&*+[F]{\text{Go}}\ar[ddlll]_-{\text{person}}\ar[ddll]_-{\text{bus}}\ar[ddl]_-{\text{city}}\ar[rr]^-{\text{is}}&{\cdot}\ar@{.}[d]\ar@{.}[u]&*+[F]{\text{Spatial-Process}}\ar[ddl]_-{\text{agent}}\ar[dd]_-{\text{inst}}\ar[ddr]_-{\text{dest}}&
\\
&&&&{\cdot}\ar@{.}[dl]&&
\\
*+[F]{\text{Person}}\ar@/_2pc/[rrrr]_-{\text{is}}&*+[F]{\text{Bus}}\ar@/_2pc/[rrrr]_-{\text{is}}&*+[F]{\text{City}}\ar@/_2pc/[rrrr]_-{\text{is}}&{\cdot}\ar@{.}[d]&*+[F]{\text{Agent}}&*+[F]{\text{Vehicle}}&*+[F]{\text{Location}}
\\
&&&&&&
}
\\
\overset{\underbrace{\rule{300pt}{0pt}}}
{\mbox{\Large\rule{0pt}{16pt}$\mathcal{P}$}}
\end{gather*}}\hspace{-3pt}
We assume that community ontology $\mathcal{C}$ and reference ontology $\mathcal{T}$ 
are combined into a portal ontology $\mathcal{P}$ 
with portal link $\mathcal{C} \xrightarrow{P} \mathcal{P}$
and alignment link $\mathcal{T} \xrightarrow{A} \mathcal{P}$.
If some other ontology $\mathcal{C}'$ is built up and aligned in the same fashion,
then $\mathcal{T}$ is being used as a common ground,
and we have a `{\sffamily W}'-shaped information system
{\begin{align}\label{dia:cg-sys}\xymatrix{
{\mathcal{C}}\ar[dr]_-{P}&&{\mathcal{T}}\ar[dl]_-{A}\ar[dr]^-{A'}&&{\mathcal{C}'}\ar[dl]^-{P'}
\\
&{\mathcal{P}}&&{\mathcal{P}'}&
}\end{align}}\hspace{-3pt}
with portals $\mathcal{P}$ and $\mathcal{P}'$ being the final parts.
This `{\sffamily W}'-shaped information system uses the sketch institution {\ttfamily Sk} for ologs.
It can be compared to the `{\sffamily W}'-shaped information system in \cite{K:CCQ},
which uses the information flow {\ttfamily IF} institution for (local) logics.


\section{More expressive ologs I}\label{sec:expressive I}

In this section and the next (\ref{sec:expressive I} and \ref{sec:expressive II}) we will introduce limits and colimits within the context of ologs. These will allow authors to build ologs that are quite expressive. For example we can declare one type to be the union or intersection of other types. We do not assume mathematical knowledge beyond that of sets and functions, which were loosely defined in Section \ref{sec:aspects}. However, the reader may benefit by consulting a reference on category theory, such as \cite{Awo}.

The basic ologs discussed in previous sections are based on the mathematical notion of categories, whereas the olog presentation language we will discuss in this section and the next are based on {\em general sketches} (see \cite{Mak}). The difference is in what can be expressed: in basic ologs we can declare types, aspects, and facts, whereas in general ologs we can express ideas like products and sums, as we will see below. 

\comment{
Many ideas can be expressed using only basic ologs, and when this is possible it is preferred. Many more ideas can be expressed with the addition of only ``layouts" (corresponding to {\em finite limit sketches}), and again an author who can restrict himself or herself to this language will benefit for it. The reason is that when authors use a richer olog presentation language, the ability to compare different ologs becomes harder in two ways. First, it becomes harder to make meaningful connections (functors or sketch-maps) between different ologs, and second the set of theorems available for transferring instance data (see Section \ref{sec:instance data}) from one olog to the other becomes restricted. 
}

We will begin by discussing layouts, which will be represented categorically by ``finite limits". As usual, the english terminology (layout) is not precise enough to express the notion we mean it to express (limit). Intuitively, a limit can be thought of as a system: it is a collection of units, each of a specific type, such that these units have compatible aspects. These will include types like \fakebox{a man and a woman with the same last name}. In Section \ref{sec:expressive II} we will discuss groupings, which will be represented by colimits. These will include types like \fakebox{a thing that is either a pear or a watermelon}. 

\subsection{Layouts}

A dictionary might define the word {\em layout} as something like ``a structured arrangement of items within certain limits; a plan for such arrangement."  In other words, we can lay out or specify the need for a set of parts, each of a given type, such that the parts fit together well. This idea roughly corresponds to the notion of limits in category theory, especially limits in the category of sets. Given a diagram of sets and functions, its limit is the set of ways to accordingly choose one element from each. For example, we could have a type \fakebox{a car and a driver}, which category-theoretically is a product, but which we are calling a ``layout" --- a compound type whose parts are ``laid out."  Of course, the term layout is insufficient to express the precise meaning of limits, but it will have to do for now.  To understand limits, one really only need understand pullbacks and products. These will be the subjects of Sections \ref{sec:pullbacks} and \ref{sec:products}, or one can see \cite{Awo} for more details. 

\subsection{Pullbacks}\label{sec:pullbacks}

Given three objects and two arrows arranged as to the left, the pullback is the commutative square to the right: $$\tn{Given:}\hspace{.3in}\parbox{.7in}{\xymatrix{&C\ar[d]^g\\B\ar[r]_f&D}}\hspace{.3in}\tn{the pullback is drawn:}\hspace{.3in}\parbox{.7in}{\xymatrix{A\ar[r]^{f'}\ar[d]_{g'}&C\ar[d]^g\\B\ar[r]_f&D.}}$$ We write $A=B\cross_DC$ and say ``$A$ is the pullback of $B$ and $C$ over $D$."  The question is, what does it signify? We will begin with some examples and then give a precise definition.

\begin{example}\label{ex:4 pbs}

We will now give four examples to motivate the definition of pullback. In the first example, (\ref{dia:pb1}), both $B$ and $C$ will be subtypes of $D$, and in such cases the pullback will be their intersection. In the next two examples (\ref{dia:pb2} and \ref{dia:pb3}), only $B$ will be a subtype of $D$, and in such cases the pullback will be the ``corresponding subtype of $C$" (as should make sense upon inspection). In the last example (\ref{dia:pb4}), neither $B$ nor $C$ will be a subtype of $D$.  In each line below, the pullback of the diagram to the left is the diagram to the right. The reader should  think of the left-hand olog as a kind of problem to which the new box $A$ in the right-hand olog is a solution.   

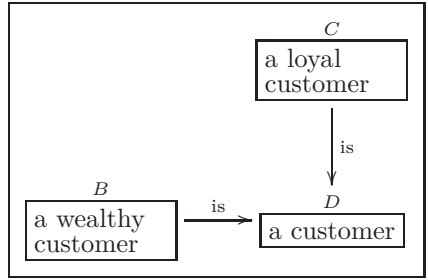
\begin{align}\label{dia:pb1}\fbox{\xymatrix{&\obox{C}{.7in}{\rr a loyal customer}\LA{d}{is}\\\obox{B}{.7in}{\rr a wealthy customer}\LA{r}{is}&\smbox{D}{a customer}}}\hsp&\fbox
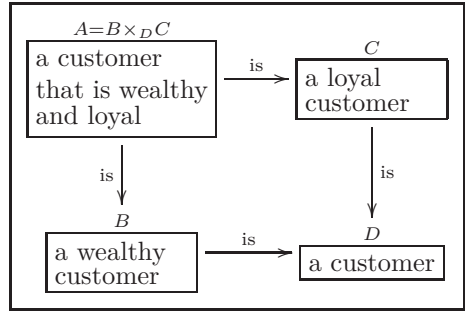
{\xymatrix{\obox{A=B\cross_DC}{.9in}{\rr a customer that is wealthy and loyal}\LAL{d}{is}\LA{r}{is}&\obox{C}{.7in}{\rr a loyal customer}\LA{d}{is}\\\obox{B}{.7in}{\rr a wealthy customer}\LA{r}{is}&\smbox{D}{a customer}}}
\\\label{dia:pb2}\fbox
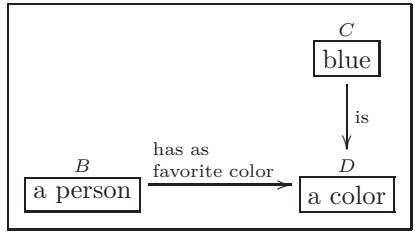
{\xymatrix{&&\smbox{C}{blue}\LA{d}{is}\\\smbox{B}{a person}\LA{rr}{\parbox{.7in}{\rr has as\\\vspace{-.05in}favorite color}}&&\smbox{D}{a color}}}\hsp&\fbox
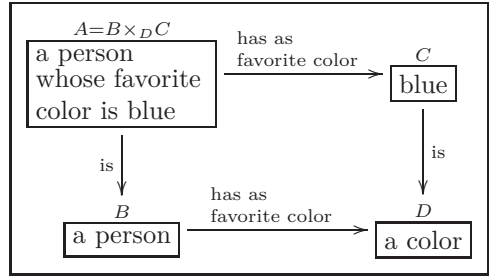
{\xymatrix{\obox{A=B\cross_DC}{.9in}{\rr a person whose favorite color is blue}\LAL{d}{is}\LA{rr}{\parbox{.7in}{\rr has as\\\vspace{-.05in}favorite color}}&&\smbox{C}{blue}\LA{d}{is}\\\smbox{B}{a person}\LA{rr}{\parbox{.7in}{\rr has as\\\vspace{-.05in}favorite color}}&&\smbox{D}{a color}}}\\
\label{dia:pb3}\fbox
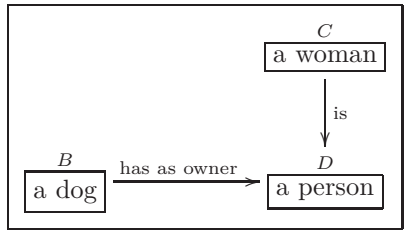
{\xymatrix{&&\smbox{C}{a woman}\LA{d}{is}\\\smbox{B}{a dog}\LA{rr}{\parbox{.7in}{\rr has as owner}}&&\smbox{D}{a person}}}\hsp&\fbox
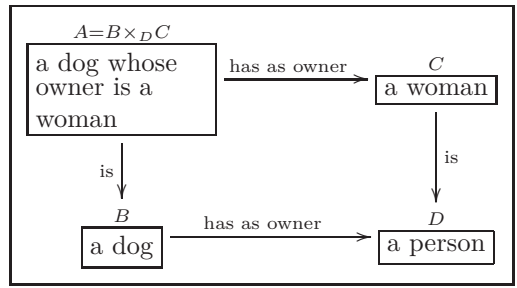
{\xymatrix{\obox{A=B\cross_DC}{.9in}{\rr a dog whose owner is a woman}\LAL{d}{is}\LA{rr}{\parbox{.7in}{\rr has as owner}}&&\smbox{C}{a woman}\LA{d}{is}\\\smbox{B}{a dog}\LA{rr}{\parbox{.7in}{\rr has as owner}}&&\smbox{D}{a person}}}
\\\label{dia:pb4}\fbox
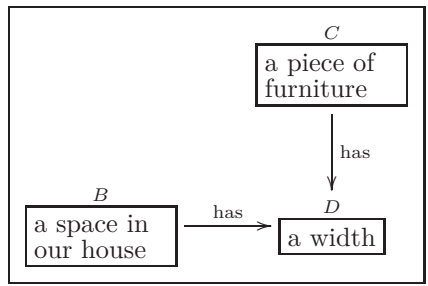
{\xymatrix{&\obox{C}{.7in}{\rr a piece of furniture}\LA{d}{has}\\\obox{B}{.7in}{\rr a space in our house}\LA{r}{has}&\smbox{D}{a width}}}\hsp&\fbox
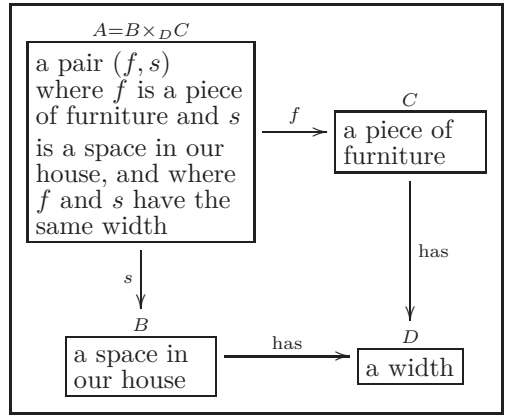
{\xymatrix{\obox{A=B\cross_DC}{1.1in}{\rr a pair $(f,s)$ where $f$ is a piece of furniture and $s$ is a space in our house, and where $f$ and $s$ have the same width}\LAL{d}{$s$}\LA{r}{$f$}&\obox{C}{.7in}{\rr a piece of furniture}\LA{d}{has}\\\obox{B}{.7in}{\rr a space in our house}\LA{r}{has}&\smbox{D}{a width}}}
\end{align}

See Example \ref{ex:justify pb example} for a justification of these, in light of Definition \ref{def:pullback}.

\end{example}

The following is the definition of pullbacks in the category of sets. For an olog, the instance data are given by sets (at least in this paper, see Section \ref{sec:instances}), so this definition suffices for now. See \cite{Awo} for more details on pullbacks.

\begin{definition}\label{def:pullback}

Let $B, C,$ and $D$ be sets, and let $f\taking B\to D$ and $g\taking C\to D$ be functions. The {\em pullback} of $B\To{f}D\From{g}C$, denoted $B\cross_DC$, is defined to be the set $$B\cross_DC:=\{(b,c)\;|\; b\in B, c\in C, \tn{ and } f(b)=g(c)\}$$ together with the obvious maps $B\cross_DC\to B$ and $B\cross_DC\to C$, which send an element $(b,c)$ to $b$ and to $c$, respectively. In other words, the pullback of $B\To{f}D\From{g}C$ is a commutative square $$\xymatrix{B\cross_DC\ar[r]\ar[d]&C\ar[d]^g\\B\ar[r]_f&D.}$$

\end{definition}

\begin{example}\label{ex:justify pb example}

In Example \ref{ex:4 pbs} we gave four examples of pullbacks. For each, we will consider $B\To{f}D\From{g}C$ to be sets and functions as in Definition \ref{def:pullback} and explain how the set $A$ follows that definition, i.e. how its label fits with the set $B\cross_DC=\{(b,c)\;|\; b\in B, c\in C, \tn{ and } f(b)=g(c)\}$.

In the case of (\ref{dia:pb1}), the set $B\cross_DC$ should consist of pairs $(w,l)$ where $w$ is a wealthy customer, $l$ is a loyal customer, and $w$ is equal to $l$ (as customers). But if $w$ and $l$ are the same customer then $(w,l)$ is just a customer that is both wealthy and loyal, not two different customers. In other words, an instance of the pullback is a customer that is both loyal and wealthy, so the label of $A$ fits.

In the case of (\ref{dia:pb2}), the set $B\cross_DC$ should consist of pairs $(p,b)$ where $p$ is a person, $b$ is the color blue, and the favorite color of $p$ is equal to $b$ (as colors). In other words, it is a person whose favorite color is blue, so the label of $A$ fits. If desired, one could instead label $A$ with \fakebox{a pair $(p,b)$ where $p$ is a person, $b$ is blue, and the favorite color of $p$ is $b$}.

In the case of (\ref{dia:pb3}), the set $B\cross_DC$ should consist of pairs $(d,w)$ where $d$ is a dog, $w$ is a woman, and the owner of $d$ is equal to $w$ (as people). In other words, it is a dog whose owner is a woman, so the label of $A$ fits. If desired, one could instead label $A$ with \fakebox{a pair $(d,w)$ where $d$ is a dog, $w$ is a woman, and the owner of $d$ is $w$}.

In the case of (\ref{dia:pb4}), the set $B\cross_DC$ should consist of pairs $(f,s)$ where $f$ is a piece of furniture, $s$ is a space in our house, and the width of $f$ is equal to the width of $s$. This is fits perfectly with the label of $A$.

\end{example}

\setcounter{subsubsection}{3}\subsubsection{Using pullbacks to classify}

To distinguish between two things, one must find a common aspect of the two things for which they have differing results. For example, a pen is different from a pencil in that they both use some material to write (a common aspect), but the two materials they use are different. Thus the material which a writing implement uses is an aspect of writing implements, and this aspect serves to segregate or classify them. We can think of three such writing-materials: graphite, ink, and pigment-wax. For each, we will make a layout in the olog below: $$\fbox
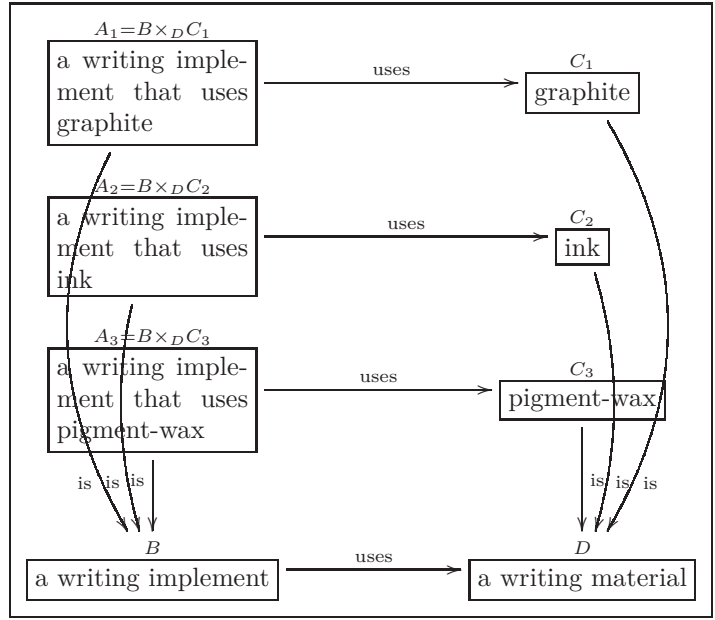
{\xymatrix@=12pt{\mebox{A_1=B\cross_DC_1}{a writing implement that uses graphite}\LA{rrrr}{uses}\ar@/_2.7pc/[dddd]_(.81){\tn{is}}&&&&\smbox{C_1}{graphite}\ar@/^2.7pc/[dddd]^(.81){\tn{is}}\\\mebox{A_2=B\cross_DC_2}{a writing implement that uses ink}\LA{rrrr}{uses}\ar@/_1pc/[ddd]_(.73){\tn{is}}&&&&\smbox{C_2}{ink}\ar@/^1pc/[ddd]^(.73){\tn{is}}\\\mebox{A_3=B\cross_DC_3}{a writing implement that uses pigment-wax}\LA{rrrr}{uses}\ar[dd]_{\tn{is}}&&&&\smbox{C_3}{pigment-wax}\ar[dd]^{\tn{is}}\\\\\smbox{B}{a writing implement}\LA{rrrr}{uses}&&&&\smbox{D}{a writing material}}}$$

One could also replace the label of box $A_1$ with ``a pencil", the label of box $A_2$ with ``a pen", and the label of box $A_3$ with ``a crayon"; in so doing, the layouts at the top would {\em define} a pencil, a pen, and a crayon to be a writing implement that uses respectively graphite, ink, and pigment-wax. 

\subsubsection{Building pullbacks on pullbacks}

There is a theorem in category theory which states the following. Suppose given two commutative squares $$
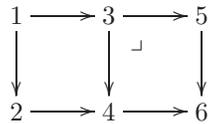
\xymatrix{1\ar[r]\ar[d]&3\ar[r]\ar[d]\ullimit&5\ar[d]\\2\ar[r]&4\ar[r]&6}$$ such that the right-hand square (3,4,5,6) is a pullback. It follows that if the left-hand square (1,2,3,4) is a pullback then so is the big rectangle (1,2,5,6). It also follows that if the big rectangle (1,2,5,6) is a pullback then so is the left-hand square (1,2,3,4). This fact can be useful in authoring ologs.

For example, the type \fakebox{a cellphone that has a bad battery} is vague, but we can lay out precisely what it means using pullbacks:\small$$\fbox
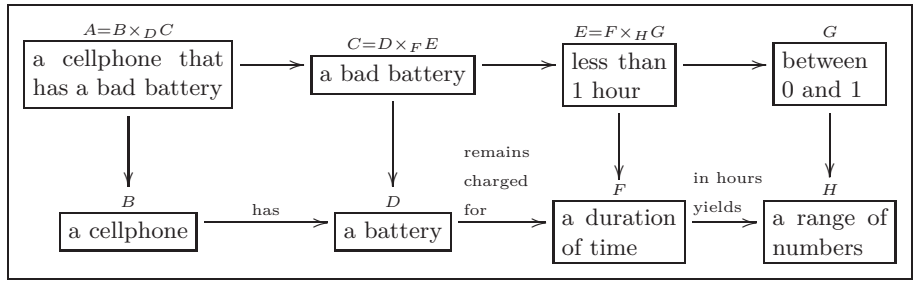
{\xymatrix{\obox{A=B\cross_DC}{1in}{a cellphone that has a bad battery}\ar[r]\ar[d]&\smbox{C=D\cross_FE}{a bad battery}\ar[r]\ar[d]&\obox{E=F\cross_HG}{.5in}{less than 1 hour}\ar[r]\ar[d]&\obox{G}{.5in}{between 0 and 1}\ar[d]\\\smbox{B}{a cellphone}\LA{r}{has}&\smbox{D}{a battery}\LA{r}{\parbox{.4in}{\rr remains charged for}}&\obox{F}{.6in}{a duration of time}\LA{r}{\hspace{.07in}\parbox{.4in}{\rr in hours yields}}&\obox{H}{.6in}{a range of numbers}}}$$\normalsize

The category-theoretic fact described above says that since $A=B\cross_DC$ and $C=D\cross_FE$, it follows that $A=B\cross_FE$. That is, we can decuce the definition ``a cellphone that has a bad battery is defined as a cellphone that has a battery which remains charged for less than one hour."  In other words, $A=B\cross_FE$.

\subsection{Products}\label{sec:products}

Given a set of types (boxes) in an olog, one can select one instance from each. All the ways of doing just that comprise what is called the product of these types. For example, if $A=\fakebox{a number between 1 and 10}$ and $B=\fakebox{a letter between x and z}$, the product includes a total of 30 elements, including $(4,z)$. We are ready for the definition.

\begin{definition}\label{def:product2}

Given sets $A,B$, their {\em product}, denoted $A\cross B$, is the set $$A\cross B=\{(a,b)\;|\; a\in A \tn{ and } b\in B\}.$$  There are two obvious {\em projection maps} $A\cross B\to A$ and $A\cross B\to B$, sending the pair $(a,b)$ to $a$ and to $b$ respectively.

\end{definition}

\begin{example}\label{ex:furniture}

In Example \ref{ex:4 pbs}, (\ref{dia:pb4}) we presented the idea of a piece of furniture that was the same width as a space in the house. What if we say that \fakebox{a nice furniture placement} is any space that is between 1 and 8 inches bigger than a piece of furniture?  We can use a combination of products and pullbacks to create the appropriate type. $$\fbox
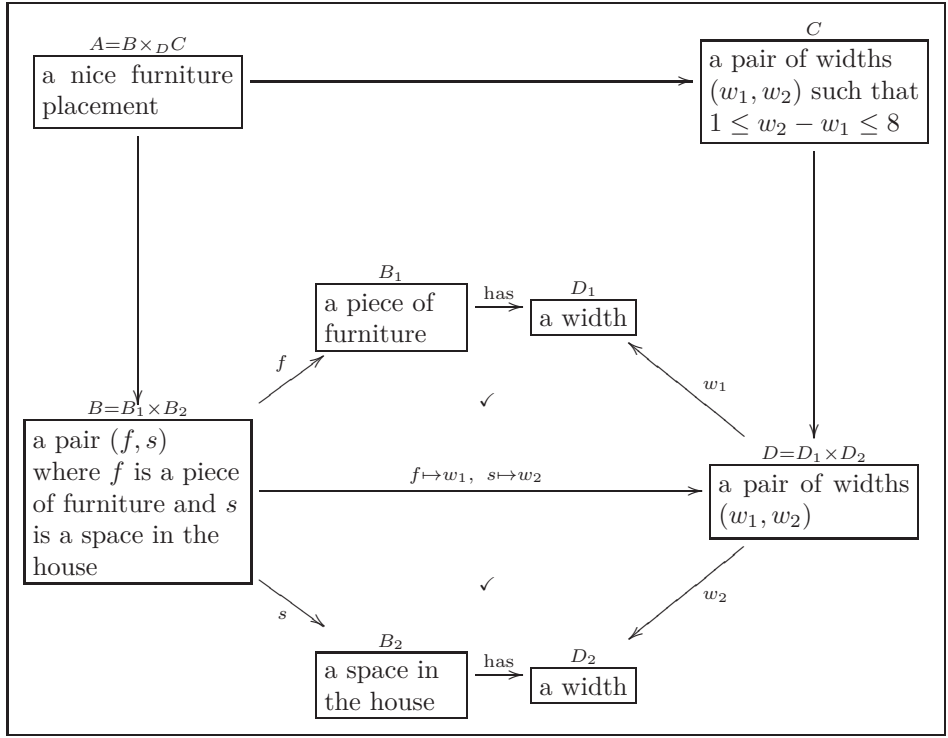
{\xymatrix@=18pt{\obox{A=B\cross_DC}{1in}{a nice furniture placement}\ar[ddd]\ar[rrr]&&&\obox{C}{1.1in}{\rr a pair of widths $(w_1,w_2)$ such that $1\leq w_2-w_1\leq 8$}\ar[ddd]\\\\&\obox{B_1}{.7in}{\rr a piece of furniture}\ar@{}[dr]|{\checkmark}\LA{r}{has}&\smbox{D_1}{a width}\\\obox{B=B_1\cross B_2}{1.1in}{\rr a pair $(f,s)$ where $f$ is a piece of furniture and $s$ is a space in the house}\ar[dr]_-s\ar[ur]^-f\ar[rrr]^{f\mapsto w_1,\;\;s\mapsto w_2}&&&\mebox{D=D_1\cross D_2}{a pair of widths $(w_1,w_2)$}\ar[ul]_{w_1}\ar[dl]^{w_2}\\&\obox{B_2}{.7in}{\rr a space in the house}\ar@{}[ur]|{\checkmark}\LA{r}{has}&\smbox{D_2}{a width}}}$$

Here $B$ and $D$ are products and $A$ is a pullback. This olog lays out what it means to be ``a nice furniture placement" using products. The bottom horizontal aspect $B\to D$ is an example of a map obtained by the ``universal property of products"; see Section \ref{sec:universal property}.

\end{example}

\setcounter{subsubsection}{2}\subsubsection{Products of more (or fewer) types}

The product of two sets $A$ and $B$ was defined in \ref{def:product2}. One may also take the product of three sets $A,B,C$ in a similar way, so the elements are triples $(a,b,c)$ where $a\in A, b\in B,$ and $c\in C$. In fact this idea holds for any number of sets. It even makes sense to take the product of one set (just $A$) or no sets!  The product of one set is itself, and the product of no sets is the singleton set $\{*\}$. For more on this, see Section \ref{sec:singleton} or \cite{Mac}.

\subsection{Declaring an injective aspect}\label{sec:injective}

A function is called {\em injective} if different inputs always yield different outputs. For example the function that doubles every integer ($x\mapsto 2x$) is injective, whereas the function that squares every integer ($x\mapsto x^2$) is not because $3^2=(-3)^2$. An example of an injective aspect is $\fakebox{a woman}\To{\tn{is}}\fakebox{a person}$ because different women are always different as people. An example of a non-injective aspect is $\fakebox{a person}\To{\tn{has as father}}\fakebox{a person}$ because different people may have the same father. 

The easiest way to indicate that an aspect is injective is to use a ``hook arrow" as in $f\taking A\inj B$, instead of a regular arrow $f\taking A\to B$, to denote it. For example, the first map is injective (and specified as such with a hook-arrow), but the second is not in the olog: $$\fbox
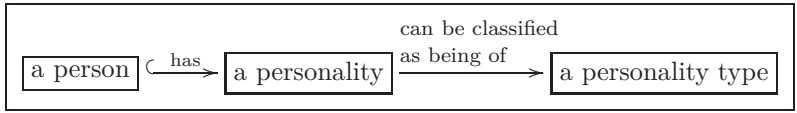
{\xymatrix{\fbox{a person}\;\ar@{^(->}[r]^-{\tn{has}}&\fbox{a personality}\ar[rr]^{\parbox{.9in}{\footnotesize\rr\tn{can be classified as being of}}\normalsize}&&\fbox{a personality type}}}$$ The author of this olog believes that no two people can have precisely the same personality (though they may have the same personality type).

We include injective aspects in this section because it turns out that injectivity can also be specified by pullbacks. See \cite{nlab-mono} for details.

\subsection{Singletons types}\label{sec:singleton}

A singleton set is a set with one element; it can be considered the ``empty product."  In other words if we denote $A^n=A\cross A\cross\cdots A$ (where $A$ is written $n$ times), then $A^0$ is the empty product and is a singleton set. One can specify that a certain type has only one instance by annotating it with $A=\{*\}$ in the olog. For example the olog $$\fbox
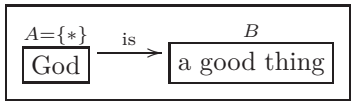
{\xymatrix{\smbox{A=\{*\}}{God}\LA{r}{is}&\smbox{B}{a good thing}}}$$ says that the author considers \fakebox{God} to be single. As a more concrete example, the intersection of $\{x\in\RR\;|\;x\geq 0\}$ and $\{y\in\RR\;|\;x\leq 0\}$ is a singleton set, as expressed in the olog $$\fbox
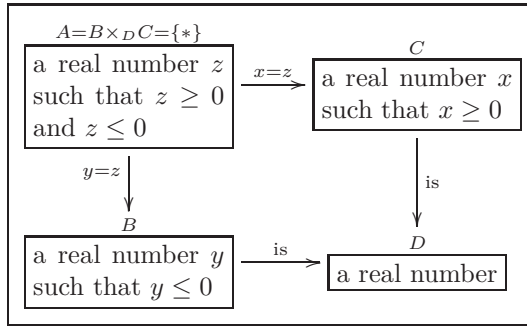
{\xymatrix{\mebox{A=B\cross_DC=\{*\}}{a real number $z$ such that $z\geq 0$ and $z\leq 0$}\ar[r]^{x=z}\ar[d]_{y=z}&\mebox{C}{a real number $x$ such that $x\geq 0$}\LA{d}{is}\\\mebox{B}{a real number $y$ such that $y\leq 0$}\LA{r}{is}&\smbox{D}{a real number}}}$$  The fact that $A=B\cross_DC$ and $A=\{*\}$ are declared indicates that there is only one possible instance of a real number that is in both $B$ and $C$.

\subsection{The universal property of layouts}\label{sec:universal property}

We cannot do the notion of universal properties justice in this paper, but the basic idea is as follows. Suppose that $\mcD$ is an olog, that $D_1,D_2$ are types in it, and that $D=D_1\cross D_2$ (together with its projection maps $p_1\taking D\to D_1$ and $p_2\taking D\to D_2$) is their product. \begin{align}\label{dia:prod1}\xymatrix@=10pt{&D_1\cross D_2\ar[ddr]^(.6){p_2}\ar[ddl]_(.6){p_1}\\\\D_1&&D_2}\end{align} The so-called universal property of products should be thought of as ``an existence and uniqueness" claim in $\mcD$. Namely, for any type $X$ with maps $f\taking X\to D_1$ and $g\taking X\to D_2$, there is exactly one possible map $m\taking X\to D$ such that the facts $f=m;p_1$ and $g=m;p_2$ hold. \begin{align}\label{dia:prod2}\xymatrix@=15pt{X\ar[rrrdd]^(.65)g\ar[ddr]_f&&&&&&X\ar[rdd]_f\ar[rrrdd]^(.65)g\ar[rr]^{m}&&D_1\cross D_2\ar[ldd]_(.6){p_1}\ar[rdd]^(.6){p_2}\\&&&&\leadsto\\&D_1&&D_2&&&&D_1&&D_2}\end{align}This may sound esoteric, but consider the following example.

The following olog is similar to the one in Example \ref{ex:furniture} $$\fbox
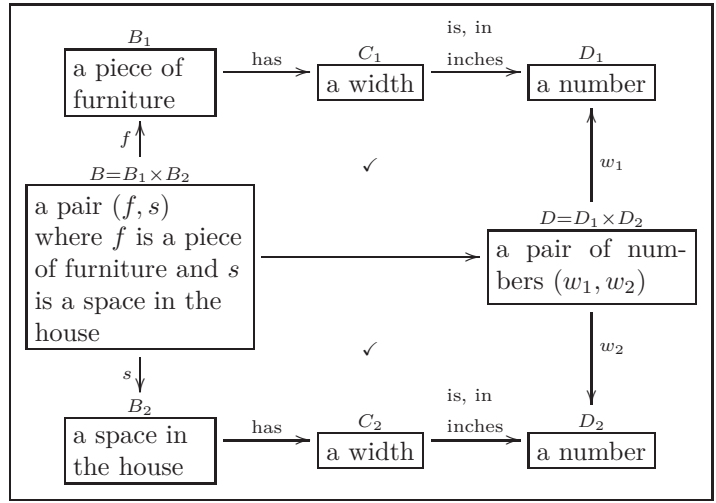
{\xymatrix@=18pt{\obox{B_1}{.7in}{\rr a piece of furniture}\LA{r}{has}&\smbox{C_1}{a width}\ar@{}[d]|{\checkmark}\LA{r}{\parbox{.3in}{\rr is, in inches}}&\smbox{D_1}{a number}\\\obox{B=B_1\cross B_2}{1.1in}{\rr a pair $(f,s)$ where $f$ is a piece of furniture and $s$ is a space in the house}\ar[d]_-s\ar[u]^-f\ar[rr]&&\mebox{D=D_1\cross D_2}{a pair of numbers $(w_1,w_2)$}\ar[u]_{w_1}\ar[d]^{w_2}\\\obox{B_2}{.7in}{\rr a space in the house}\LA{r}{has}&\smbox{C_2}{a width}\ar@{}[u]|{\checkmark}\LA{r}{\parbox{.3in}{\rr is, in inches}}&\smbox{D_2}{a number}}}$$  Here the only unlabeled map is the horizontal one $B\to D$; how can we get away with leaving it unlabeled?  How does a piece of furniture and a space in the house yield a pair of numbers?  The answer is that $B$ has a map to $D_1$ (the path across the top) and a map to $D_2$ (the path across the bottom), and hence the universal property of products gives a unique arrow $B\to D$ such that the two facts indicated by checkmarks hold. (In terms of (\ref{dia:prod1}) and (\ref{dia:prod2}) we are using $X=B$.)  In other words, there is exactly one way to take a piece of furniture and a space in the house and yield a pair of numbers if we enforce that the first number is the width in inches of the piece of furniture and the second number is the width in inches of the space in the house.

At this point we hope it is clear that the universal property of products is a useful and constructive one. We will not describe the other universal properties (either for pullbacks, singletons, or any colimits); as mentioned above they can be found in \cite{Awo}.


\section{More expressive ologs II}\label{sec:expressive II}

In this section we will describe various colimits, which are in some sense dual to limits. Whereas limits allow one to ``lay out" a team consisting of many different interacting or non-interacting parts, colimits allow one to ``group" different types together. For example, whereas the product of \fakebox{a number between 1 and 10} and \fakebox{a letter between x and z} has 30 elements (such as $(3,y)$), the coproduct of these two types has 13 elements (including 4). Just as ``layout" is a too weak a word to capture the essence of limits, ``grouping" is too weak a word to capture the essence of colimits, but it will have to do.

We will start by describing coproducts or ``disjoint unions" in Section \ref{sec:coproducts}. Then we will describe pushouts in Section \ref{sec:pushouts}, wherein one can declare some elements in a union to be equivalent to others. There is a category-theoretic duality between coproducts and products and between pushouts and pullbacks. It extends to a duality between surjections and injections and a duality between empty types and singleton types, the subject of Sections \ref{sec:surjective} and \ref{sec:empty}. The interested reader can see \cite{Awo} for details.

\subsection{Coproducts}\label{sec:coproducts}

Coproducts are also called ``disjoint unions."  If $A$ and $B$ are sets with no members in common, then the coproduct of $A$ and $B$ is their union. However, if they have elements in common, one must include both copies in $A\amalg B$ and differentiate between them. Here is a definition.

\begin{definition}\label{def:coproduct}

Given sets $A$ and $B$, their {\em coproduct}, denoted $A\amalg B$, is the set $$A\amalg B=\{(a,``A")\;|\;a\in A\} \cup \{(b,``B")\;|\;b\in B\}.$$  There are two obvious {\em inclusion maps} $A\to A\amalg B$ and $B\to A\amalg B$, sending $a$ to $(a,``A")$ and $b$ to $(b,``B")$, respectively.

\end{definition}

If $A$ and $B$ have no elements in common, then the one can drop the $``A"$ and ``$B$" labels without changing the set $A\amalg B$ in a substantial way. Here are two examples that should make the coproduct idea clear.

\begin{example}\label{ex:coproduct1}

In the following olog the types $A$ and $B$ are disjoint, so the coproduct $C=A\amalg B$ is just the union. $$\fbox{\xymatrix{\smbox{A}{a person}\LA{r}{is}&\smbox{C=A\amalg B}{a person or a cat}&\smbox{B}{a cat}\LAL{l}{is}}}$$

\end{example}

\begin{example}\label{ex:coproduct2}

In the following olog, $A$ and $B$ are not disjoint, so care must be taken to differentiate common elements. $$\fbox
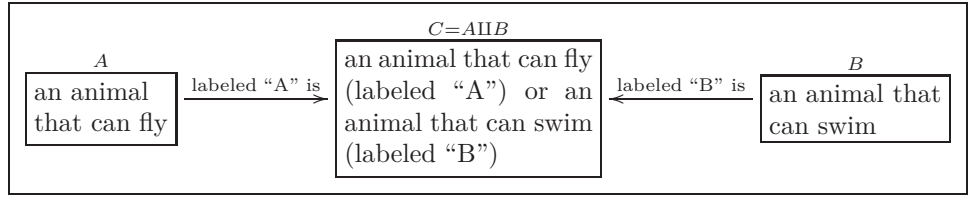
{\xymatrix{\obox{A}{.7in}{\rr an animal that can fly}\LA{rr}{labeled ``A" is}&&\obox{C=A\amalg B}{1.3in}{an animal that can fly (labeled ``A") or an animal that can swim (labeled ``B")}&&\obox{B}{.9in}{\rr an animal that can swim}\LAL{ll}{labeled ``B" is}}}$$  Since ducks can both swim and fly, each duck is found twice in $C$, once labeled as a flyer and once labeled as a swimmer. The types $A$ and $B$ are kept disjoint in $C$, which justifies the name ``disjoint union."

\end{example}

\subsection{Pushouts}\label{sec:pushouts}

Pushouts can express unions in which an overlap is declared. They can also express ``quotients," where different objects can be declared equivalent. Given three objects and two arrows arranged as to the left, the pushout is drawn as the commutative square to the right: $$\tn{Given:}\hspace{.3in}\parbox{.7in}
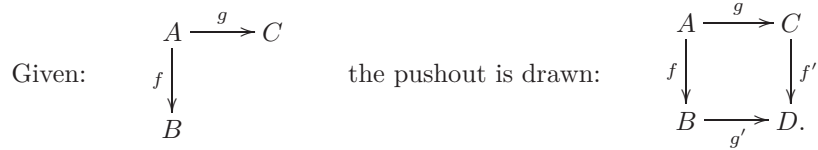
{\xymatrix{A\ar[r]^g\ar[d]_f&C\\B}}\hspace{.3in}\tn{the pushout is drawn:}\hspace{.3in}\parbox{.7in}{\xymatrix{A\ar[r]^{g}\ar[d]_{f}&C\ar[d]^{f'}\\B\ar[r]_{g'}&D.}}$$ We write $D=B\amalg_AC$ and say ``$D$ is the pushout of $B$ and $C$ along $A$."  The question is, what does it signify?  

The idea is that an instance of the pushout $B\amalg_AC$ is any instance of $B$ or any instance of $C$, but where some instances are considered equivalent to others. That is, for any instance of $A$, its $B$-aspect is considered the same as its $C$-aspect. This is formalized in Definition \ref{def:pushout} after being exemplified in Example \ref{ex:pushout}.

\begin{example}\label{ex:pushout}

In each example below, the diagram to the right is the pushout of the diagram to the left. The new object, $D$, is the union of $B$ and $C$, but instances of $A$ are equated to their $B$ and $C$ aspects. This will be discussed after the two diagrams.

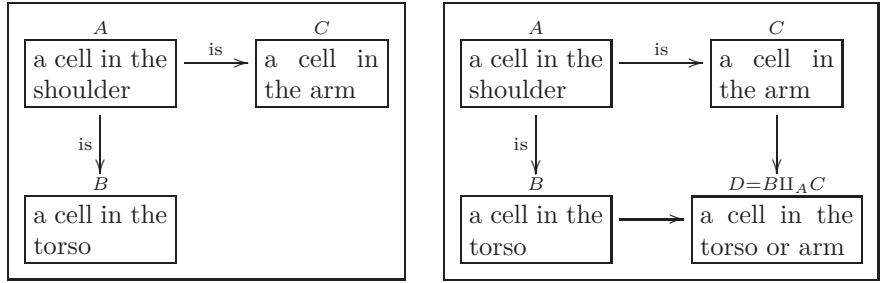
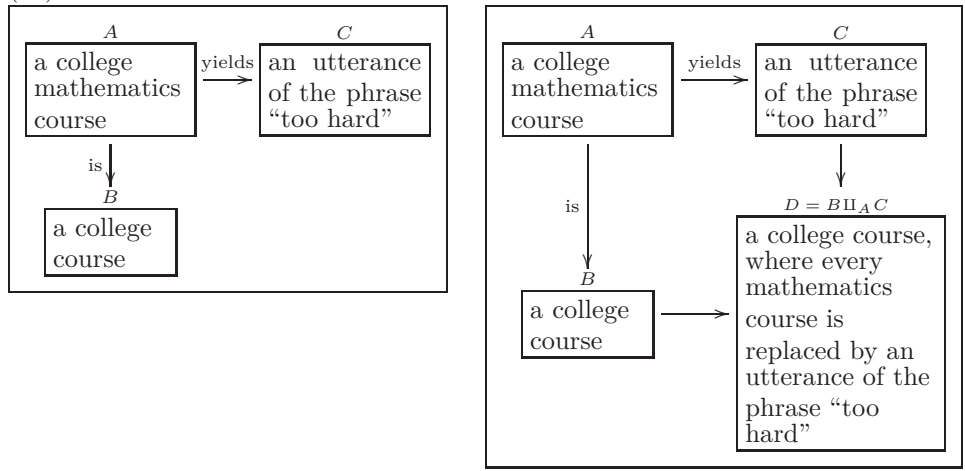
\begin{align}
\label{dia:po1}\fbox{\xymatrix{\obox{A}{.7in}{a cell in the shoulder}\LA{r}{is}\LAL{d}{is}&\obox{C}{.6in}{a cell in the arm}\\\obox{B}{.7in}{a cell in the torso}}}\hsp&\fbox{\xymatrix{\obox{A}{.7in}{a cell in the shoulder}\LA{r}{is}\LAL{d}{is}&\obox{C}{.6in}{a cell in the arm}\LA{d}{}\\\obox{B}{.7in}{a cell in the torso}\LA{r}{}&\obox{D=B\amalg_AC}{.8in}{a cell in the torso or arm}}}
\\\label{dia:po2}\fbox{\xymatrix@=18pt{\obox{A}{.8in}{\rr a college mathematics course}\LA{r}{yields}\LAL{d}{is}&\obox{C}{.8in}{an utterance of the phrase ``too hard"}\\\obox{B}{.6in}{\rr a college course}}}\hsp&\fbox{\xymatrix@=18pt{\obox{A}{.8in}{\rr a college mathematics course}\LA{r}{yields}\LAL{d}{is}&\obox{C}{.8in}{an utterance of the phrase ``too hard"}\LA{d}{}\\\obox{B}{.6in}{\rr a college course}\LA{r}{}&\obox{\parbox{.6in}{\vspace{.1in}\tiny$D=B\!\amalg_A\!C$}}{1in}{\rr a college course, where every mathematics course is replaced by an utterance of the phrase ``too hard"}}}
\end{align}

In Olog (\ref{dia:po1}), the shoulder is seen as part of the arm and part of the torso. When taking the union of these two parts, we do not want to ``double-count" the shoulder (as would be done in the coproduct $B\amalg C$, see Example \ref{ex:coproduct2}). Thus we create a new type $A$ for cells in the shoulder, which are considered the same whether viewed as cells in the arm or cells in the body. In general, if one wishes to take two things and glue them together, the glue serves as $A$ and the two things serve as $B$ and $C$, and the union (or grouping) is the pushout $B\amalg_AC$.

In Olog (\ref{dia:po2}), if every mathematics course is simply ``too hard," then when reading off a list of courses, each math course will not be read aloud but simply read as ``too hard."  To form $D$ we begin by taking the union of $B$ and $C$, and then we consider everything in $A$ to be the same whether one looks at it as a course or as the phrase ``too hard."  The math courses are all blurred together as one thing. Thus we see that the power to equate different things can be exercised with pushouts.

\end{example}

\begin{definition}\label{def:pushout}

Let $A, B,$ and $C$ be sets and let $f\taking A\to B$ and $g\taking A\to C$ be functions. The {\em pushout} of $B\From{f}A\To{g}C$, denoted $B\amalg_AC$, is the quotient of $B\amalg C$ (see Definition \ref{def:coproduct}) by the equivalence relation generated by declaring $b\sim c$ (i.e. $b$ is equivalent to $c$) if: $b\in B, c\in C$, and there exists $a\in A$ with $f(a)=b$ and $g(a)=c$.

\end{definition}

\subsection{Declaring a surjective aspect}\label{sec:surjective}

A function $f\taking A\to B$ is called {\em surjective} if every value in $B$ is the image of something in the domain $A$. For example, the function which subtracts 1 from every integer ($x\mapsto x-1$) is surjective, because every integer has a successor; whereas the function that doubles every integer ($x\mapsto 2x$) is not surjective because odd numbers are not mapped to. The aspect is $\fakebox{a published paper}\To{\tn{was published in}}\fakebox{an established journal}$ is surjective because every established journal has had at least one paper published in it. The aspect is $\fakebox{a published paper}\To{\tn{has as first author}}\fakebox{a person}$ is not surjective because not every person is the first author of a published paper.

The easiest way to indicate that an aspect is surjective is to denote it with a ``two-headed arrow" as in $f\taking A\surj B$. For example, the second map is surjective (and indicated with a two-headed arrow) in the olog $$\fbox
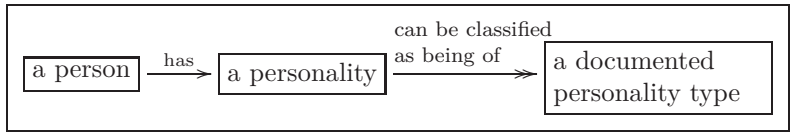
{\xymatrix{\fbox{a person}\ar[r]^-{\tn{has}}&\fbox{a personality}\ar@{->>}[rr]^{\parbox{.9in}{\footnotesize\rr\tn{can be classified as being of}}\normalsize}&&\obox{}{1.1in}{\rr a documented personality type}}}$$  Here the first aspect is not considered surjective, presumably because the author imagines personalities had by no person.

We include surjective aspects in this section because it turns out that surjectivity can also be specified by pushouts. See \cite{nlab-epi} for details.

\subsection{Empty types}\label{sec:empty}

The empty set is a set with no elements; it can be considered the ``empty coproduct."  In other words if we denote $n*A=A\amalg A\amalg\cdots \amalg A$ (where $A$ is written $n$ times), then $0*A$ is the empty coproduct and is the empty set. One can declare a type to be empty by annotating it with $A=\emptyset$ in the olog. For example the olog $$\fbox{$\smbox{A=\emptyset}{a supernatural being}$}$$ says that the set of supernatural beings is empty. As a more concrete example, the intersection of positive numbers and negative numbers is empty, as expressed in the olog $$\fbox
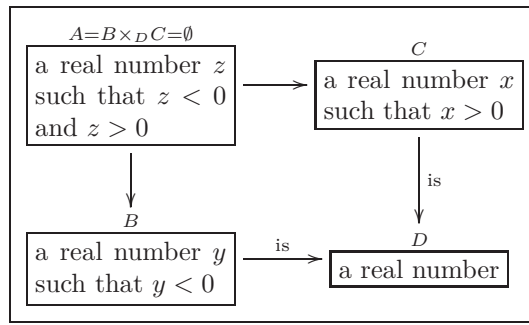
{\xymatrix{\mebox{A=B\cross_DC=\emptyset}{a real number $z$ such that $z<0$ and $z>0$}\ar[r]\ar[d]&\mebox{C}{a real number $x$ such that $x>0$}\LA{d}{is}\\\mebox{B}{a real number $y$ such that $y<0$}\LA{r}{is}&\smbox{D}{a real number}}}$$

\subsection{Images}

In what remains of Section \ref{sec:expressive II}, we will discuss how the ideas of this section and the previous (Section \ref{sec:expressive I}) can be used together to create quite expressive ologs. First we will discuss how each aspect $f\taking A\to B$ has an ``image," the subset of $B$ that are ``hit" by $f$. Then, in Sections \ref{sec:app recursion} and \ref{sec:app mathematics}, we will discuss how ologs can express all primitive recursive functions and many other mathematical concepts.

Consider the olog \begin{align}\label{olog:person computer}\fbox{\xymatrix{&\labox{X}{a pair $(p,c)$ where $p$ is a person, $c$ is a computer, and $p$ owns $c$}\ar[ld]_-{p}\ar[rd]^-c\\\smbox{Y}{a person}&&\smbox{Z}{a computer}}}\end{align}  Some people own more than one computer, and some computers are owned by more than one person. Some computers are not owned by a person, and some people do not own a computer.  The purpose of this section is to show how to use ologs to capture ideas such as ``a person who owns a computer" and ``a computer that is owned by a person". These are called the images of $p$ and $c$ respectively. 

Every aspect has an image, and these are quite important for human understanding. For example the image of the map $\fakebox{a person}\To{\tn{has as father}}\fakebox{a person}$ is the type \fakebox{a father}. In other words, a father is defined to be a person $x$ for which there is some other person $y$ such that $x$ is the father of $y$.  

The image of a function $f\taking A\to B$ is a commutative diagram (fact) $$\xymatrix{A\ar[rr]^f\ar@{->>}[dr]_{f_s}&\ar@{}[d]|(.4){\checkmark}&B\\&\im(f)\ar@{^(->}[ur]_{f_i}}$$ where $f_s$ is surjective and $f_i$ is injective (see Sections \ref{sec:surjective} and \ref{sec:injective}). We indicate that a type is the image of a map $f$ by annotating it with {\bf Im}$(f)$, as in the following olog: $$\fbox{\xymatrix{\smbox{A}{a child}\LA{rr}{has as parents}\ar@{->>}[dd]\LAL{ddrr}{$f$}&&\obox{B}{1.5in}{a pair $(w,m)$ where $w$ is a woman and $m$ is a man}\ar@{}[ld]|(.6){\checkmark}\ar[dd]^m\\&&\\\smbox{C={\bf Im}(f)}{a father}\ar@{^(->}[rr]_{\tn{is}}\ar@{}[ur]|(.7){\checkmark}&&\smbox{D}{a man}}}$$  Hopefully it is also clear that \fakebox{a person who owns a computer} and \fakebox{a computer that is owned by a person} are the images of $p\taking X\to Y$ and $c\taking X\to Z$ (respectively) in Olog (\ref{olog:person computer}).

Using the label {\bf Im}$(f)$ is the easiest way to indicate an image, although one can also do so categorically using limits and colimits. See \cite[Chapter VIII]{Mac} for details.

\subsection{Application: Primitive recursion}\label{sec:app recursion}

We have already seen how ologs can be used to express a conceptual understanding of a situation (all the ologs thus far exemplify this idea). In this section we hope to convince the reader that ologs are also able to express certain computations. In particular we will show by example that primitive recursive functions (like factorial or fibonacci) can be expressed by ologs. In this way, we can to put computation and knowledge representation together into the same framework. It would be quite valuable to strengthen this connection by showing that Ologs (or an extension thereof) can express any recursive function (i.e. simulate Turing machines). This is an open research possibility.

\begin{example}\label{ex:factorial}

In this example we will present an olog that can represent the ``Factorial function," often denoted $n\mapsto n!$, where for example the factorial of $4$ is $24$. Recall that a {\em natural number} is any nonnegative whole number: $0,1,2,3,4,\ldots$. 

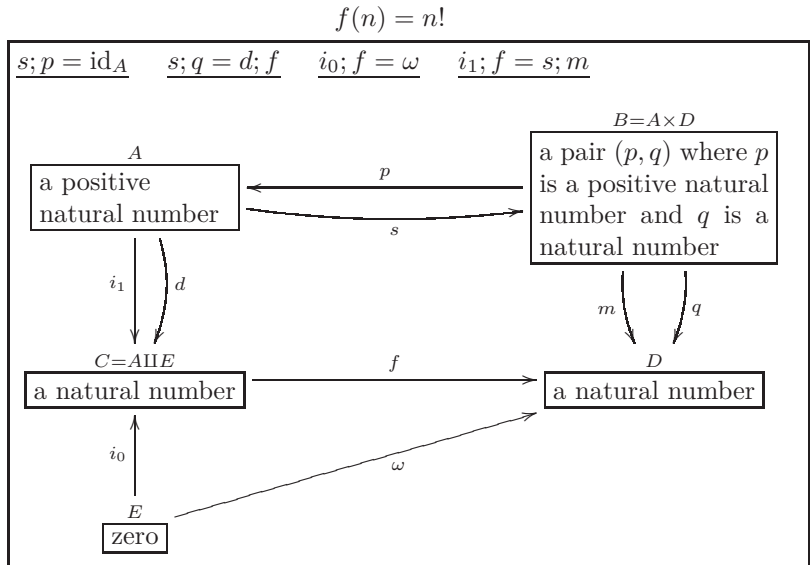
$$\parbox{4in}{\begin{center} $f(n)=n!$\end{center}\fbox{\parbox{4.1in}{$\underline{s;p=\id_A}\hsp\underline{s;q=d;f}\hsp\underline{i_0;f=\omega}\hsp\underline{i_1;f=s;m}$\\\\\xymatrix@=30pt{\obox{A}{1in}{\rr a positive natural number}\ar@/_1pc/[rrr]_{s}\ar@/^1pc/[d]^{d}\ar[d]_{i_1}&&&\obox{B=A\cross D}{1.2in}{a pair $(p,q)$ where $p$ is a positive natural number and $q$ is a natural number}\ar[lll]_-{p}\ar@/^1pc/[d]^-{q}\ar@/_1pc/[d]_-m\\\smbox{C=A\amalg E}{a natural number}\ar[rrr]^f&&&\smbox{D}{a natural number}\\\smbox{E}{zero}\ar[u]^{i_0}\ar[urrr]_{\omega}}}}}$$

The idea of this olog is to convey the factorial function as follows. A natural number is either zero or positive. Every positive natural number $n$ has a decrement, $n-1$. The factorial of zero is 1. The factorial of a positive number $n$ is obtained by multiplying $n$ by the factorial of $n-1$. 

To more explicitly describe the above olog, we must describe its intended instances. Hopefully the instances of each type ($A$ through $E$) are self-explanatory, so we will describe the grouping, the layout, the aspects, and the facts. The set of natural numbers is the disjoint union of zero and the set of positive natural numbers and the maps $i_0$ and $i_1$ are the inclusions into the coproduct, which explains the grouping $C=A\amalg E$. The layout $B=A\cross D$ is self-explanatory, and the maps $p$ and $q$ are the projections from the product. The map $d$ is the decrement map $n\mapsto n-1$, the map $\omega$ sends $0$ to $1$, the map $m$ is multiplication $(n,n')\mapsto n*n'$. Once $m$, $d$, and $\omega$ are so-defined, the first two facts ($s;p=\id_A$ and $s;q=d;f$) specify that $s$ sends $n$ to the pair $(n,f(d(n)))$, and the second two facts specify that $f$ sends $0$ to $1$ and sends a positive number $n$ to  $m(s(n))=m(n,f(d(n)))$, i.e. $n$ goes to the product $n*(n-1)!$.

The above olog defines the factorial function ($f$) in terms of itself, which is the hallmark of primitive recursion. Note, however, that this same olog can compute many things besides the factorial function. That is, nothing about the olog says that the instances of \fakebox{Zero} is the set $\{0\}$, that $\omega$ sends $0$ to $1$, that $d$ is the decrement function, or that $m$ is multiplication --- changing any of these will change $f$ as a function. For example, the same olog can be used to compute ``triangle numbers" (e.g. f(4)=1+2+3+4=10) by simply changing the instances of $\omega$ and $m$ in the obvious ways (use $\omega=0, m=+$ rather than $\omega=1,m=*)$). For a radical departure, fix any forest (set of graphical trees) $F$, let $E=\fakebox{zero}$ represent its set of roots, $A$ the other nodes, $\omega$ the constant 0 function, $d$ the parent function, and $m$ sending $(p,d(p))$ to $f(d(p))+1$. Then for each tree in $F$ and each node $n$ in that tree, the function $f$ will send $n$ to its height on the tree.

\end{example}

Primitive recursion is a powerful technique for deriving new functions from the repetition of others using a kind of ``while loop."  The general form of primitive recursive functions can be found in \cite{BBJ}, and it is not hard to imitate Example \ref{ex:factorial} for the general case.

\subsection{Application: defining mathematical concepts}\label{sec:app mathematics}

In this subsection we hope to convince the reader that many mathematical concepts can be defined by ologs. This should not seem like much of a stretch: ologs describe relationships between sets, so we rely on the maxim that all of mathematics can be formulated within set theory. To make the idea explicit, however, we will recall the definition of pseudo-metric space (in \ref{def:pseudo}) and then provide an olog with the same content (in \ref{olog:metric}).

\begin{definition}\label{def:pseudo}

Let $\RR_{\geq 0}$ denote the set of non-negative real numbers. A {\em pseudo-metric space} is a pair $(X,\delta)$ where $X$ is a set and $\delta\taking X\cross X\to\RR_{\geq 0}$ is a function with the following properties for all elements $x,y,z\in X$:\begin{enumerate}\item $\delta(x,x)=0$;\item $\delta(x,y)=\delta(y,x)$; and\item $\delta(x,z)\leq\delta(x,y)+\delta(y,z)$.\end{enumerate}

\end{definition}

\begin{align}\label{olog:metric}\fbox{\parbox{4.6in}{\underline{$d_0;y=d_1;y$}\hsp\underline{$d_2;z=d_0;z$}\hsp\underline{$d_1;z=d_0;z$}\hsp\underline{$\phi;z=y$}\hsp\underline{$\phi;y=z$}\\\underline{$s;y=\id_7$}\hsp\underline{$s;z=\id_7$}\hsp\underline{$s$;$\delta$=$g$;0}\hsp\underline{$\delta$ $;\phi$=$\delta$ }\\\underline{$d_2;$$\delta$ =$f;a$}\hsp\underline{$d_0$;$\delta$ =$f;b$}\hsp\underline{$d_1$;$\delta$ $=f;c$}\\\xymatrix{\smbox{1=5\cross_75\cross_75}{a triple $(x_0,x_1,x_2)$ of points in $X$}\ar@<3.5ex>[d]_{d_0}\ar[d]_{d_1}\ar@<-3.5ex>[d]_{d_2}\ar[r]^-f&\labox{2}{a triple $(a,b,c)$ of non-negative real numbers such that $c\leq a+b$}\ar@<2.5ex>[d]_c\ar[d]_b\ar@<-2.5ex>[d]_a\\\smbox{3=5\cross_75}{a pair $(y,z)$ of points in $X$}\ar`l[]`d[][]_(0)\phi\LA{r}{$\delta$ }\ar@<2.5ex>[d]^z\ar@<-2.5ex>[d]_y&\smbox{4}{a non-negative real number}\\\smbox{5}{a point in $X$}\ar[d]_X\ar[u]_s\ar[r]_g&\smbox{6=\{*\}}{$\{0\}$}\ar[u]_0\\\smbox{7}{a pseudo-metric space}}}}\end{align} 

As long as the instances for the right-hand side of this olog are mathematically correct (i.e. we assign $4$ the set of non-negative real numbers), this olog has the same content as Definition \ref{def:pseudo}. One can use ologs to define usual metric spaces (in which Property (1) in Definition \ref{def:pseudo} is strengthened), but it would have taken too much space here. 

It should be clear that ologs provide a more precise and explicit description of any concept, relying less on the grammar of English and more on the mathematical ``grammar" of sets and functions. Assumptions are exposed as all the working parts of an object need to be explicitly documented. Thus an olog is likely to be instantly readable by a theorem prover such as Coq (\cite{Coq}), at least if one creates the olog within an appropriate Olog-Coq interface API. Moreover, various parts of this olog may be reusable in other contexts, and hence connect pseudo-metric spaces into a web of neighboring definitions and theorems. 

In fact, once a corpus of mathematics has been written in olog form, evidence of conjectures not yet proven could be written down as instance data. For example, one could record every known prime as instances of a type \fakebox{prime} and a machine could automatically check that Goldbach's conjecture (written as an olog containing \fakebox{prime} as a type) holds for all example ``so far."  With definitions, theorems, and examples all written in the same computer-readable language of ologs, one may hope for much more advanced searching and knowledge retrieval by humans. For example, one could formulate very precise questions as database queries and use SQL on the database corresponding to a given olog (see Section \ref{sec:relationship olog db}).




\section{Further directions}\label{sec:further}

Ologs are basically categories which have text labels to explain their intended semantic. As such there are many directions to explore ranging from quite theoretical to quite practical. Here we consider three main classes: extending the theory of ologs, studying communication with ologs, and implementing ologs in the real world.

\subsection{Extending the theory of ologs}

In this paper we began by discussing basic ologs, which are rich enough to capture the semantic of many situations. In Sections \ref{sec:expressive I} and \ref{sec:expressive II} we added more expressivity to ologs to allow one to encode ideas such as intersections, unions, and images. However, ologs could be even more expressive. One could add ``function types" (also known as exponentials); add a ``subobject classifier type," which could allow for negation and complements as well as power-sets; or even add fixed sets (like the set of Strings) to the language of ologs. This is not too hard (using sketches, see \cite{Mak}); the reason we did not include them in this paper was more because of space than any other reason. 

Another generalization would be to allow the instances of an olog to take values in a category other than $\Set$. For example, one could have an instance-space rather than an instance-set, e.g. it is clear that the instances of the type \fakebox{a point on the unit circle} constitute a topological space. One could similarly argue that the instances of the type \fakebox{a human invention} have a topology or metric as well (e.g. as an invention, the cellphone is closer to the telephone than it is to artificial flavoring).  Instance data on an olog $\mcC$ corresponds to a functor $\mcC\to\Set$ in this paper, but it is quite easy to replace $\Set$ with a different category such as ${\bf Top}$ (the category of topological spaces), and this may have interesting uses in data modeling.

In Section \ref{sec:app mathematics}, we explicitly showed that pseudo-metric spaces (and we stated further that metric spaces) can be presented by ologs. It would be interesting to see if theorems could also be proven entirely within the context of ologs. If so, a teacher could first sketch a mathematical proof as a small or sparse olog $\mcC$, and then use a functor $\mcC\to\mcD$ to rigorously ``zoom in" on that proof so that the sketch becomes a full-fledged proof (as the maps in $\mcC$ are factored into understandable units in $\mcD$). 

If ologs are to be viable venues in which to discuss results in mathematics, then they should be capable of describing all recursion, not just primitive recursion (as in Section \ref{sec:app recursion}). We do not yet have an understanding for how this can be done. If recursion can be fully defined with the ologs described above, it would be interesting to see it written out; if not, it would be interesting to understand what basic idea could be gracefully added to ologs so that recursion becomes expressible.

In a different direction, one could test the expressive power of ologs by defining simple games, like Tic Tac Toe or Chess, using ologs. It would be impressive to define a vocabulary for writing games and a program which could automatically convert an olog-defined game into a playable computer game. This would show that the same theory that we have seen express ideas about fatherhood and factorials can also be used to invent games and program computers.

\subsection{Studying communication with ologs}\label{sec:studying comm}

As discussed in Section \ref{sec:connecting ologs}, ologs can be connected by functors into networks that are not just 2-way, but $n$-way. These communication networks should be studied: what kinds of information can pass, how reliable is it, how quickly can it spread, etc. This may be applicable in fields from economics to psychology to sociology. Such research may use results from established mathematics such as Network Coding Theory (see \cite{YLC}).

In \cite{SA}, we study how two or more entities (described as ologs) can communicate new ideas (not just new instance data) to each other. It would be interesting to see how well this ``communication protocol" works in practice, and whether it can be theoretically automated. Furthermore, this communication protocol and any theoretical automation of it should be implemented on a computer to see if different database schemas can be meaningfully integrated with minimal human assistance. 

It may be possible to train children to create ologs about their interests or about a given lesson. These ologs would show how the child actually perceives something, which would probably be fascinating. By our experience and that of people we have taught, the process of building an olog usually leads to a clarification of the concepts involved. Moreover, a class project to connect the ologs of different students and between the students and the teacher, may have excellent pedagogical benefits. 

Finally, it may be interesting to study ``local truth" vs. \!\!\!``global truth" in a network of ologs. Functorial connections between ologs can allow for translation of ideas between members of a group, but there may be ideas which do not extend globally, just as a M\"{o}bius band does not admit a global orientation. That is, given three parties on the M\"{o}bius band, any pair can agree on a compass orientation, but there is no choice that the three can simultaneously agree on. Similarly, whether or not it is possible to construct a global language which extends all the existing local ones could be determined if these local languages and their connections were entered into a computer olog system.

\subsection{Implementing ologs in the real world}

Once ologs are implemented on computers, and once people learn how to author good ologs, much is possible. One advantage comes in searching the information space. Currently when we search for a concept (say in Google or on our hard drive), we can only describe the concept in words and hope that those words are found in a document describing the concept. That is, search is always text-based. Better would be if the concept is meaningfully interconnected in a web of concepts (an olog) that could be navigated in a meaningful (as opposed to text-based) way. 

Indeed, this is the semantic web vision: When internet data is machine-readable, search becomes much more powerful. Currently, we rely on RDF scrapers that scour web pages for $\langle$subject, predicate, object$\rangle$ sentences and store them in RDF format, as though each such sentence is a fact. Since people are inputting their data as prose text, this may be the best available method for now; however, it is quite inaccurate (e.g. often 15\% of the facts are wrong, a number which can lead to degeneration of deductive reasoning -- see \cite{MBCH}). If ideas could be put on the internet such that they compatibly made sense to both human and computer, it would give a huge boost to the semantic web. We believe that ologs can serve as such a human-computer interface.

While it is often assumed that because we all speak the same language we all must mean the same things by it, this is simply not true. The age-old question about whether ``blue for me" is the same as ``blue for you" is applicable to every single word and idiom in our language. There is no easy way to sync up different people's perceptions. If communication is to be efficient, agreements must be fairly explicit and precise, and this precision demands a rigor that is simply unavailable in English prose. It is available in a network of ologs (as described in Section \ref{sec:connecting ologs}). 

For example, the laws of the United States are hopelessly complex. Residents of the US are required to obey the laws. However, unlike the rules of the Scholastic Aptitude Test (SAT), which take 10 minutes for the proctor to read aloud, the laws of the US are never really expressed --- the most important among them are hopefully picked up by cultural osmosis. If an olog was created which had enough detail that laws could be written in that format, then a woman could research for  herself whether her landlord was required to fix her refrigerator or whether this was her responsibility. It may prove that the olog of laws is internally inconsistent, i.e. that it is impossible for a person to satisfy all the laws --- such an analysis, if performed, could fundamentally change our outlook on the legal system.

The same goes for science; information written up in articles is much less accessible than information that is entered into an ontology. However, the dream of a single universal ontology is untenable (\cite{Min}). Instead we must allow each lab or institute to create its own ontology, and then require citations to be functorial olog connections, rather than mere silo-to-silo pointers. Thus, a network of ologs should be created to represent the understanding of the modern scientific community as a multi-faceted whole.

Another impetus for a scientist to write an olog about the study at hand is that, once an olog is made, it can be instantly converted to a database schema which the scientist can use to input all the data pertaining to this study. Indeed, if some data did not fit within this schema, then the olog must have been insufficient to begin with and should be modified to fully describe the experiment. If scientists work this way, then the separation between them and database modelers can be reduced or eliminated (the scientist assumes the database modeling role with little additional burden). Moreover, if functorial connections are established between the ologs of different labs, then data can be meaningfully shared along those connections, and ideas written in the language of one lab's olog can be translated automatically into the language of the other's. The speed and accuracy of scientific research should improve.


\bibliographystyle{amsalpha}

\end{document}